\documentclass[12pt,prb,aps]{revtex4-1}
\usepackage{amsmath}           		          	
\usepackage{graphicx,epstopdf}					
\usepackage{amssymb}
\usepackage{fullpage}
\usepackage{color}
\usepackage{esint}
\pdfoutput = 1 
\newcommand {\bxi}{\mbox{\boldmath$\xi$}}
\allowdisplaybreaks

\begin{document}
\title{Investigation of  Tearing Mode Stability Near Ideal Stability Boundaries Via Asymptotic Matching Techniques}
\author{Richard Fitzpatrick\,\footnote{rfitzp@utexas.edu}}
\affiliation{Institute for Fusion Studies, Department of Physics, University of Texas at Austin, Austin TX 78712}

\begin{abstract}
A number of improvements to the TJ toroidal tearing mode code [Phys.\ Plasmas {\bf 31}, 102507 (2024)] are documented. The TJ
code is also successfully benchmarked against the STRIDE toroidal tearing mode code [Phys.\ Plasmas {\bf 25}, 082502 (2018)]. Finally, the 
 new capabilities of the TJ code are used to investigate the stability of tearing modes in tokamak plasmas as
an ideal stability boundary, associated with either an external-kink or an internal-kink mode, is  approached. 
All elements of the tearing stability matrix are found to tend to infinity as an ideal stability boundary is approached. 
Furthermore, as the stability boundary is approached, the eigenfunctions of the various tearing modes in the plasma, which are decoupled by sheared plasma rotation,
are all found to morph into that of the marginally-stable ideal mode. However, the growth-rates and real frequencies of the various ``ideal-tearing-modes'' are different from one another.
Moreover, the growth-rate of the ideal-tearing-mode that reconnects magnetic flux at the rational surface that lies closest to the
edge of the plasma is the one that tends to a very large value as the stability boundary is approached. A relatively simple test for
ideal stability that is capable of detecting stability boundaries for external-kink and internal-kink modes, even in the presence of
a very close-fitting ideal wall, is described and verified.

\end{abstract}
\maketitle

\section{Introduction} 
The calculation of the stability of a tokamak plasma to tearing perturbations is most efficiently formulated as  an asymptotic
matching problem in which the  plasma is  divided into two distinct regions.\cite{fkr}  In the ``outer region'', which comprises most
of the plasma, the tearing perturbation is described by the equations of linearized, marginally-stable, ideal magnetohydrodynamics (a.k.a.\ the ``ideal-MHD'' equations). 
However, these equations become singular on   ``rational'' magnetic flux-surfaces at which the perturbed magnetic field resonates with the equilibrium field. In the ``inner region'', which
consists of a set of narrow layers centered on the various rational surfaces, non-ideal-MHD effects such as plasma inertia, plasma resistivity, 
plasma viscosity,  diamagnetic flows, the ion sound radius,  and perpendicular and parallel energy transport become important.\cite{ara,hkm,fw,cole,diff}  The resistive layer
solutions in the various segments of the inner region must be simultaneously asymptotically matched to the ideal-MHD solution in the outer region to produce a matrix 
dispersion relation that determines the growth-rates and  rotation frequencies of the various tearing modes to which the plasma is subject.\cite{con0,cht} (See Sect.~\ref{disp}.)

In general, the  determination of the ideal-MHD contributions to the elements of the  tearing mode dispersion relation is an exceptionally challenging computational task.\cite{connor,nish,gal,pletz,pletz1,tokuda,brennan,ham,ham1,ham2,am1,am2,am3,aglas,aglas1,aglas2}
One way of greatly reducing the complexity of this task is to employ an inverse aspect-ratio expanded plasma equilibrium.\cite{greene,gim,inverse} In such an equilibrium,
the metric elements of the flux-coordinate system can be expressed analytically in terms of a relatively small number of  flux-surface functions,
which represents a major simplification.\cite{con0} Another significant advantage of an inverse aspect-ratio expanded equilibrium is that the magnetic perturbation in the plasma can be efficiently 
matched to an exterior vacuum solution  that is expressed as an expansion in toroidal functions.\cite{am1} (See Sect.~\ref{vacuum}.) The aspect-ratio expansion approach to the tearing mode stability problem  was previously implemented in the T3\,\cite{connor} and T7\,\cite{am1} codes, and has recently been reimplemented in an improved form in the TJ code,
which is described in Ref.~\onlinecite{tj}.

The first aim of this paper is to document three major improvements to the TJ code. 

As is well known, tearing modes in tokamak plasmas become extremely
unstable as an ideal stability boundary is approached.\cite{brennan,bren1,bren2} When investigating this phenomenon, it is imperative to know the exact location
of the  stability boundary. Appendix~\ref{ideal} explains how the the ideal-MHD solutions in the outer region, calculated by the TJ code, can be repurposed to
construct a complete set of marginally-stable ideal eigenfunctions, and how these eigenfunctions can then be used to calculate the total ideal perturbed potential
energy, $\delta W$, of all possible ideal modes.  Of course, the ideal stability boundary corresponds to the point at which the smallest possible value of $\delta W$ passes through
zero.\cite{freidberg,ideal}

Previously, the TJ code either allowed for fixed boundary calculations, or calculations in which the plasma is surrounded by a vacuum region. Appendix~\ref{vacuum} describes a
new capability that allows for the presence of a perfectly conducting wall in the vacuum region. The radius of the wall, relative to the plasma, can be varied. 

Previously, the TJ code only calculated the ``tearing stability matrix'' that emanates from the ideal-MHD solution in the outer region. However, this is only half of the tearing stability problem. 
In order to determine the growth-rates and rotation frequencies of possible tearing modes, it is necessary to combine the data contained in the tearing stability matrix with data obtained from resistive layer solutions
in the various segments of the inner region. Appendices~\ref{lmodel} and \ref{asym}   describe a new resistive layer model that has been implemented in the TJ code. The model takes into account 
plasma resistivity, plasma inertia, electron and ion diamagnetic flows, the decoupling of the electron and ion fluids on lengthscales below the ion sound radius, perpendicular
momentum transport, parallel and perpendicular energy transport, and the stabilizing effect of average magnetic field-line curvature.

The second aim of the paper is to benchmark the results of the TJ toroidal tearing mode code against results from the STRIDE\,\cite{aglas1} toroidal tearing mode code. This benchmarking
exercise is described in Sect.~\ref{benchmark}. 

The final aim of the paper is to combine all of the new features of the TJ code in an investigation of the stability of tearing modes in tokamak plasmas as
an ideal stability boundary is approached. We are particularly interested in understanding the similarities and differences when the ideal stability boundary in question is associated with
an {\em external-kink}\/ mode and an {\em internal-kink}\/ mode. This investigation is described in Sect.~\ref{invest}.

Finally, the paper is summarized, and conclusions are drawn, in Sect.~\ref{conc}. 

\section{Benchmark of TJ Code against STRIDE Code}\label{benchmark}
\subsection{Introduction}
The STRIDE toroidal tearing mode code\,\cite{aglas1} performs the same task as the RDCON toroidal tearing mode code,\cite{aglas2} but in a generally more reliable fashion. Both codes are built on top of the
DCON ideal stability code.\cite{dcon} The three latter codes are all  part of the GPEC package.\cite{gpec} This section compares the tearing stability matrix, $E_{kk'}$ (see Sect.~\ref{disp}), where $k$ and $k'$ index the various rational surfaces in the
plasma (so if there are $K$ rational surfaces then the innermost is the $k=1$ surface, the next innermost is the $k=2$ surface, et cetera), calculated by the TJ code with the equivalent symmetrized tearing stability matrix, $\hat{\mit\Delta}_{kk'}$, calculated by the STRIDE code. (Appendix~\ref{comp} explains how
$\hat{\mit\Delta}_{kk'}$ is related to the actual tearing stability matrix, ${\mit\Delta}_{kk'}$, calculated by STRIDE.)

\subsection{Plasma Equilibrium}
In the TJ code, the plasma equilibrium is determined by the shape of the plasma boundary, which is taken to be circular in this paper, as well as two profile functions. 
The lowest-order safety-factor profile is written\,\cite{tj}
\begin{equation}
q(r) = \frac{q_0\,\nu\,(r/a)^2}{1-[1-(r/a)^2]^\nu},
\end{equation}
and the normalized (see Sect.~\ref{norm}) pressure profile takes the form
\begin{equation}
p(r) = \frac{\beta_0}{2}\left[1-\left(\frac{r}{a}\right)^2\right]^{p_p}.
\end{equation}
Here, $r$ is the flux-surface label with the dimensions of length that is defined in Sect.~\ref{coord}.  Moreover, the magnetic axis corresponds to $r=0$, and the plasma
boundary to $r=a$.  Note that $a$ is the effective inverse aspect-ratio of the plasma (given that all lengths are normalized to the major radius of the magnetic axis, $R_0$---see Sect.~\ref{norm}.)
Furthermore, $q_0=q(0)$ is the safety-factor on the magnetic axis, and  $\beta_0$ is the central plasma beta. 
The value of $\nu$ is adjusted to obtain the desired value of the safety-factor at the plasma boundary, $q_a=q(a)$. 

In the  STRIDE code, we make use of a slightly modified form of the \verb|lar| 
large-aspect-ratio, circular cross-section analytic equilibrium. This equilibrium is also characterized by two profile functions. 
The 
 normalized parallel current profile, $\sigma={\bf J}\cdot{\bf B}/B^{\,2}$,  is written
\begin{equation}
\sigma(r) = \frac{2}{q_0}\left[1-\left(\frac{r}{a}\right)^2\right]^{p_\sigma}.
\end{equation}
whereas the  normalized pressure profile takes the form
\begin{equation}
p(r)= \frac{\beta_0}{2}\left[1-\left(\frac{r}{a}\right)^2\right]^{p_p}.
\end{equation}
In order to obtain (almost) the same equilibrium in the two codes (each of which is based on an expansion of the Grad-Shafranov equation in terms of the inverse aspect-ratio of the plasma), we choose common values of $q_0$, $p_\sigma$, $\beta_0$, and $p_p$, and then
adjust the parameter $\nu$ in the TJ code until both codes have the same value of the safety-factor at the plasma boundary, $q_a$. This procedure ensures that the locations of the various rational surfaces in the plasma are
almost identical in both codes. All of the benchmark tests  described in this section calculate the stability of $n=1$ tearing modes  in free-boundary plasma equilibria (i.e., plasmas surrounded by a vacuum region that contains no ideal wall). The TJ calculations include all poloidal harmonics in the range $m=-10$ to $m=+20$. 

\subsection{Single Rational Surface}
We, first,  consider the stability of $n=1$ tearing modes 
in a zero pressure plasma equilibrium with a circular poloidal boundary that only contain a single rational surface: namely, the $m=2/n=1$ surface. 

\subsubsection{Benchmark Test 1}
\begin{figure}
\centerline{\includegraphics[width=\textwidth]{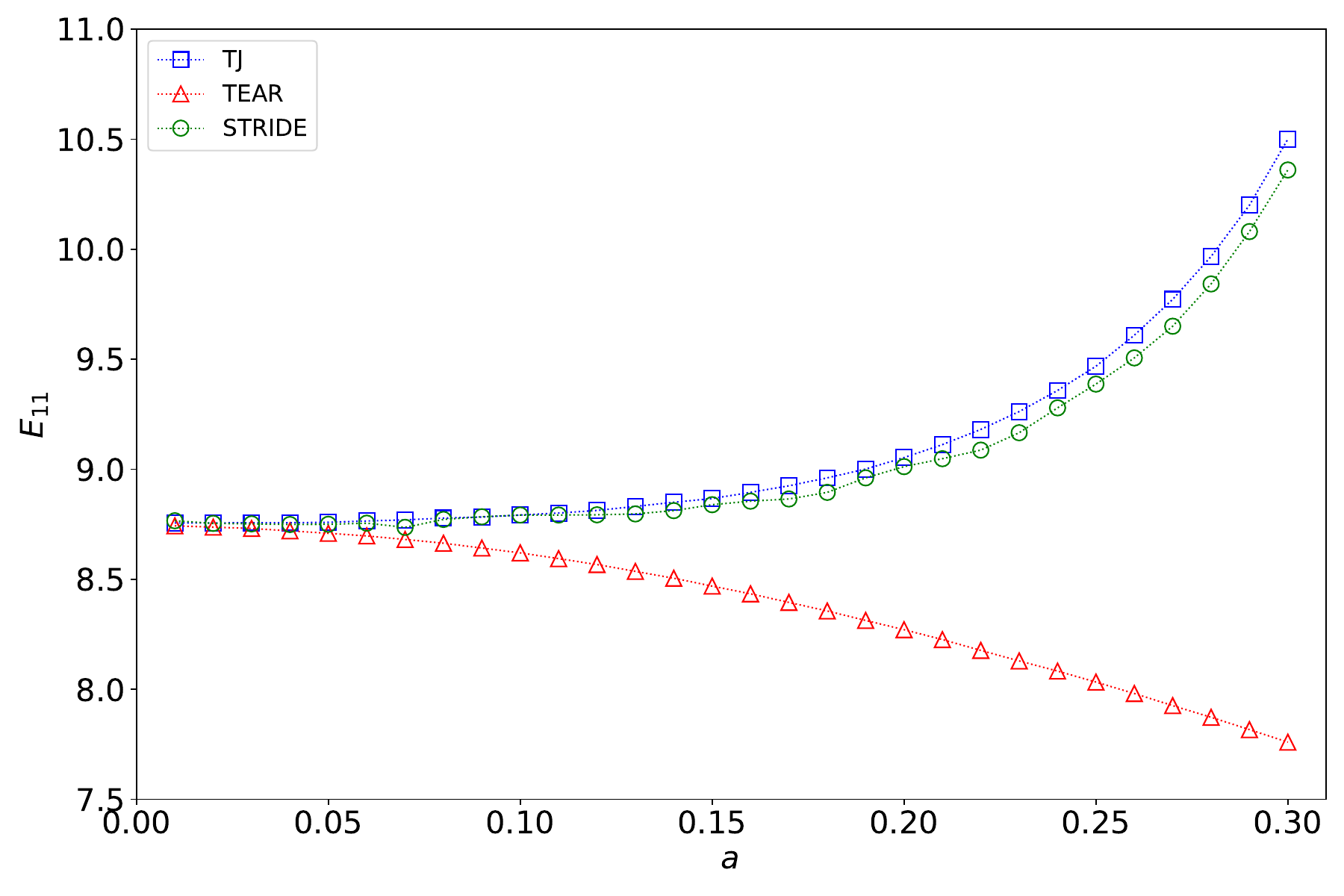}}
\caption{TJ/STRIDE Benchmark Test 1. Circular cross-section plasma with  $q_0=1.1$, $p_\sigma=1.36$,  and $\beta_0=0.0$. Variation of the  free-boundary $m=2/n=1$ tearing stability index with the inverse-aspect ratio, $a$. \label{fig1}}
\end{figure}

The first benchmark  test has plasma equilibria characterized by $q_0=1.1$ and $p_\sigma=1.36$, with a range of different inverse aspect-ratios, $a$. 
Figure~\ref{fig1} compares the $E_{11}$ (i.e.,
the tearing stability index of the  $m=2/n=1$ tearing mode) values calculated by the TJ code,\cite{tj} the TEAR code (which is a cylindrical tearing mode code), and
the STRIDE code. It can be seen that the tearing stability indices calculated by  TJ and STRIDE are in very good agreement, and that both asymptote to the stability index calculated by the TEAR code
in the cylindrical limit, $a\rightarrow 0$. Given that  $E_{kk'}$ is necessarily Hermitian,\cite{tj} the imaginary part of $E_{11}$ should be zero. This is found to be the case, to a high
degree of accuracy, in both the TJ and the STRIDE codes. 

\subsubsection{Benchmark Test 2}
\begin{figure}
\centerline{\includegraphics[width=\textwidth]{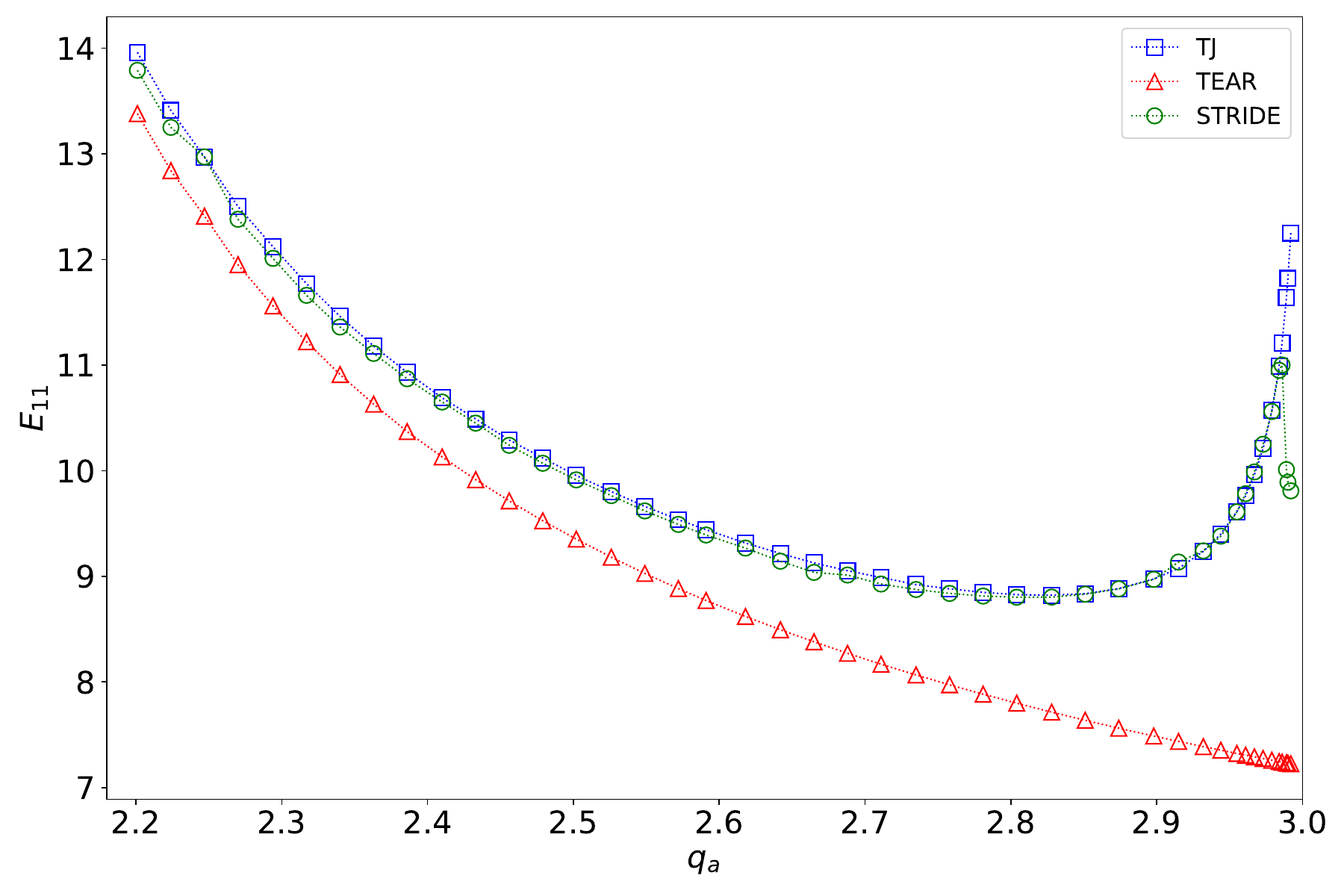}}
\caption{TJ/STRIDE Benchmark Test 2. Circular cross-section plasma with  $q_0=1.1$, $a=0.2$, and $\beta_0=0.0$. Variation of the free-boundary  $m=2/n=1$ tearing stability index with  the edge safety-factor, $q_a$.\label{fig2} }
\end{figure}

The second benchmark test has plasma equilibria characterized by $q_0=1.1$ and $a=0.2$, with a range of different edge safety-factor values, $q_a$. 
Figure~\ref{fig2} compares the $E_{11}$ (i.e.,
the tearing stability index of the $m=2/n=1$ tearing mode) values calculated by the TJ code, the TEAR code, and
the STRIDE code. It can be seen that the tearing stability indices calculated by the TJ and STRIDE are in good agreement, except when the $q=3$ surface very closely approaches the
plasma boundary (i.e., $q_a\rightarrow 3$). 
Note that, according to TJ and STRIDE,  $E_{11}$ becomes very large as $q_a\rightarrow 3$ because the plasma approaches an ideal stability boundary. On the other hand,
the $E_{11}$ values calculate by TEAR shows no such increase, because the ideal stability boundary is associated with the $m=3/n=1$ external-kink mode, and this mode does not couple to the $m=2/n=1$ tearing mode in a cylindrical plasma. 

\subsection{Two Rational Surfaces}
 Next, we consider the stability of $n=1$ tearing modes 
in a  plasma equilibrium with a circular poloidal boundary  that  contains two rational surfaces: namely, the $m=2/n=1$ surface and the $m=3/n=1$ surface. 

\subsubsection{Benchmark Test 3}
\begin{figure}
\centerline{\includegraphics[width=0.9\textwidth]{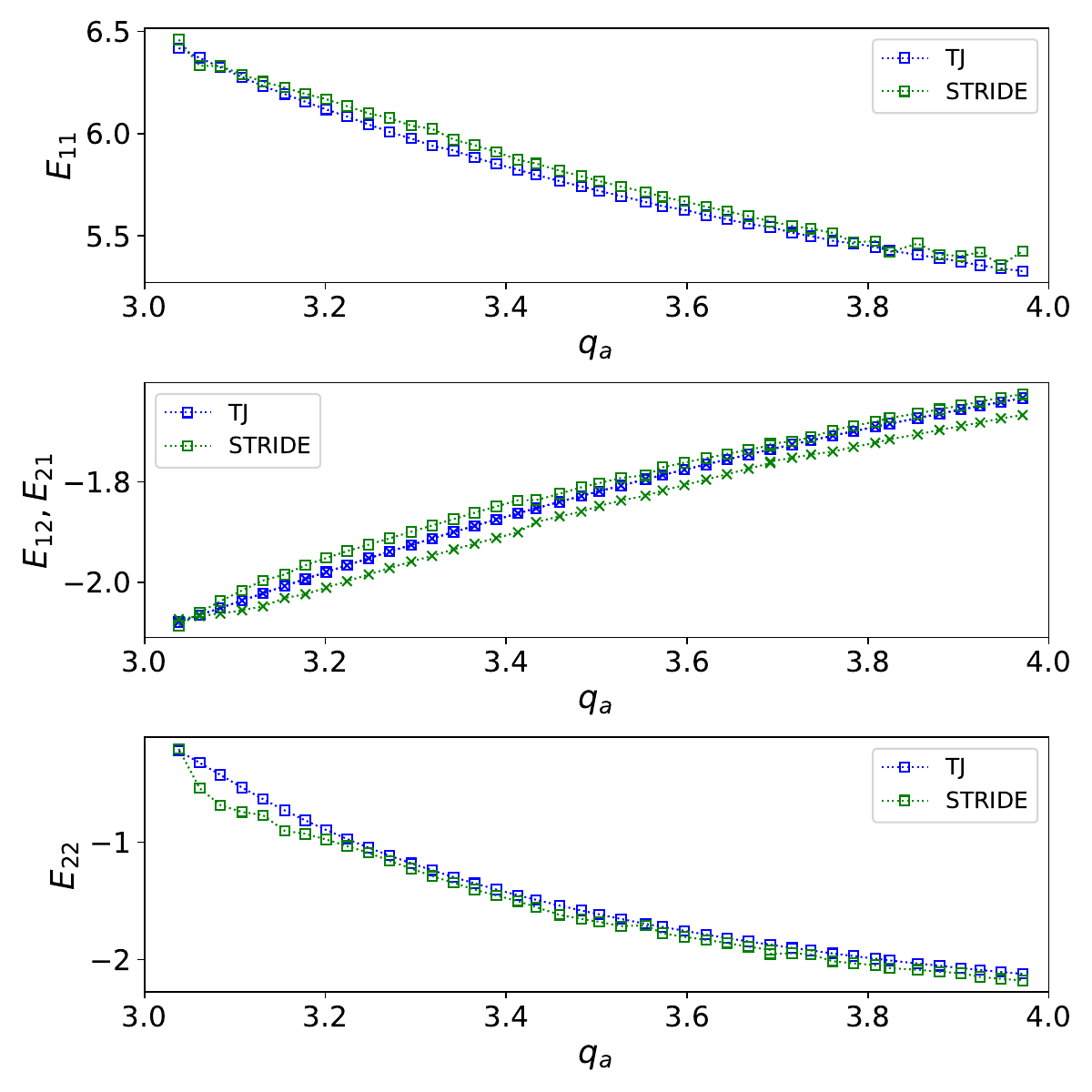}}
\caption{TJ/STRIDE Benchmark Test 3. Circular cross-section plasma with $q_0=1.1$, $a=0.2$, and $\beta_0=0.0$. Variation of the elements of the free-boundary $n=1$ tearing stability matrix with $q_a$. In the middle panel, the squares
indicate  $E_{12}$ and the crosses indicate $E_{21}$. \label{fig3}}
\end{figure}

The third benchmark test has plasma equilibria characterized by $q_0=1.1$ and $a=0.2$, with a range of different edge safety-factor values, $q_a$. 
Figure~\ref{fig3} compares the elements of the tearing stability matrix, $E_{kk'}$,  calculated by the TJ and the STRIDE codes. It can be seen that the elements calculated by
the two codes are in good agreement. In particular, both codes predict that $E_{21}=E_{12}$, as should be the case because $E_{kk'}$ is necessarily Hermitian.\cite{tj} 
Both codes also predict that the imaginary parts of $E_{11}$ and $E_{22}$ are zero to a high degree of accuracy (again, as should be the case, because $E_{kk'}$ is Hermitian),
and that the imaginary part of $E_{12}$ and $E_{21}$ are also zero (as should be the case, because the plasma equilibrium is up-down symmetric). 

\subsubsection{Benkmark Test 4}
\begin{figure}
\centerline{\includegraphics[width=0.9\textwidth]{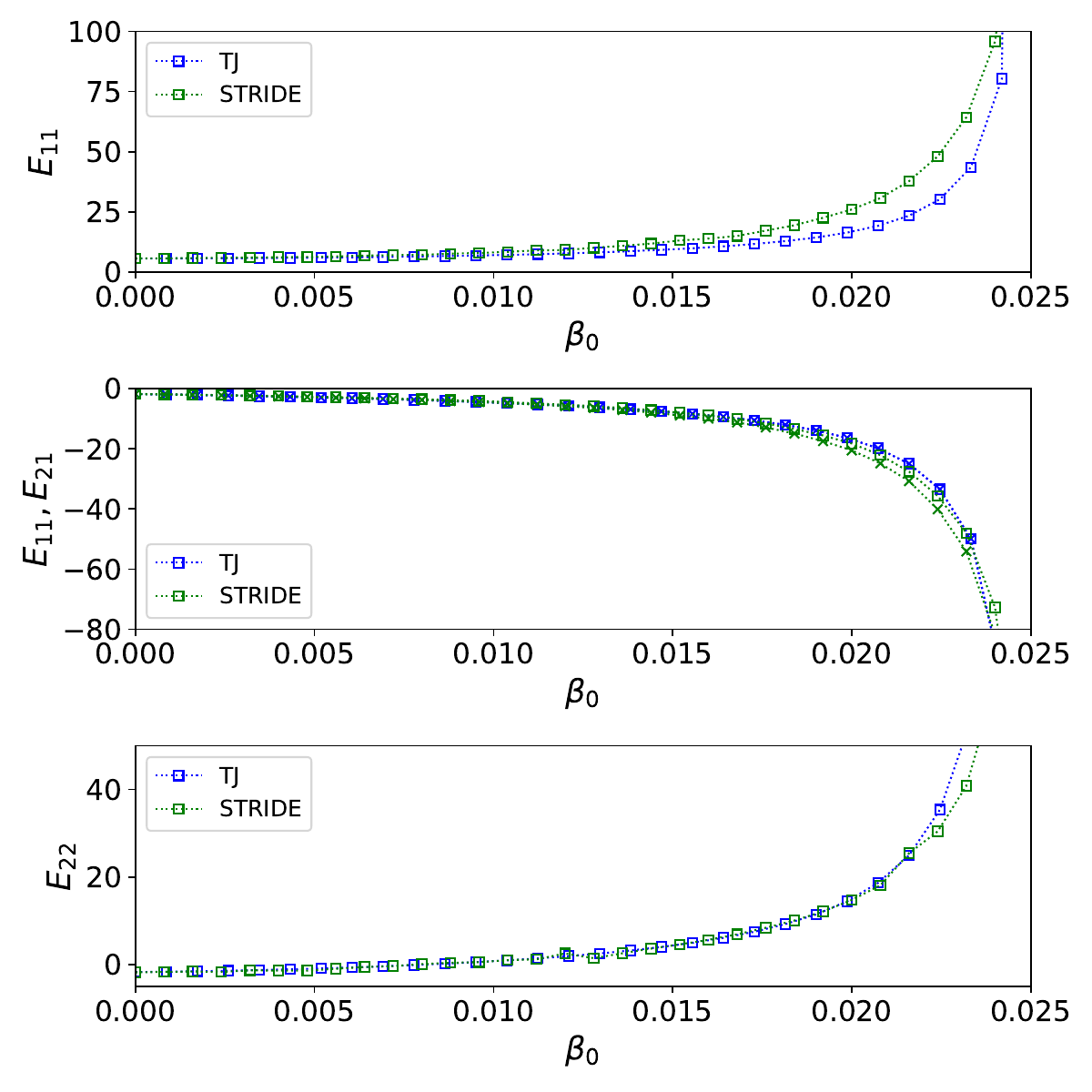}}
\caption{TJ/STRIDE Benchmark Test 4. Circular cross-section plasma with $q_0=1.1$, $a=0.2$, $p_\sigma=2.1$,  and $p_p=2.0$. Variation of the elements of the free-boundary $n=1$ tearing stability matrix with the central plasma beta,  $\beta_0$. In the middle panel, the squares
indicate  $E_{12}$ and the crosses indicate $E_{21}$. The $\beta_0$ values from TJ have  been rescaled by a factor $1.08$.\label{fig4}}
\end{figure}

The fourth, and final,  benchmark test has plasma equilibria characterized by $q_0=1.1$, $a=0.2$, $p_\sigma = 2.1$, and $p_p=2.0$, with a range of different $\beta_0$ values.  
Figure~\ref{fig4} compares the elements of the tearing stability matrix calculated by the TJ and the STRIDE codes. The $\beta_0$ values from TJ have been rescaled
by a factor $1.08$ to compensate for the slightly different $\beta$-limits in the slightly different TJ and STRIDE equilibria. (The STRIDE $\beta$-limit, at which all of the
elements of the $E_{kk'}$ matrix become infinite,  lies at
$\beta_0=0.025$.)
 Once this correction has been made, it can be seen that the elements calculated by
TJ and STRIDE are in reasonably good agreement.  

\subsection{Conclusions}
The elements of the tearing stability matrix calculated by the TJ and STRIDE codes are in good agreement with one another. Given that that the RDCON code,
which has been benchmarked against the STRIDE code, has been previously benchmarked against the MARS-F and PEST-III codes,\cite{am2,aglas2} we can be confident  that the TJ code is operating as designed. 

\section{Tearing Mode Stability Near Ideal Stability Boundaries}\label{invest}

\subsection{Tearing Mode Dispersion Relation}
Consider the stability of a tokamak plasma to a tearing perturbation that is characterized by the toroidal mode number $n$. (See Sect.~\ref{perturbed}.)
Suppose that there are $K$ rational surfaces in the plasma at which the perturbation resonates with the equilibrium magnetic field. (See Sect.~\ref{rational}.)
Let ${\mit\Psi}_k$, ${\mit\Delta\Psi}_k$, and ${\mit\Delta}_k$ be the reconnected helical magnetic flux, the helical current sheet density, and the
tearing-parity layer response parameter, respectively, at the $k$th rational surface. (See Sects.~\ref{rational} and \ref{disp}.) Note that all of these
quantities are, in general, complex. The matrix tearing-mode dispersion relation takes the form (see Sect.~\ref{disp}) 
\begin{equation}\label{tmdisp}
\sum_{k'=1,K}(E_{kk'}-{\mit\Delta}_k\,\delta_{kk'})\,{\mit\Psi}_{k'} = 0
\end{equation}
for $k=1,K$. Here, $E_{kk'}$ is the Hermitian tearing stability matrix determined from the ideal-MHD solution in the outer region. 
Moreover, ${\mit\Delta\Psi}_k = {\mit\Delta}_k\,{\mit\Psi}_k$ for $k=1,K$. Note that, in this paper, we are neglecting the twisting-parity responses of the
various segments of the inner region to the ideal-MHD solution in the outer region in favor of the, generally, much larger tearing-parity responses.\cite{am3}
In fact, the neglect of the twisting-parity responses   is not a good approximation in very high-$\beta$ plasmas,\cite{bren2} but is justified in the relatively low-$\beta$
plasmas considered in this paper. 

The tearing-parity layer response parameter at the $k$th rational surface takes the form (see Sect.~\ref{match})
\begin{equation}
{\mit\Delta}_k = S_k^{\,1/3}\,\hat{\mit\Delta}_k(\gamma_k) + {\mit\Delta}_{k\,{\rm crit}},
\end{equation}
where
\begin{equation}\label{dcrit}
{\mit\Delta}_{k\,{\rm crit}} =- \sqrt{2}\,\pi^{3/2}\,D_{R\,k}\,\frac{r_k}{\delta_{d\,k}}.
\end{equation}
Here, $r_k$ (see Sect.~\ref{coord}), $S_k$ (see Sect.~\ref{rpara}), $D_{R\,k}$ [see Eq.~(\ref{dr})], and $\delta_{d\,k}$ (see Sect.~\ref{match}) are the minor radius,  Lundquist number, resistive interchange parameter, and critical layer width above
which parallel thermal transport forces the electron temperature to be a flux-surface function, respectively, at the $k$th rational surface.  Moreover, $\hat{\mit\Delta}_k(\gamma_k)$ is
the ${\cal O}(1)$ complex layer response parameter obtained by solving  resistive layer equations in the segment of the inner region centered on the $k$th rational
surface. (See Sect.~\ref{lmodel}.) This quantity is a function of the complex growth-rate of the tearing mode, $\gamma_k$,  as seen in the local E-cross-B frame at the surface. 

In general, the elements of the $E_{kk'}$ matrix, the $\hat{\mit\Delta}_k$ values, and  the ${\mit\Delta}_{k\,{\rm crit}}$ values are all ${\cal O}(1)$, whereas the
$S_k^{\,1/3}$ values are much greater than unity. In this case, the only way of finding a solution of the tearing mode dispersion relation, (\ref{tmdisp}), is
to choose $\gamma_k$, for some particular value of $k$,  in such a manner that $|\hat{\mit\Delta}_k|\ll 1$. However, because of sheared E-cross-B and diamagnetic rotation in  the plasma, 
this choice of $\gamma_k$ is generally such that $|\hat{\mit\Delta}_{k'}|\sim{\cal O}(1)$ for $k\neq k'$.\cite{am1} In this situation, the $K\times K$ 
matrix tearing mode dispersion relation, (\ref{tmdisp}), decouples to give $K$ independent dispersion relations of the form\,\cite{am1}
\begin{equation}\label{decouple}
S_k^{\,1/3}\,\hat{\mit\Delta}_k(\gamma_k) = E_{kk}-{\mit\Delta}_{k\,{\rm crit}}
\end{equation}
for $k=1,K$. Here, the $k$th decoupled dispersion relation corresponds to a tearing mode that only reconnects helical magnetic flux at the
$k$th rational surface. The complex growth-rate of this mode only depends on the resistive layer physics at the $k$th rational surface. The
real quantity $E_{kk}$ is the effective tearing stability index of the mode.\cite{fkr} Moreover, the real quantity ${\mit\Delta}_{k\,{\rm crit}}$ is the critical
value of $E_{kk}$ above which the mode becomes unstable. 

The marginal stability point of the $k$th tearing mode is $E_{kk}=  {\mit\Delta}_{k\,{\rm crit}}$, rather than $E_{kk}=0$,\cite{fkr,ara} because of the
stabilizing effect of average equilibrium magnetic field-line curvature in tokamak plasmas.\cite{ggj,ggj1} Note that our expression, (\ref{dcrit}),  for
${\mit\Delta}_{k\,{\rm crit}}$ differs from that of Refs.~\onlinecite{ggj} and \onlinecite{ggj1} because  $\delta_{d\,k}$ is the critical layer width above which
parallel thermal transport dominates perpendicular transport, which has no dependence on $S_k$, rather that the much smaller resistive
layer width, which scales as $S_k^{\,-1/4}$.\cite{lut,con1} This fact resolves a mystery in tearing mode stability theory. According to Refs.~\onlinecite{ggj} and \onlinecite{ggj1},
large hot tokamak plasmas should be much more stable to tearing modes than small cold tokamak plasmas, because the latter plasmas are characterized by much
larger Lundquist numbers, and the critical value of the tearing stability index, $E_{kk}$,  above which the tearing mode that reconnects
magnetic flux at the $k$th rational surface becomes unstable scales as ${\mit\Delta}_{k\,{\rm crit}}\propto S_k^{\,1/4}$. In fact, large  tokamak plasmas are only moderately more
stable to tearing modes than small  plasmas. The reason for this is that when parallel transport is taken into account, ${\mit\Delta}_{k\,{\rm crit}}$, in large tokamak
plasmas is of order unity, rather than much greater than unity, and does not depend on $S_k$. Thus, the stabilizing effect of average magnetic field-line curvature is a moderate effect in
large tokamak plasmas, rather than a dominant effect. Nevertheless, this moderate stabilizing effect is important. For instance, it is virtually impossible to find a
sensible tokamak equilibrium in which $E_{kk}$ for the $m=2/n=1$ tearing mode is not positive. However,  the classical $m=2/n=1$ tearing mode is usually observed to
be stable in large tokamak plasmas. The reason for this is that $E_{kk}$ does not generally exceed ${\mit\Delta}_{k\,{\rm crit}}$ for the $m=2/n=1$  mode in such  plasmas. 

\subsection{Approach to an Ideal Stability Boundary}\label{approach}
The matrix tearing mode dispersion relation, (\ref{tmdisp}), can be written in the alternative form\,\cite{am1}
\begin{equation}
\sum_{k'=1,K} F_{kk'}\,{\mit\Delta\Psi}_{k'} = {\mit\Psi}_k
\end{equation}
for $k=1,K$, where $F_{kk'}$ is the inverse of $E_{kk'}$. Now, a marginally-stable ideal mode is characterized by ${\mit\Psi}_k=0$ for $k=1,K$. In other words, the
mode does not reconnect helical magnetic flux at any rational surface in the plasma.  (The mode is marginally-stable
because our ideal-MHD solution in the outer region neglects plasma inertia.) Thus, for such a mode, 
\begin{equation}
\sum_{k'=1,K} F_{kk'}\,{\mit\Delta\Psi}_{k'} =0
\end{equation}
for $k=1,K$. The only way of finding a non-trivial solution of the previous equation is if the determinant of $F_{kk'}$ is zero. In general, the elements of
the $F_{kk'}$ matrix remain finite as ${\rm det}(F_{kk'})\rightarrow 0$.\cite{am1} Hence, we deduce that, as an ideal stability boundary is approached
[i.e., as ${\rm det}(F_{kk'})\rightarrow 0$], the elements of the $E_{kk'}= {\rm adj}(F_{kk'})/{\rm det}(F_{kk'})$ matrix  all become {\em infinite}. This implies, from Eq.~(\ref{decouple}), that all
of the decoupled tearing modes in the plasma simultaneously become extremely unstable.\cite{brennan,bren1,bren2} 

\subsection{Determination of Ideal Stability Boundary}\label{determine}
The DCON ideal stability code\,\cite{dcon} implements two tests for ideal stability. The first test is for fixed-boundary ideal modes (i.e., modes in which the
radial plasma displacement is constrained to be zero at the plasma boundary). This test involves calculating the perturbed ideal plasma potential energy
matrix, $\underline{\underline{W}}_{\,p}(r)$  (see Sect.~\ref{penergy}), in the region lying between the magnetic axis and the equilibrium magnetic
flux-surface whose label is $r$. If ${\rm det}(\underline{\underline{W}}_{\,p}^{\,-1}) =0$
for any value of $r$ lying in the range $0\leq r\leq a$ then the plasma is unstable to a fixed-boundary ideal mode. The second test is for
free-boundary ideal modes (i.e., modes in which the radial plasma displacement is not constrained to be zero at the plasma boundary 
because the magnetic perturbation extends into the vacuum region surrounding the plasma). This test involves calculating the Hermitian perturbed ideal plasma potential energy
matrix, $\underline{\underline{W}}_{\,p}$,  for the whole  region lying  between the magnetic axis and the plasma boundary, and adding to it the Hermitian vacuum energy matrix,  $\underline{\underline{W}}_{\,v}$,  in order to
form the Hermitian  total energy matrix, $\underline{\underline{W}} =\underline{\underline{W}}_{\,p} +  \underline{\underline{W}}_{\,v}$. If any of the real eigenvalues
of the total energy matrix are negative then the plasma is unstable to a free-boundary ideal mode.  As described in Sect.~\ref{ideal}, the TJ code
implements the relatively straightforward second test, rather than the much more complicated first test. One question that we would like to address in
this paper is whether the second test is sufficient to detect ideal stability boundaries associated with $m=1/n=1$ internal-kink modes. 

\subsection{Approach to External-Kink Ideal Stability Boundary} \label{external}
\subsubsection{Plasma Equilibria}
In this section, we shall investigate the behavior of $n=1$ tearing modes as an ideal stability boundary associated with an  external-kink mode is approached. 

For the sake of simplicity and clarity, all of the calculations described in this section are performed for circular poloidal cross-section plasma equilibria that contain two rational surfaces: namely, the $m=2/n=1$ and the $m=3/n=1$ surfaces (a.k.a. the $q=2$ and the $q=3$ surfaces). 
The equilibria are characterized by $q_0=1.5$, $q_a=3.6$, $p_p=2.0$, and $a=0.2$. The calculations include all poloidal harmonics in the range $m=-10$ to $m=+20$. 

Resistive layer quantities are calculated with the following plasma parameters (see Sect.~\ref{ppara}) that are broadly characteristic of a high-field tokamak fusion reactor:
$B_0=12\,{\rm T}$, $R_0=2\,{\rm m}$, $n_0=3\times 10^{20}\,{\rm m}^{-3}$, $T_{e\,{\rm edge}}=100\,{\rm eV}$, $\alpha =0.5$, $Z_{\rm eff}=2$, $M=2$, 
and $\chi_\perp=0.2\,{\rm m}^{-3}$. Note that the electron and ion temperatures in the plasma are assumed to be equal to one another, as is likely to be the case in a
fusion reactor. Moreover, the
perpendicular momentum/energy diffusivity, $\chi_\perp$, is assumed to be spatially uniform, for the sake of simplicity. 

\subsubsection{Ideal Stability of Free-Boundary Modes}
\begin{figure}
\centerline{\includegraphics[width=0.9\textwidth]{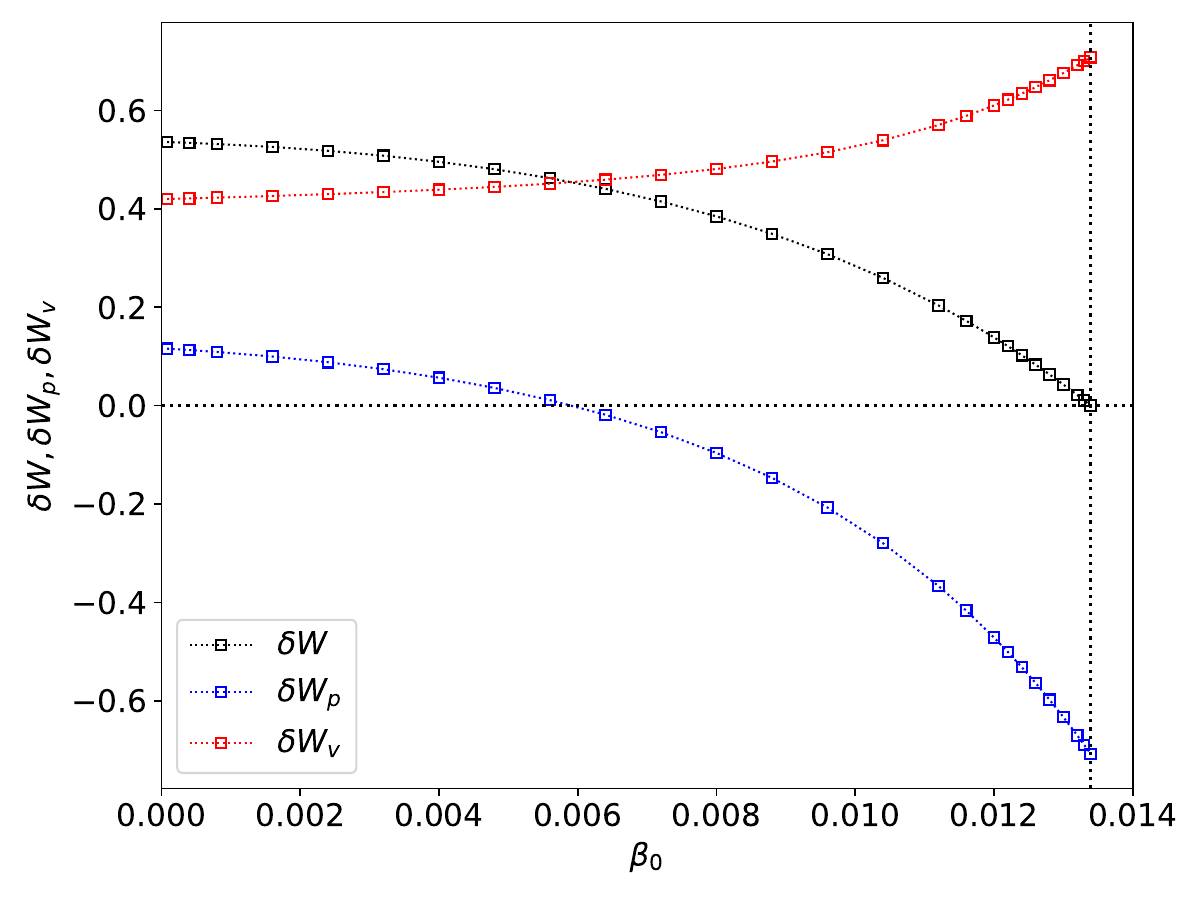}}
\caption{Variation of the smallest eigenvalue of the free-boundary, $n=1$, total ideal energy matrix, $\delta W$, and the corresponding eigenvalues of the plasma and vacuum energy matrices, 
$\delta W_p$ and $\delta W_v$, with the central plasma beta, $\beta_0$, in a circular cross-section plasma equilibrium characterized by $q_0=1.5$, $q_a= 3.6$, $p_p=2.0$, and
$a=0.2$. The vertical dotted line corresponds to the ideal stability boundary (at which $\delta W=0$), $\beta_0=0.0134$.  \label{fig5}}
\end{figure}

Figure~\ref{fig5} shows the variation of the smallest eigenfunction of the free-boundary (i.e., with no ideal wall in the vacuum region surrounding the plasma), $n=1$, 
total ideal energy matrix, $\delta W$, and the corresponding eigenvalues of the plasma and vacuum energy matrices, $\delta W_p$, and $\delta W_v$, with the
central plasma beta, $\beta_0$. It can be seen that $\delta W$ decreases monotonically with increasing $\beta_0$. This decrease is due
to a corresponding decrease in the plasma energy, $\delta W_p$, rather than the positive definite vacuum energy, $\delta W_v$ (which actually increases). 
At low $\beta_0$ values, the total energy is positive, indicating that the plasma equilibrium is ideally stable.\cite{freidberg,ideal} However,  the
total energy passes through zero at $\beta_0=0.0134$, which implies that the plasma equilibrium is ideally unstable for $\beta_0>0.0134$,
and that the marginal stability boundary corresponds to $\beta_0=0.0134$. 

\subsubsection{Free-Boundary Tearing Stability Matrix}
\begin{figure}
\centerline{\includegraphics[width=0.9\textwidth]{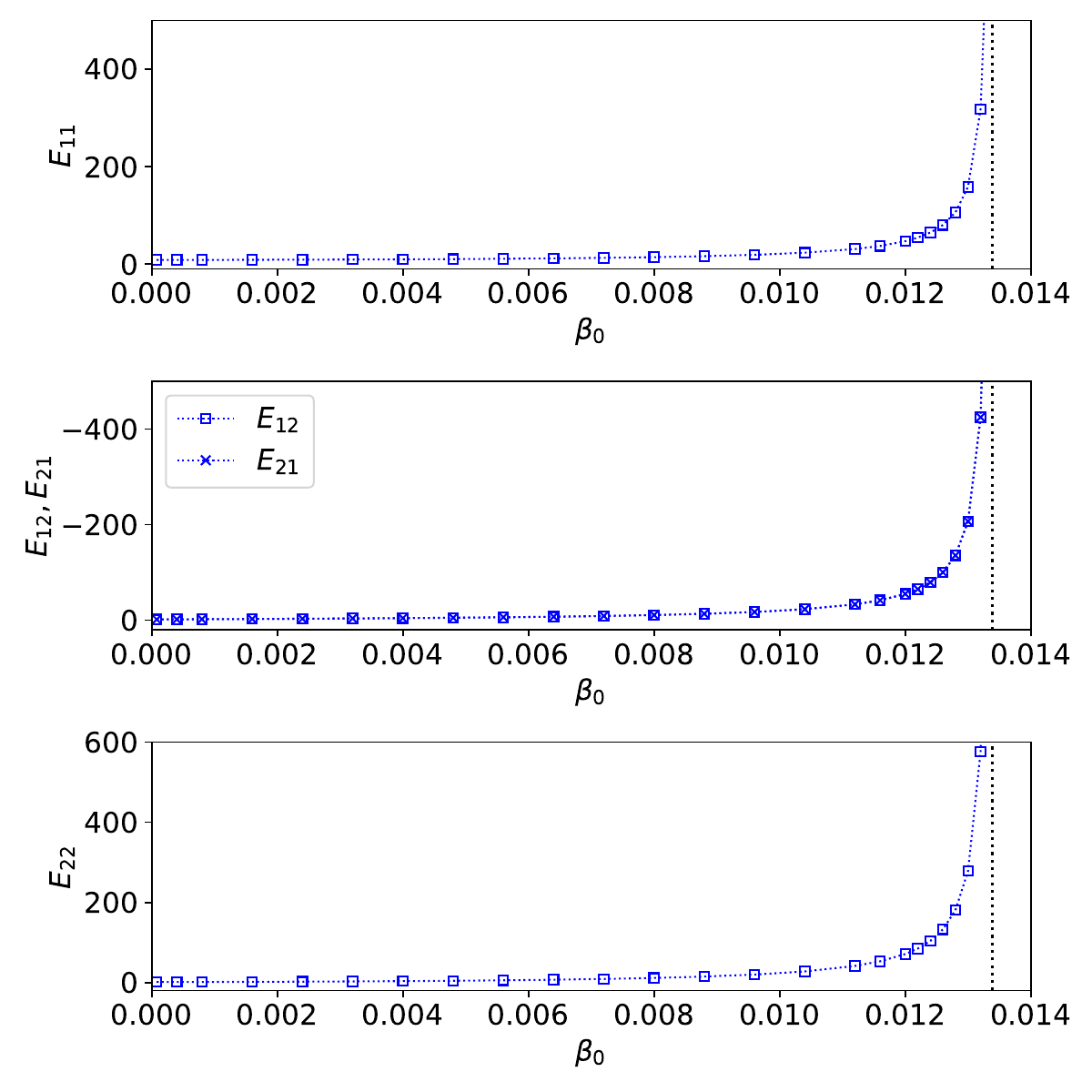}}
\caption{Variation of the elements of the free-boundary, $n=1$, tearing stability matrix with the central plasma beta, $\beta_0$, in a circular cross-section plasma equilibrium  characterized by $q_0=1.5$, $q_a= 3.6$, $p_p=2.0$, and
$a=0.2$. The vertical dotted line corresponds to the ideal stability boundary, $\beta_0=0.0134$. \label{fig6}}
\end{figure}

Figure~\ref{fig6} shows the variation of the elements of the free-boundary, $n=1$, tearing stability matrix with $\beta_0$. In accordance with the discussion in
Sect.~\ref{approach}, the elements all tend to infinity as the ideal stability boundary is approached. This implies that the decoupled tearing modes that
only reconnected magnetic flux at the $q=2$ surface (a.k.a.\ the $2/1$ mode) and the $q=3$ surface  (a.k.a.\ the $3/1$ mode)  simultaneously become very unstable as the
stability boundary is approached. Furthermore, the coupling between the two modes (which is due to the off-diagonal elements of the matrix)
becomes very large as the  boundary is approached. 

\subsubsection{Growth and Rotation of Free-Boundary Tearing Modes}
\begin{figure}
\centerline{\includegraphics[width=0.9\textwidth]{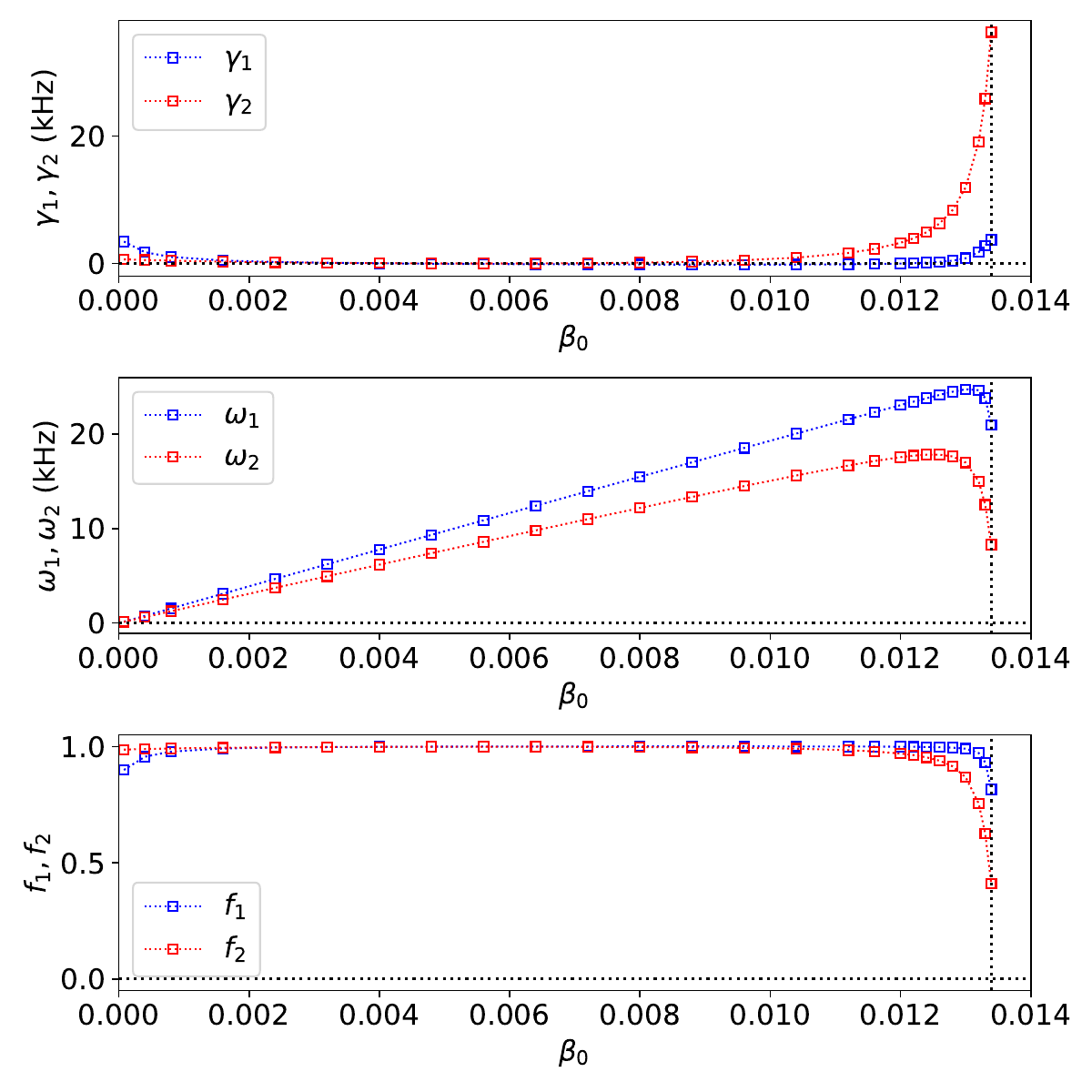}}
\caption{Variation of the real growth-rates and real frequencies of the free-boundary, $n=1$, tearing modes that only reconnect magnetic flux at the  $q=2$ (labelled $1$) and $q=3$ (labelled $2$)  surfaces with the central plasma beta, $\beta_0$, in a circular cross-section plasma equilibrium characterized by $q_0=1.5$, $q_a= 3.6$, $p_p=2.0$, and
$a=0.2$. The vertical dotted line corresponds to the ideal stability boundary, $\beta_0=0.0134$. The $f_k$ parameters would take the values $+1$/0/$-1$ if the tearing modes were to co-rotate with the
electron/E-cross-B/ion fluids at the surface at which they reconnect magnetic flux.\label{fig7}}
\end{figure}

Figure~\ref{fig7} shows the variations  with $\beta_0$ of the real growth-rates and real frequencies of free-boundary $2/1$ and $3/1$  modes.  At low $\beta_0$ values, both modes are
unstable. However, as $\beta_0$ increases, the $2/1$ mode is eventually stabilized by
magnetic curvature effects. (Unfortunately, this is a little hard to see in the figure.) This is because $E_{11}$, despite being positive, falls below ${\mit\Delta}_{1\,{\rm crit}}\propto\beta_0$.  As $\beta_0$ further increases, 
$E_{11}$ eventually becomes  large enough to overcome curvature stabilization, and the $2/1$ mode becomes unstable again. As the ideal stability
boundary is approached, the growth-rate of the $3/1$ mode becomes extremely large, whereas that of the $2/1$ mode remains finite. 
This is partly because $E_{22}>E_{11}$ (i.e., the $3/1$ mode is more intrinsically unstable than the $2/1$ mode), but mostly because
the plasma at the $q=3$ surface is significantly colder than that at the $q=2$ surface, so the resistive diffusion time at the former surface is much  shorter than
that at the latter. The real frequencies of the two modes (which are measured in the local E-cross-B frames at the surfaces at which they reconnect magnetic flux) initially increase linearly with $\beta_0$, because the frequencies are diamagnetic in nature, and therefore scale linearly with the plasma
pressure.\cite{ara} However, the frequencies of both modes exhibit sharp downturns as the ideal stability boundary is approached. The orgin
of these downturns is elucidated  in the bottom panel of the figure. The parameters $f_1$ and $f_2$ are designed to take the values $+1$/0/$-1$ if the $2/1$ and
$3/1$ tearing modes, respectively,  co-rotate with the
electron/E-cross-B/ion fluids at the surfaces at which they reconnect magnetic flux,  and to take appropriate linear intermediate values. It can be seen that, 
at intermediate values of $\beta_0$,  $f_1$ and $f_2$ are both close to $+1$. This indicates that the modes in question are ``drift-tearing modes'' that co-rotate
with the electric fluids at the associated rational surfaces.\cite{ara} However, as the ideal stability boundary is approached, and the growth-rates of
both modes become very large, there is a shift in the frequency of each mode in the ion direction. This indicates that the mode are transitioning to
``resistive-kink modes'', which we expect to co-rotate with the ion fluids at the associated rational surfaces.\cite{ara}

\subsubsection{Structure of Free-Boundary Tearing Modes}
\begin{figure}
\centerline{\includegraphics[width=0.9\textwidth]{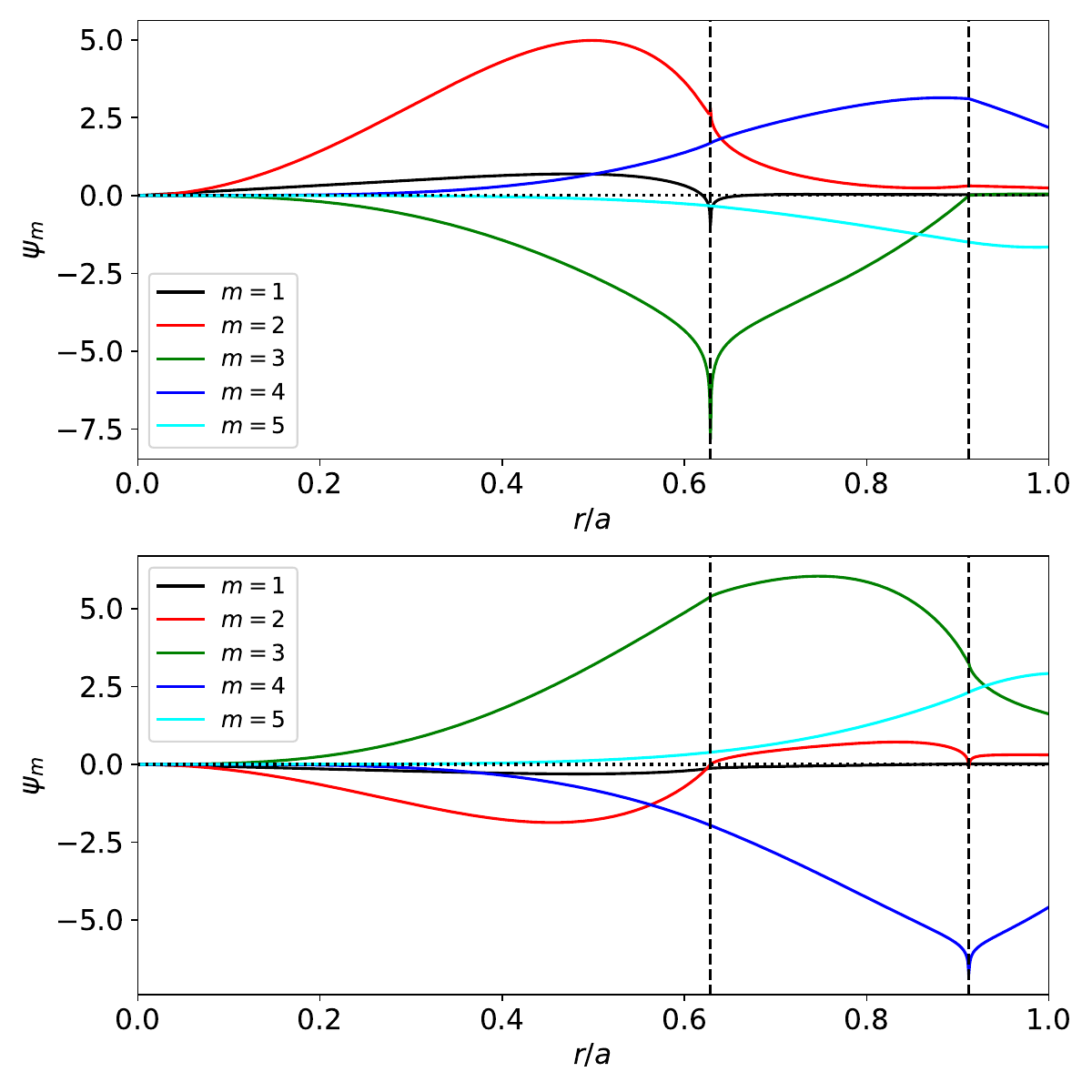}}
\caption{Poloidal harmonics of the perturbed magnetic flux, $\psi(r,\theta)$,  associated with free-boundary, $n=1$, tearing modes that only reconnect magnetic flux at the $q=2$  (top panel) and the $q=3$ (bottom panel) surfaces in a 
circular cross-section plasma equilibrium characterized by $q_0=1.5$, $q_a= 3.6$, $p_p=2.0$,
$a=0.2$, and $\beta_0 =0.0064$. The vertical dashed lines show the locations of the $q=2$ and $q=3$ surfaces. \label{fig8}}
\end{figure}

\begin{figure}
\centerline{\includegraphics[width=\textwidth]{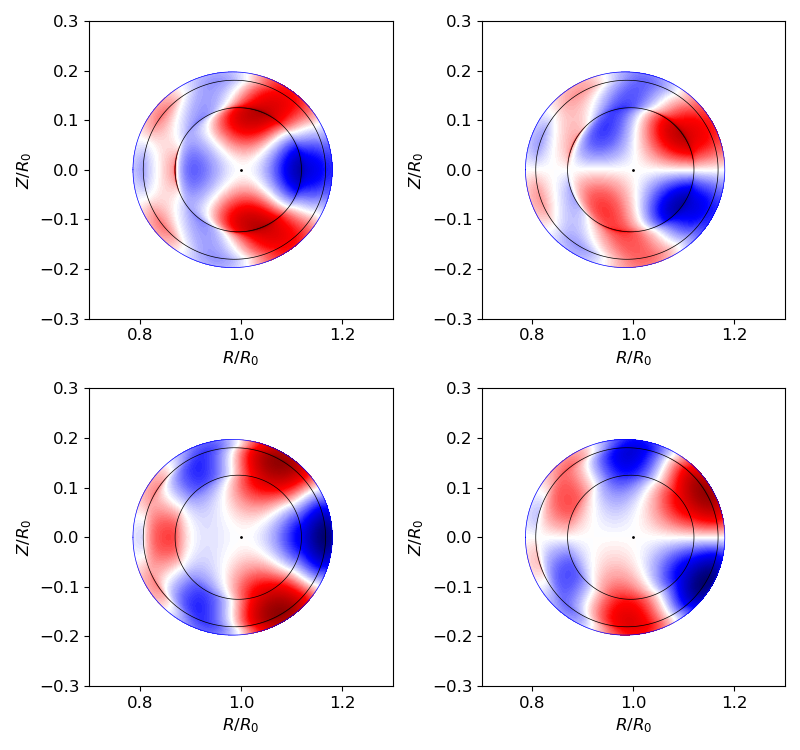}}
\caption{Contour plots of the real (left columns) and imaginary (right columns) parts the perturbed magnetic flux, $\psi(r,\theta)$, associated with free-boundary, $n=1$, tearing modes that only reconnect magnetic flux at the $q=2$  (top panels) and the $q=3$ (bottom panels) surfaces in a 
circular cross-section plasma equilibrium characterized by $q_0=1.5$, $q_a= 3.6$, $p_p=2.0$,
$a=0.2$, and $\beta_0 =0.0064$. Here, red/blue/white denotes positive/negative/zero values. The black dot shows the location of the magnetic axis, and the
black circles show the locations of the $q=2$ and $q=3$ surfaces. \label{fig9}}
\end{figure}

Figures~\ref{fig8} and \ref{fig9} show the poloidal harmonics and contours of the perturbed magnetic flux, $\psi(r,\theta)$ [see Eq.~(\ref{a3})], associated
with the free-boundary  $2/1$ and $3/1$ modes in a  plasma equilibrium that is not close to the ideal stability boundary. It is clear from Fig.~\ref{fig9} that both modes are essentially $m=2$ modes inside the $q=2$ surface, $m=3$ modes between
the $q=2$ and $q=3$ surfaces, and $m=4$ modes outside the $q=3$ surface. Although both modes look fairly similar to
one another (modulo a multiplicative constant), there are subtle differences. The $2/1$ mode  has $\psi_{m=2}\neq 0$ at the $q=2$ surface, and $\psi_{m=3}=0$ at the $q=3$ surface.
On the other hand, the $3/1$ mode  has $\psi_{m=2}= 0$ at the $q=2$ surface, and $\psi_{m=3}\neq 0$ at the $q=3$  surface. The  $3/1$ mode also has a much larger $m=4$ component at the plasma boundary than the $2/1$ mode. 

\begin{figure}
\centerline{\includegraphics[width=0.9\textwidth]{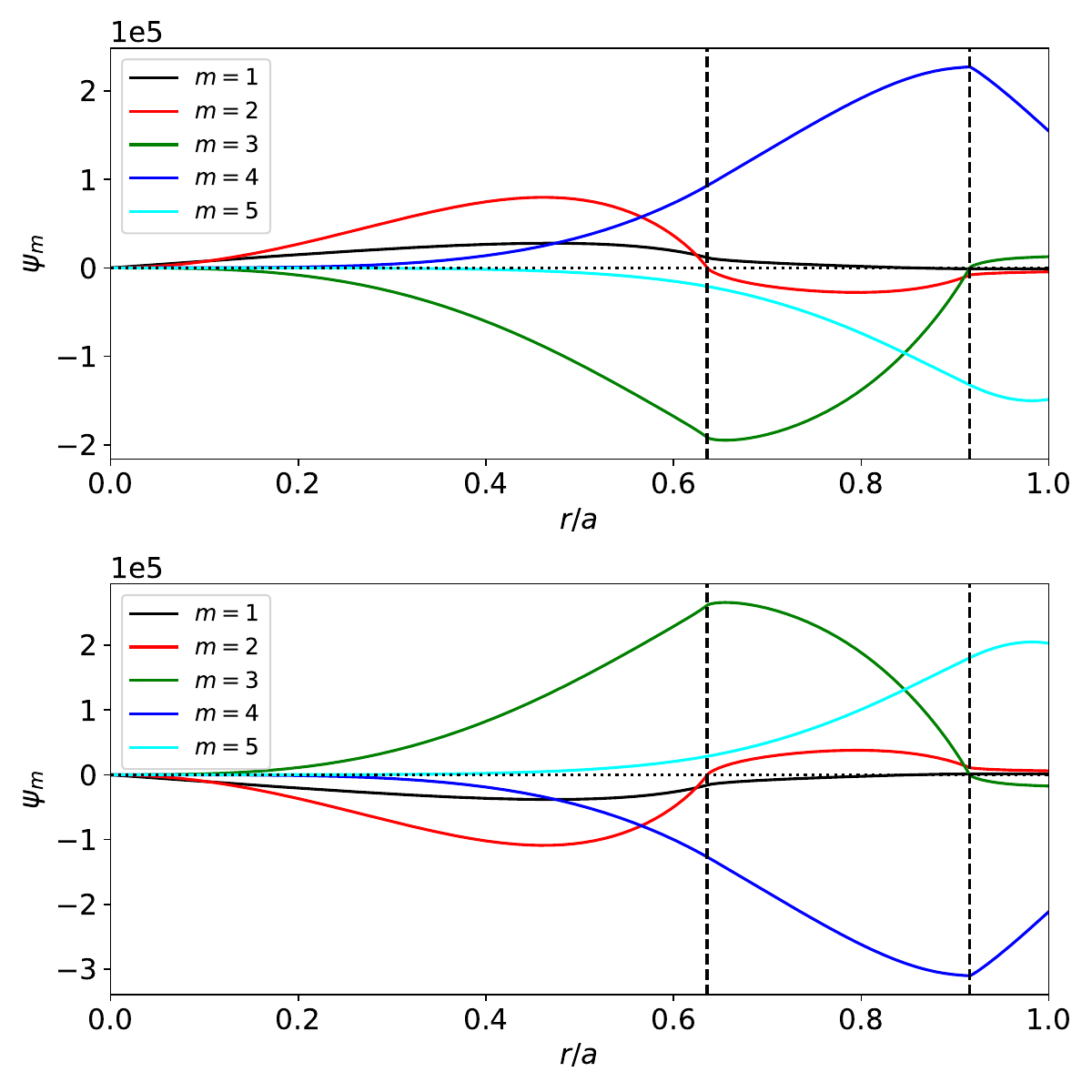}}
\caption{Poloidal harmonics of the perturbed magnetic flux, $\psi(r,\theta)$,  associated with free-boundary, $n=1$, tearing modes that only reconnect magnetic flux at the $q=2$  (top panel) and the $q=3$ (bottom panel) surfaces in a marginally-ideally-stable, 
circular cross-section, plasma equilibrium characterized by $q_0=1.5$, $q_a= 3.6$, $p_p=2.0$, and
$a=0.2$, and  $\beta_0=0.0134$. See caption to Fig.~\ref{fig8}.\label{fig10}}
\end{figure}

\begin{figure}
\centerline{\includegraphics[width=0.9\textwidth]{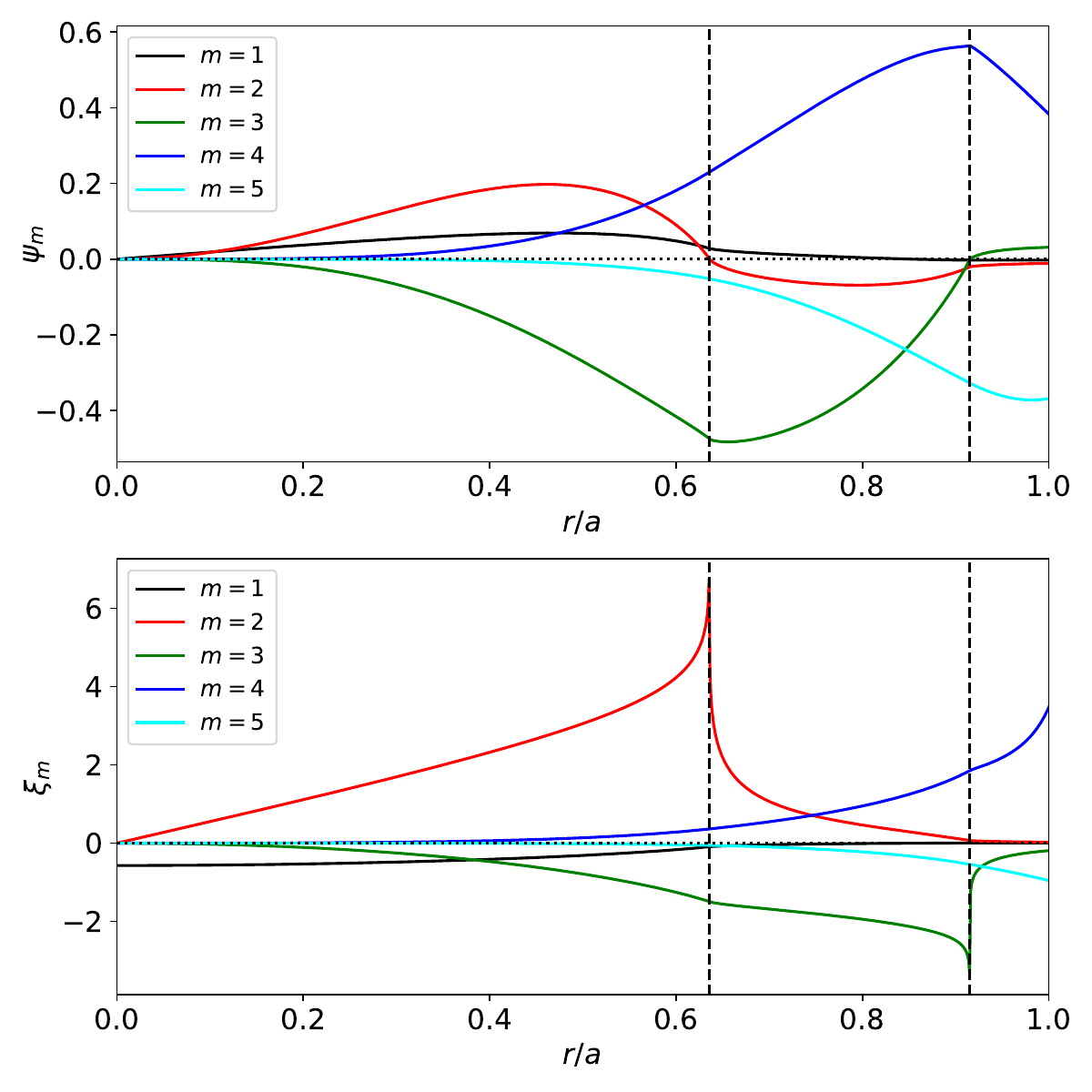}}
\caption{Poloidal harmonics of the perturbed magnetic flux, $\psi(r,\theta)$ (top panel), and the radial plasma displacement, $\xi^r(r,\theta)$,  associated with a free-boundary, 
marginally-stable, $n=1$, ideal mode in a circular cross-section plasma equilibrium characterized by $q_0=1.5$, $q_a= 3.6$, $p_p=2.0$, 
$a=0.2$, and $\beta_0=0.0134$. See caption to Fig.~\ref{fig8}.\label{fig11}}
\end{figure}

\begin{figure}
\centerline{\includegraphics[width=\textwidth]{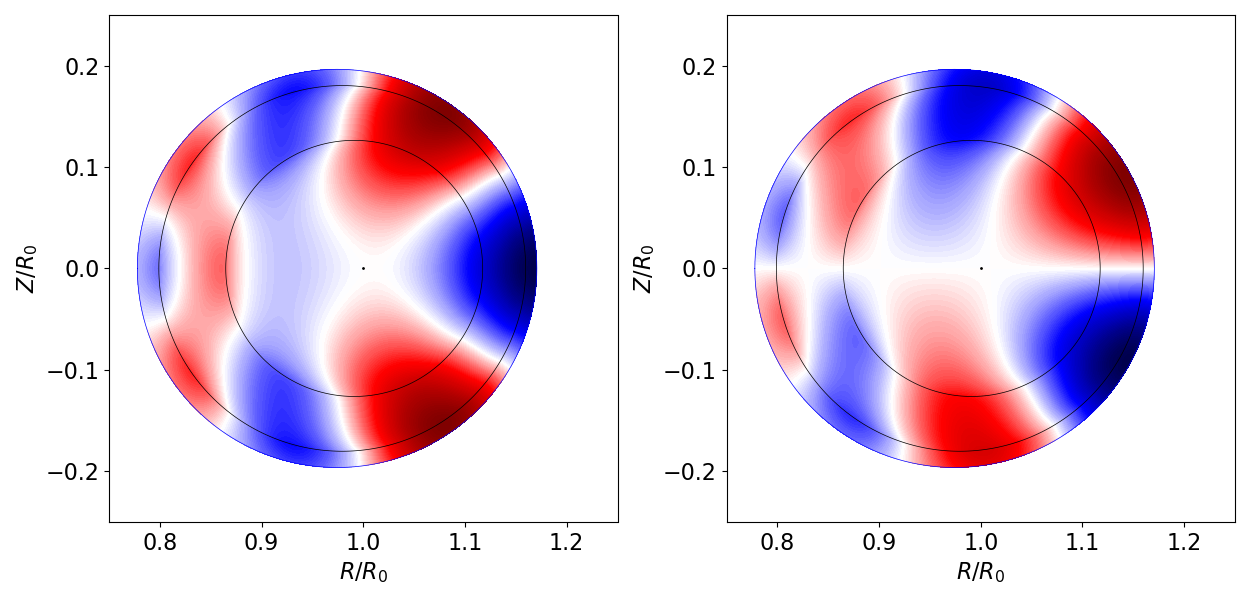}}
\caption{Contour plots of the real (left panel) and imaginary (right panel) parts of the perturbed magnetic flux, $\psi(r,\theta)$, associated with a free-boundary,
marginally-stable, $n=1$, ideal mode in a 
circular cross-section plasma equilibrium characterized by $q_0=1.5$, $q_a= 3.6$, $p_p=2.0$, 
$a=0.2$, and $\beta_0=0.0134$. See caption to Fig.~\ref{fig9}. \label{fig12}}
\end{figure}

Figure~\ref{fig10} shows the poloidal harmonics of the perturbed magnetic flux associated
with the free-boundary $2/1$ and $3/1$ modes in a  plasma equilibrium 
that lies exactly at the ideal stability boundary. Figures~\ref{fig11} and \ref{fig12} show the poloidal harmonics and contours of the perturbed magnetic flux, as well
as the poloidal harmonics of the radial
plasma displacement, $\xi^r(r,\theta)$ [see Eq.~(\ref{e10})], associated with the marginally-stable, $n=1$, ideal mode at the stability boundary. It is
clear, from a comparison of Fig.~\ref{fig10} and the top panel of Fig.~\ref{fig11}, that at the ideal stability boundary the eigenfunctions of the two
tearing modes are both  identical to that of the marginally-stable ideal mode (modulo an arbitrary multiplicative constant). Thus, we deduce that, as
an ideal stability boundary is approached, all of the decoupled tearing modes in the plasma (with the same toroidal mode number as the ideal mode)
effectively morph into the ideal mode. However, the growth-rate and real frequency of a given ``ideal-tearing-mode'' is determined by the 
resistive layer solution at the rational surface at which it reconnects magnetic flux. In other words, at an ideal stability boundary, there are as
many independent marginally-stable ideal modes as there are associated rational surfaces in the plasma, each possessing a different growth-rate and
real frequency. As is clear from Fig.~\ref{fig7}, in the present case, the ideal mode that reconnects flux closest to the edge of the plasma has the largest growth-rate. 

Finally, according to the bottom panel of Fig.~\ref{fig11} and Fig.~\ref{fig12}, the marginally-stable ideal mode  is predominately an
$m=4$ mode at the edge of the plasma. In other words, the ideal mode is essentially an $m=4/n=1$ external-kink mode, as we would expect from
cylindrical theory.\cite{freidberg,wesson} 

\subsubsection{Stabilizing Effect of an Ideal Wall}\label{stabw1}
\begin{figure}
\centerline{\includegraphics[width=0.9\textwidth]{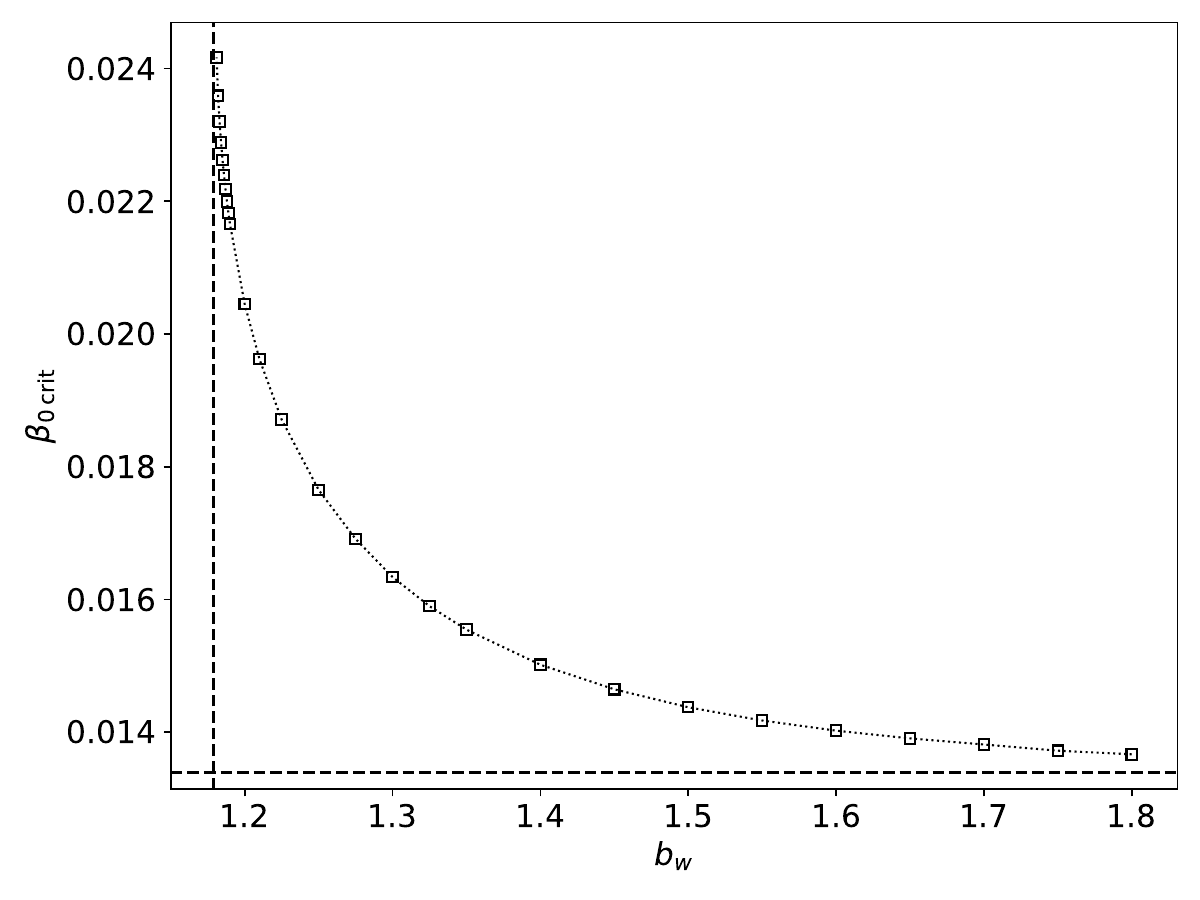}}
\caption{Variation of the critical central beta value at marginal ideal stability, $\beta_{0\,{\rm crit}}$, with the relative radius, $b_w$, of the ideal wall surrounding the plasma in 
a circular cross-section plasma equilibrium characterized by $q_0=1.5$, $q_a= 3.6$, $p_p=2.0$, and
$a=0.2$. The horizontal dashed line corresponds to the no-wall  marginal stability boundary, $\beta_{0\,{\rm crit}}=0.0134$. The vertical dashed line corresponds to $b_w=1.179$.  \label{fig13}}
\end{figure}

Figure~\ref{fig13} shows how the critical central beta value at marginal ideal stability, $\beta_{0\,{\rm crit}}$, varies with the
relative radius, $b_w$, of an ideal wall that is situated in the vacuum region surrounding the plasma. It is clear that the
presence of the wall has a stabilizing effect on the ideal mode.  It is plausible, from
the figure, that $\beta_{0\,{\rm crit}}$ approaches its free-boundary value as $b_w\rightarrow \infty$. On the other hand,
as the wall approaches the plasma boundary (i.e., as $b_w\rightarrow 1$), the critical beta value becomes larger, and seems
to asymptote to infinity at $b_w=1.179$. This implies that a close-fitting ideal wall is capable of completely stabilizing the
$m=4/n=1$ ideal external-kink mode. Again, this is in accordance with cylindrical theory.\cite{freidberg,wesson}

\subsection{Approach to Internal-Kink Ideal Stability Boundary}
\subsubsection{Plasma Equilibria}
In this section, we shall investigate the behavior of $n=1$ tearing modes as an ideal stability boundary associated with an  $m=1/n=1$ internal-kink mode is approached. 

All of the calculations described in this section are performed for circular poloidal cross-section plasma equilibria that contain two rational surfaces: namely, the $m=1/n=1$ and the $m=2/n=1$ surfaces (a.k.a.\ the $q=1$ and $q=2$ surfaces). 
The equilibria are  characterized by $q_0=0.8$, $q_a=2.8$, $p_p=2.0$, and $a=0.2$. The calculations include all poloidal harmonics in the range $m=-10$ to $m=+20$. 

The same  plasma parameters  are used to calculate resistive layer quantities as those used in Sect.~\ref{external}. 

\subsubsection{Ideal Stability of Free-Boundary Modes}
\begin{figure}
\centerline{\includegraphics[width=0.9\textwidth]{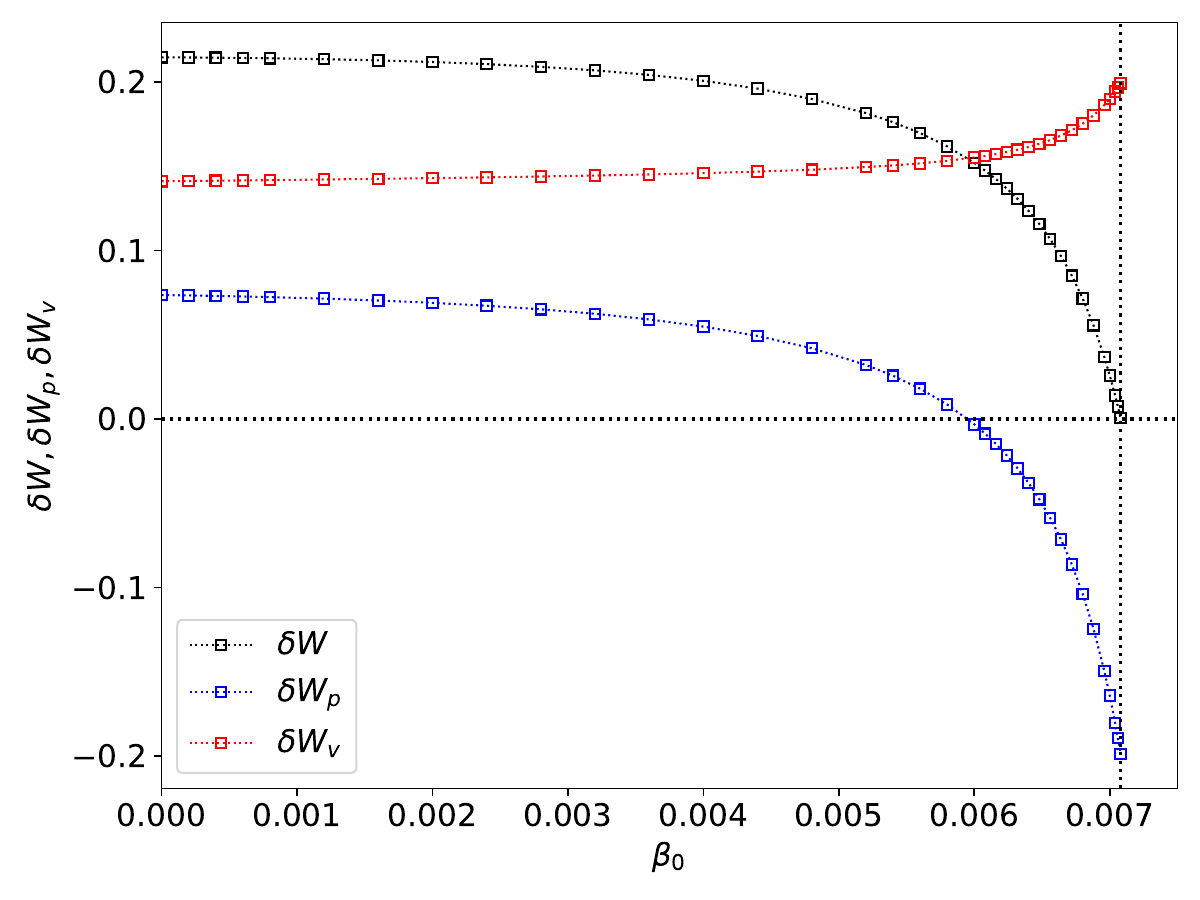}}
\caption{Variation of the smallest eigenvalue of the free-boundary, $n=1$, total ideal energy matrix, $\delta W$, and the corresponding eigenvalues of the plasma and vacuum energy matrices, 
$\delta W_p$ and $\delta W_v$, with the central plasma beta, $\beta_0$, in a circular cross-section plasma equilibrium characterized by $q_0=0.8$, $q_a= 2.8$, $p_p=2.0$, and
$a=0.2$. The vertical dotted line corresponds to the ideal stability boundary, $\beta_0=0.00708$.  \label{fig14}}
\end{figure}

Figure~\ref{fig14} shows the variation of the smallest eigenfunction of the free-boundary, $n=1$, 
total ideal energy matrix, $\delta W$, and the corresponding eigenvalues of the plasma and vacuum energy matrices, $\delta W_p$, and $\delta W_v$, with the
central plasma beta, $\beta_0$. Similarly to Fig.~\ref{fig5}, $\delta W$ decreases monotonically with increasing $\beta_0$  due
to a decrease in the plasma energy, $\delta W_p$, rather than the positive definite vacuum energy, $\delta W_v$. 
 The marginal stability boundary corresponds to $\beta_0=0.00708$.  

\subsubsection{Free-Boundary Tearing Stability Matrix}
\begin{figure}
\centerline{\includegraphics[width=0.9\textwidth]{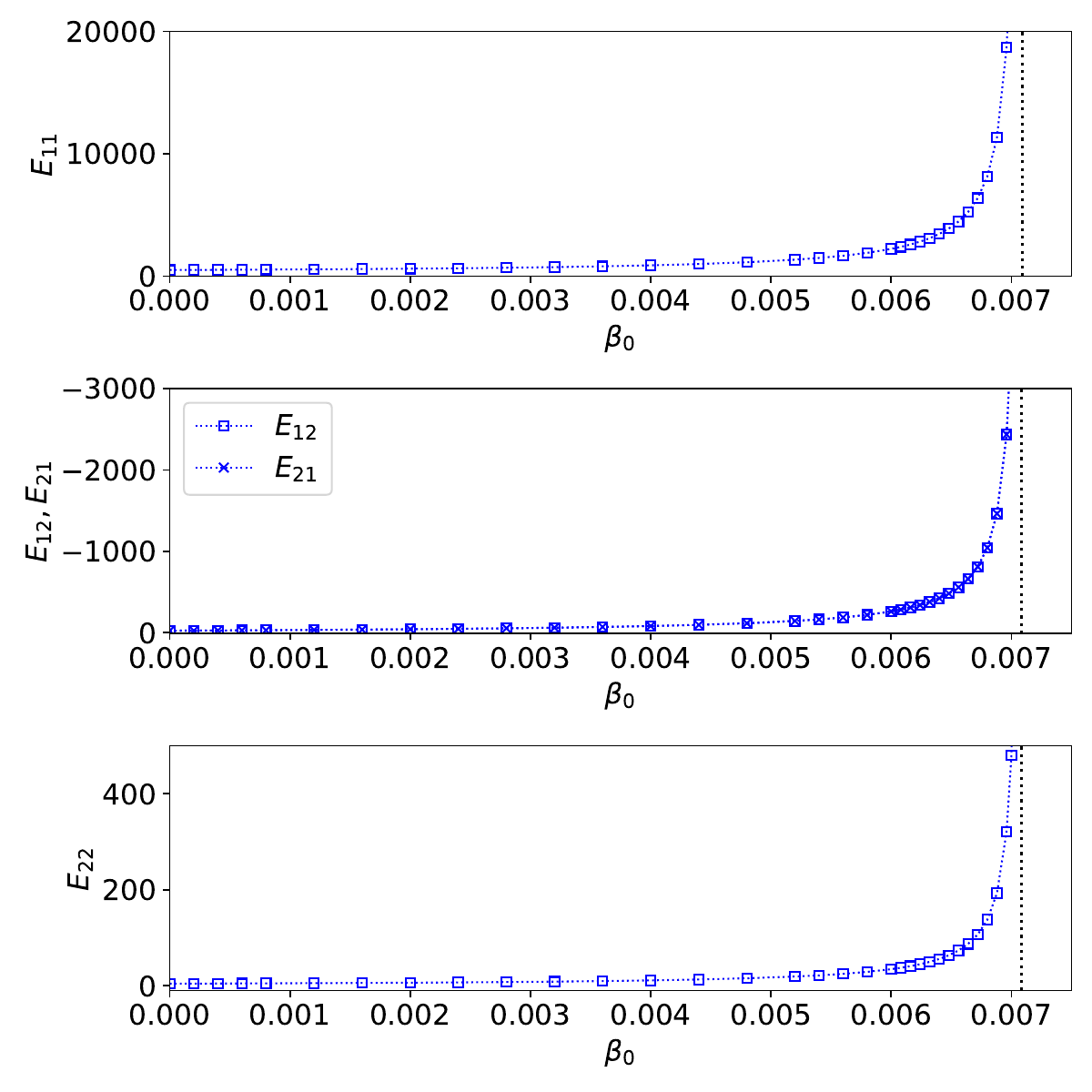}}
\caption{Variation of the elements of the free-boundary, $n=1$, tearing stability matrix with the central plasma beta, $\beta_0$, in a circular cross-section plasma equilibrium  characterized by $q_0=0.8$, $q_a= 2.8$, $p_p=2.0$, and
$a=0.2$. The vertical dotted line corresponds to the ideal stability boundary, $\beta_0=0.00708$. \label{fig15}}
\end{figure}

Figure~\ref{fig15} shows the variation of the elements of the free-boundary, $n=1$, tearing stability matrix with $\beta_0$. As before, in accordance with the discussion in
Sect.~\ref{approach}, the elements all tend to infinity as the ideal stability boundary is approached. This implies that the decoupled tearing modes that
only reconnected magnetic flux at the $q=1$ surface (a.k.a.\ the $1/1$ mode) and the $q=2$ surface  (a.k.a.\ the $2/1$ mode)   simultaneously become very unstable as the
stability boundary is approached. Note that $E_{11}$ (i.e., the tearing stability index for the $1/1$ internal-kink mode) is
considerably larger that $E_{22}$ (i.e., the tearing stability index for the $2/1$ tearing mode), in accordance with cylindrical theory.\cite{freidberg,wesson}

\subsubsection{Growth and Rotation of Free-Boundary Tearing Modes}
\begin{figure}
\centerline{\includegraphics[width=0.9\textwidth]{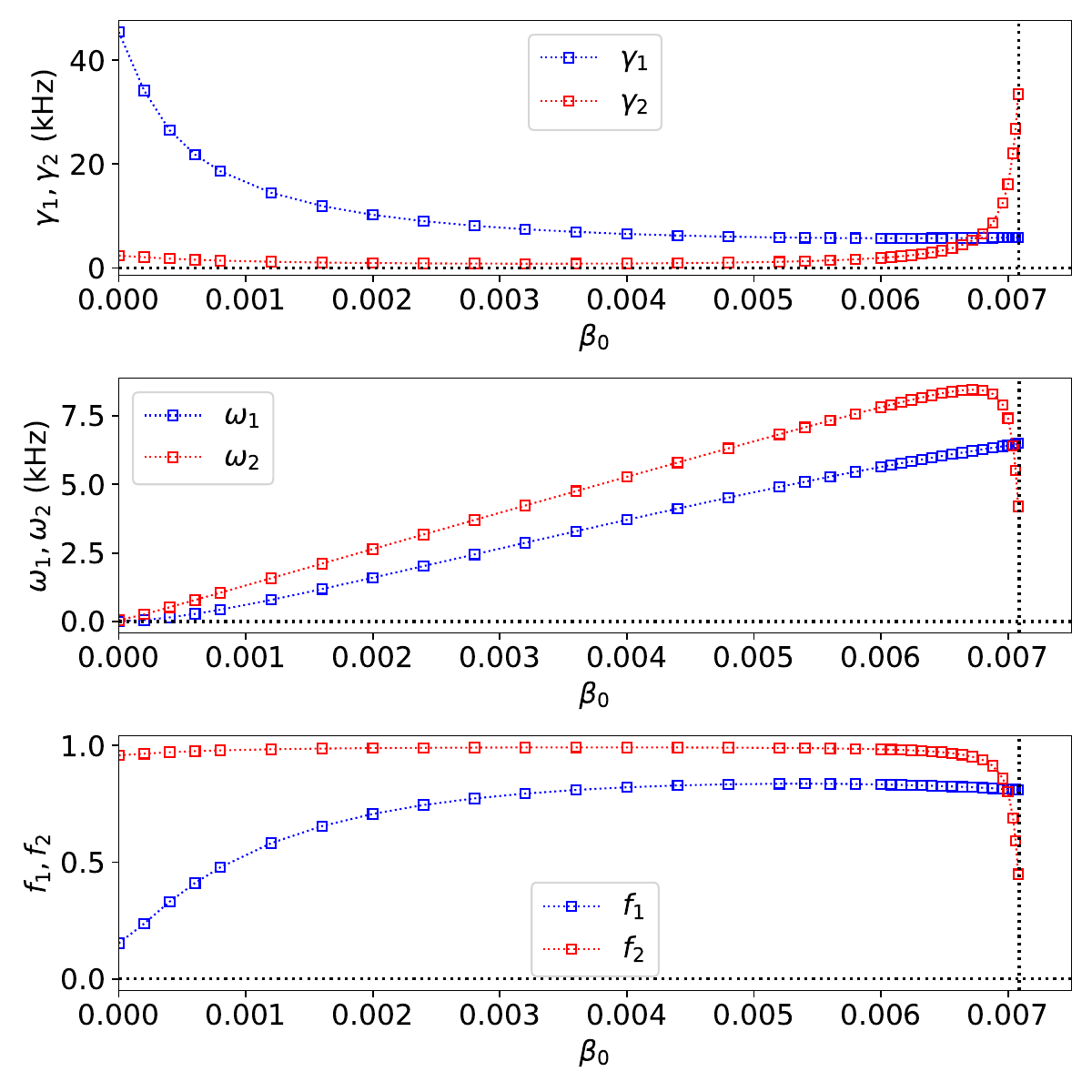}}
\caption{Variation of the real growth-rates and real frequencies of the free-boundary, $n=1$, tearing modes that only reconnect magnetic flux at the  $q=1$ (labelled $1$) and $q=2$ (labelled $2$)  surfaces with the central plasma beta, $\beta_0$, in a circular cross-section plasma equilibrium characterized by $q_0=0.8$, $q_a= 2.8$, $p_p=2.0$, and
$a=0.2$. The vertical dotted line corresponds to the ideal stability boundary, $\beta_0=0.00708$. See caption to Fig.~\ref{fig7}.\label{fig16}}
\end{figure}

Figure~\ref{fig16} shows the variations  with $\beta_0$ of the real growth-rates and real frequencies of the free-boundary $1/1$ and $2/1$ modes.  At low values of $\beta_0$, the growth-rates of both modes decrease with increasing
$\beta_0$. This is because  the plasma temperature is proportional to $\beta_0$, and the decrease in the resistive diffusion times at the $q=1$ and $q=2$ surfaces
with increasing temperature offsets the increase in $E_{11}$ and $E_{22}$ with increasing $\beta_0$. 
At low to intermediate values of $\beta_0$, the growth-rate
of the $1/1$ mode is considerably larger than that of the $2/1$ mode, because the former mode is much more intrinsically
unstable than the latter (i.e., $E_{11}\gg E_{22}$). The fact that $f_1$ is less than unity, whereas $f_2=+1$,  suggests that the former mode is transitioning to a resistive-kink mode,
whereas the latter mode remains a drift-tearing mode.\cite{ara} Somewhat surprisingly, as the ideal stability boundary is approached, the growth-rate of the
$2/1$ mode, rather than the $1/1$ mode, becomes very large.  This is because
the plasma at the $q=2$ surface is much colder than that at the $q=1$ surface, so the resistive diffusion time at the former surface is very much  shorter than
that at the latter. As the stability boundary is approached, the frequency of the  $2/1$ mode shifts  in the ion direction as it
transitions from a drift-tearing to a resistive-kink mode.\cite{ara}

\subsubsection{Structure of Free-Boundary Tearing Modes}
\begin{figure}
\centerline{\includegraphics[width=0.9\textwidth]{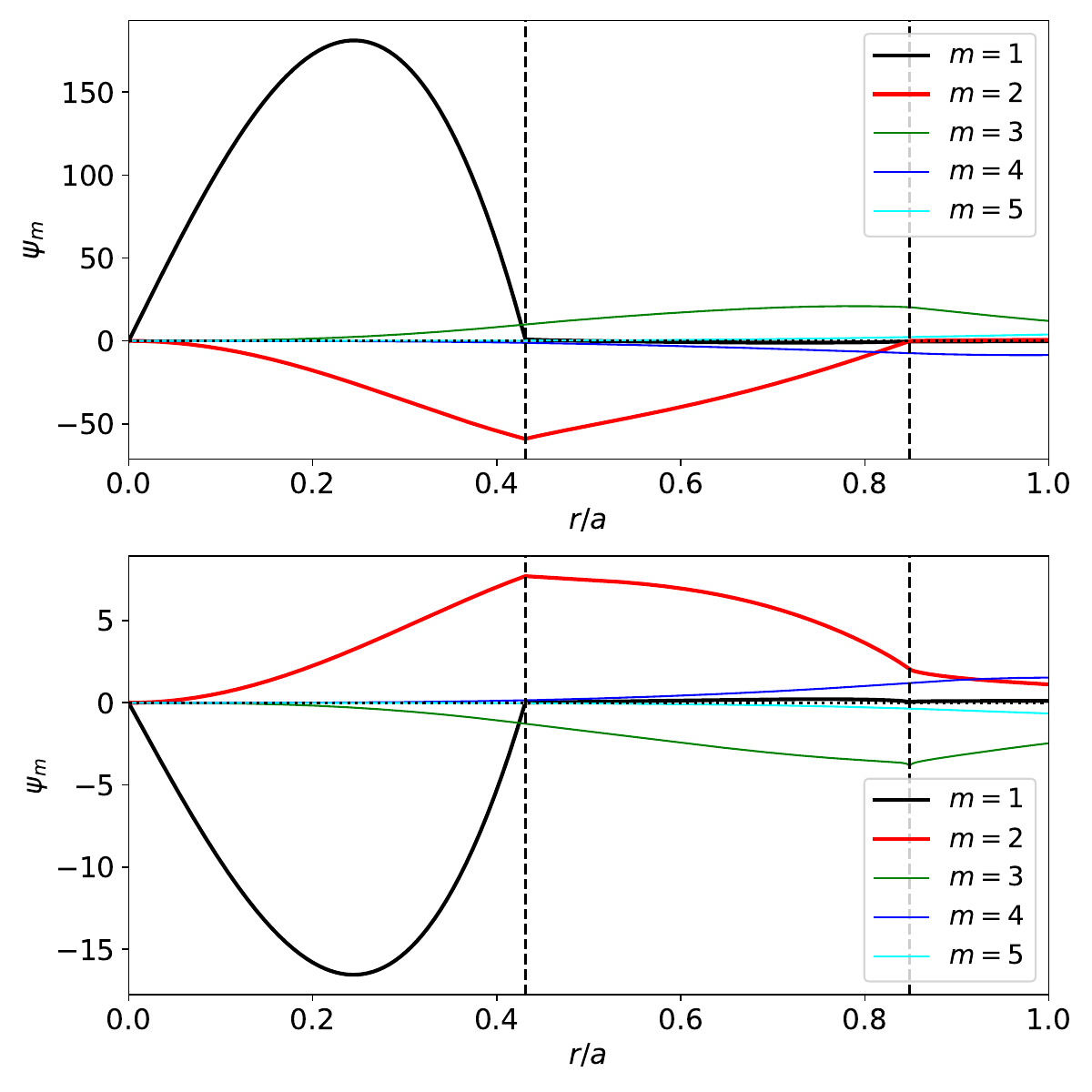}}
\caption{Poloidal harmonics of the perturbed magnetic flux, $\psi(r,\theta)$,  associated with free-boundary, $n=1$, tearing modes that only reconnect magnetic flux at the $q=1$  (top panel) and the $q=2$ (bottom panel)  surfaces in a 
circular cross-section plasma equilibrium characterized by $q_0=0.8$, $q_a= 2.8$, $p_p=2.0$,
$a=0.2$, and $\beta_0 =0.004$. The vertical dashed lines show the locations of the $q=1$ and $q=2$ surfaces. \label{fig17}}
\end{figure}

\begin{figure}
\centerline{\includegraphics[width=\textwidth]{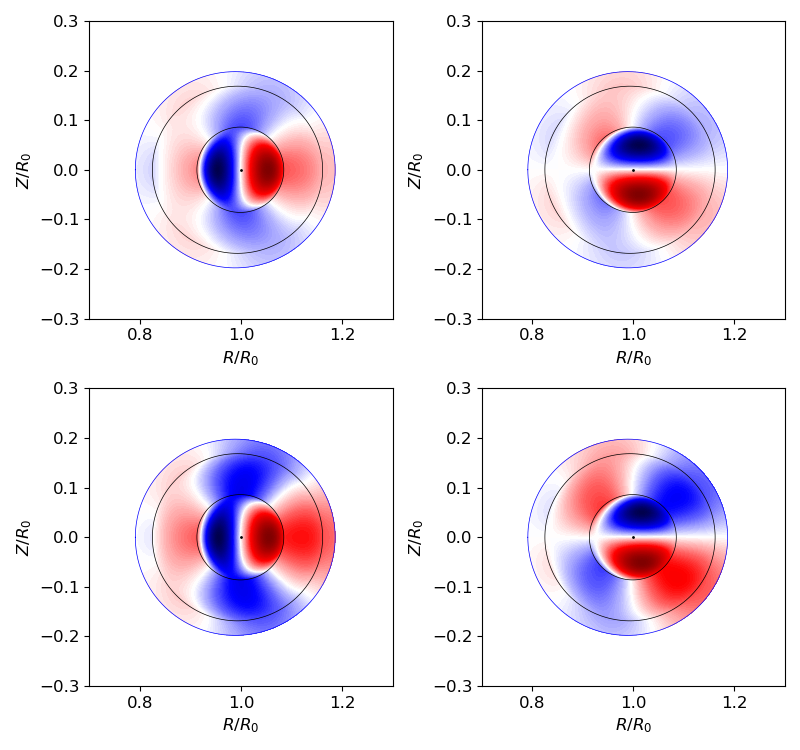}}
\caption{Contour plots of the real (left columns) and imaginary (right columns) parts the perturbed magnetic flux, $\psi(r,\theta)$, associated with free-boundary, $n=1$, tearing modes that only reconnect magnetic flux at the $q=1$  (top panels) and the $q=2$ (bottom panels)  surfaces in a 
circular cross-section plasma equilibrium characterized by $q_0=0.8$, $q_a= 2.8$, $p_p=2.0$,
$a=0.2$, and $\beta_0 =0.004$.  The
black circles show the locations of the $q=1$ and $q=2$  surfaces. See caption to Fig.~\ref{fig9}.\label{fig18}}
\end{figure}

Figures~\ref{fig17} and \ref{fig18} show the poloidal harmonics and contours of the perturbed magnetic flux, $\psi(r,\theta)$, associated
with the free-boundary $1/1$ and the $2/1$ modes in a  plasma equilibrium that is not close to the ideal stability boundary. It is clear from Fig.~\ref{fig17} that both modes are essentially $m=1$ modes inside the $q=1$ surface, $m=2$ modes between
the $q=1$ and $q=2$  surfaces, and $m=3$ modes outside the $q=2$  surface. Although both modes look fairly similar to
one another (modulo a multiplicative constant), there is a slight difference in that the $2/1$ mode has $\psi_{m=2}\neq 0$ at the $q=2$ rational surface, whereas the $1/1$ mode has 
$\psi_{m=2}=0$ .

\begin{figure}
\centerline{\includegraphics[width=0.9\textwidth]{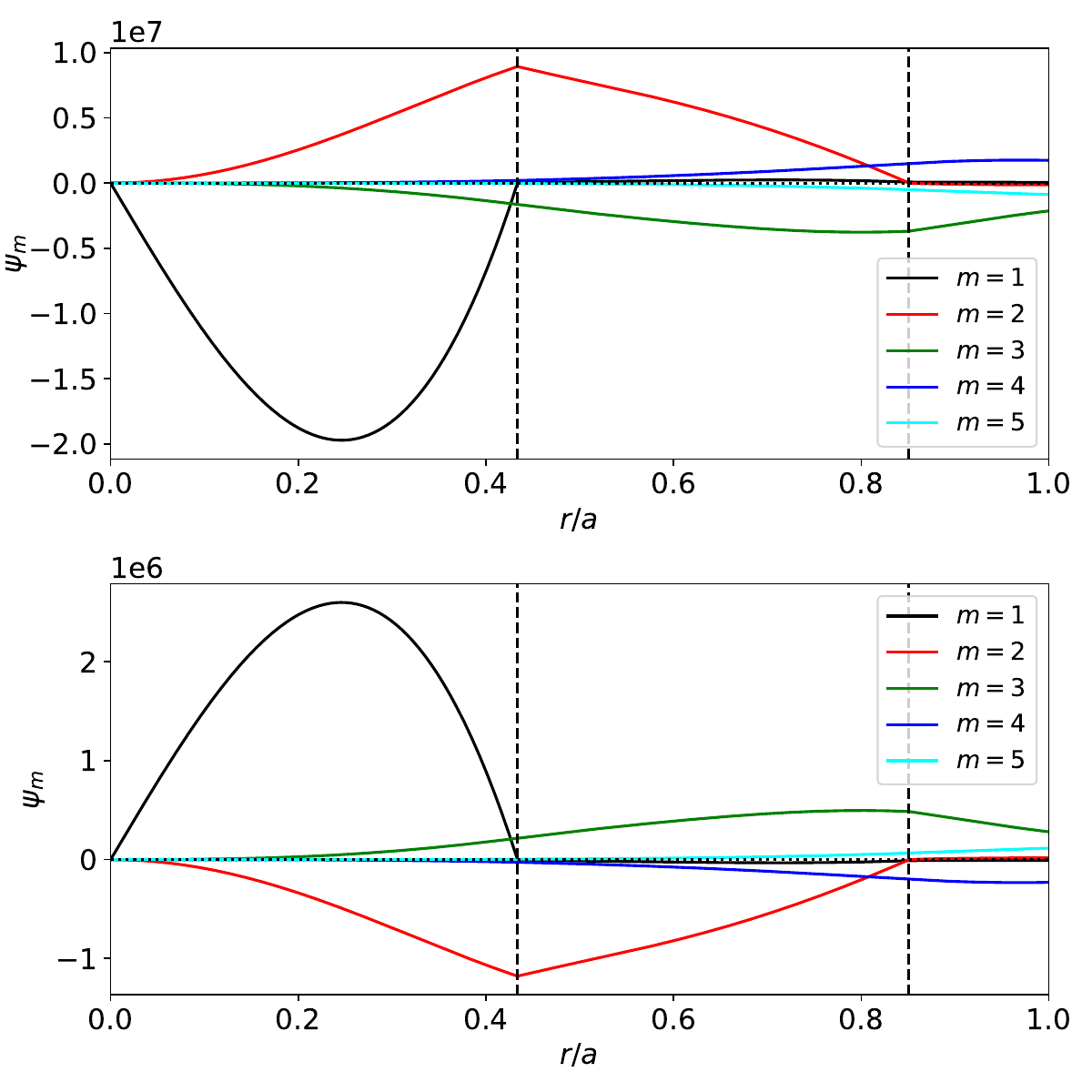}}
\caption{Poloidal harmonics of the perturbed magnetic flux, $\psi(r,\theta)$,  associated with free-boundary, $n=1$, tearing modes that only reconnect magnetic flux at the $q=2$  (top panel) and the $q=2$ (bottom panel) surfaces in a marginally-ideally-stable, 
circular cross-section,  plasma equilibrium characterized by $q_0=0.8$, $q_a= 2.8$, $p_p=2.0$, 
$a=0.2$, and $\beta_0=0.00708$. See caption to Fig.~\ref{fig17}.\label{fig19}}
\end{figure}

\begin{figure}
\centerline{\includegraphics[width=0.9\textwidth]{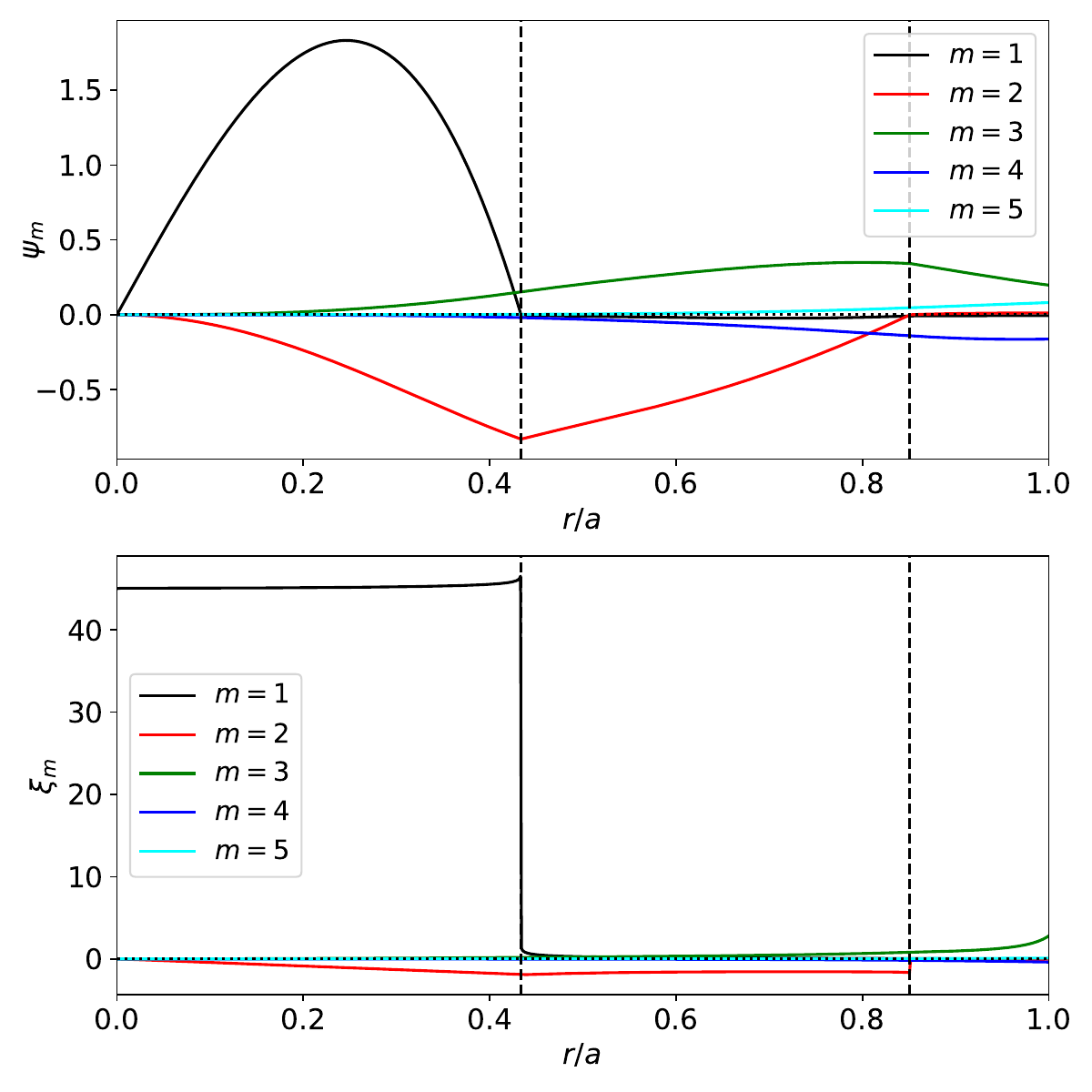}}
\caption{Poloidal harmonics of the perturbed magnetic flux, $\psi(r,\theta)$ (top panel), and the radial plasma displacement, $\xi^r(r,\theta)$,  associated with a free-boundary,
marginally-stable, $n=1$, ideal mode in a 
circular cross-section plasma equilibrium characterized by $q_0=0.8$, $q_a= 2.8$, $p_p=2.0$, 
$a=0.2$, and $\beta_0=0.00708$. See caption to Fig.~\ref{fig17}.\label{fig20}}
\end{figure}

\begin{figure}
\centerline{\includegraphics[width=\textwidth]{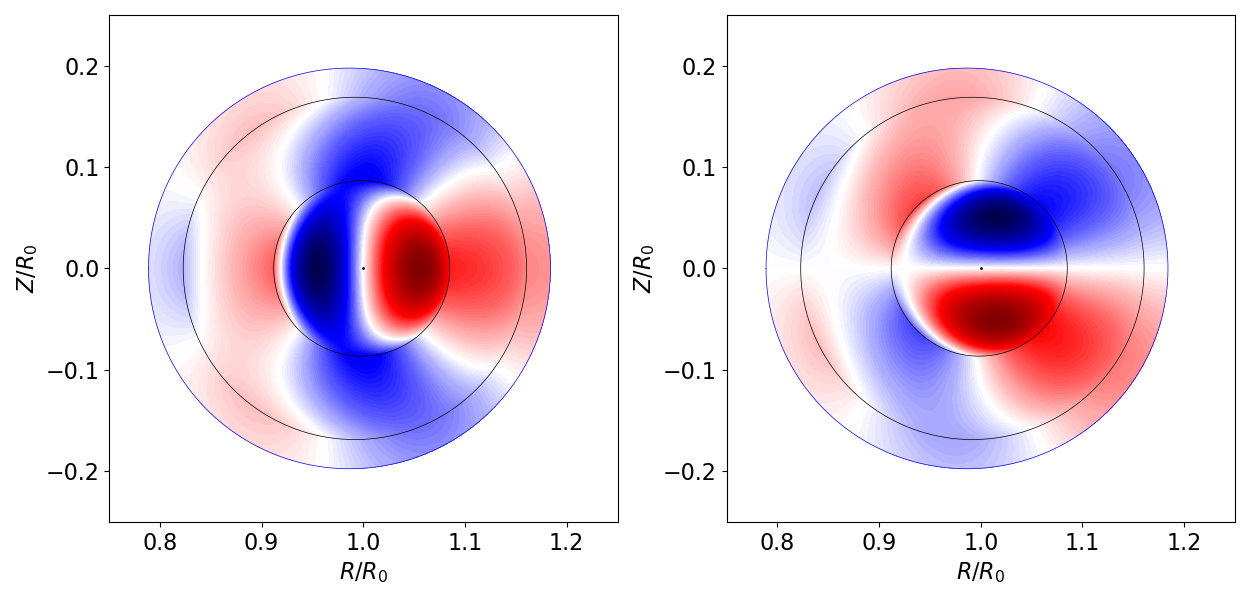}}
\caption{Contour plots of the real (left panel) and imaginary (right panel) parts the perturbed magnetic flux, $\psi(r,\theta)$, associated with a free-boundary, marginally-stable, $n=1$,
ideal mode mode  in a 
circular cross-section plasma equilibrium characterized by $q_0=0.8$, $q_a= 3.6$, $p_p=2.0$, 
$a=0.2$, and $\beta_0=0.00708$. See captions to Figs.~\ref{fig9} and \ref{fig21}. \label{fig21}}
\end{figure}

Figure~\ref{fig19} shows the poloidal harmonics of the perturbed magnetic flux associated
with the free-boundary $1/1$ and the $2/1$ modes  in a  plasma equilibrium 
that lies exactly at the ideal stability boundary. Figures~\ref{fig20} and \ref{fig21} show the poloidal harmonics and contours of the perturbed magnetic flux, as well
as the poloidal harmonics of the radial
plasma displacement, $\xi^r(r,\theta)$, associated with the marginally-stable, $n=1$, ideal mode at the stability boundary. It is
clear, from a comparison of Fig.~\ref{fig19} and the top panel of Fig.~\ref{fig20}, that at the ideal stability boundary the eigenfunctions of the two
tearing modes have both become identical with that of the marginally-stable ideal mode (modulo an arbitrary multiplicative constant). Thus, we again deduce that, as
an ideal stability boundary is approached, all of the decoupled tearing modes in the plasma 
effectively morph into the ideal mode, but that the growth-rate and real frequency of a given ideal-tearing-mode is determined by the 
resistive layer solution at the rational surface at which it reconnects magnetic flux. As is clear from Fig.~\ref{fig16}, it is again the case that the ideal mode that reconnects flux closest to the edge of the plasma has the largest growth-rate. 

Finally, according to the bottom panel of Fig.~\ref{fig20} and Fig.~\ref{fig21}, the marginally-stable ideal mode at the stability boundary is a $1/1$
internal-kink mode in which the $m=1$ plasma displacement is almost constant inside the $q=1$ surface, and essentially zero
outside the surface.\cite{freidberg,wesson} Note, in particular, that the harmonics of the plasma displacement are all very small at the 
plasma boundary.

\subsubsection{Stabilizing Effect of an Ideal Wall}
\begin{figure}
\centerline{\includegraphics[width=0.9\textwidth]{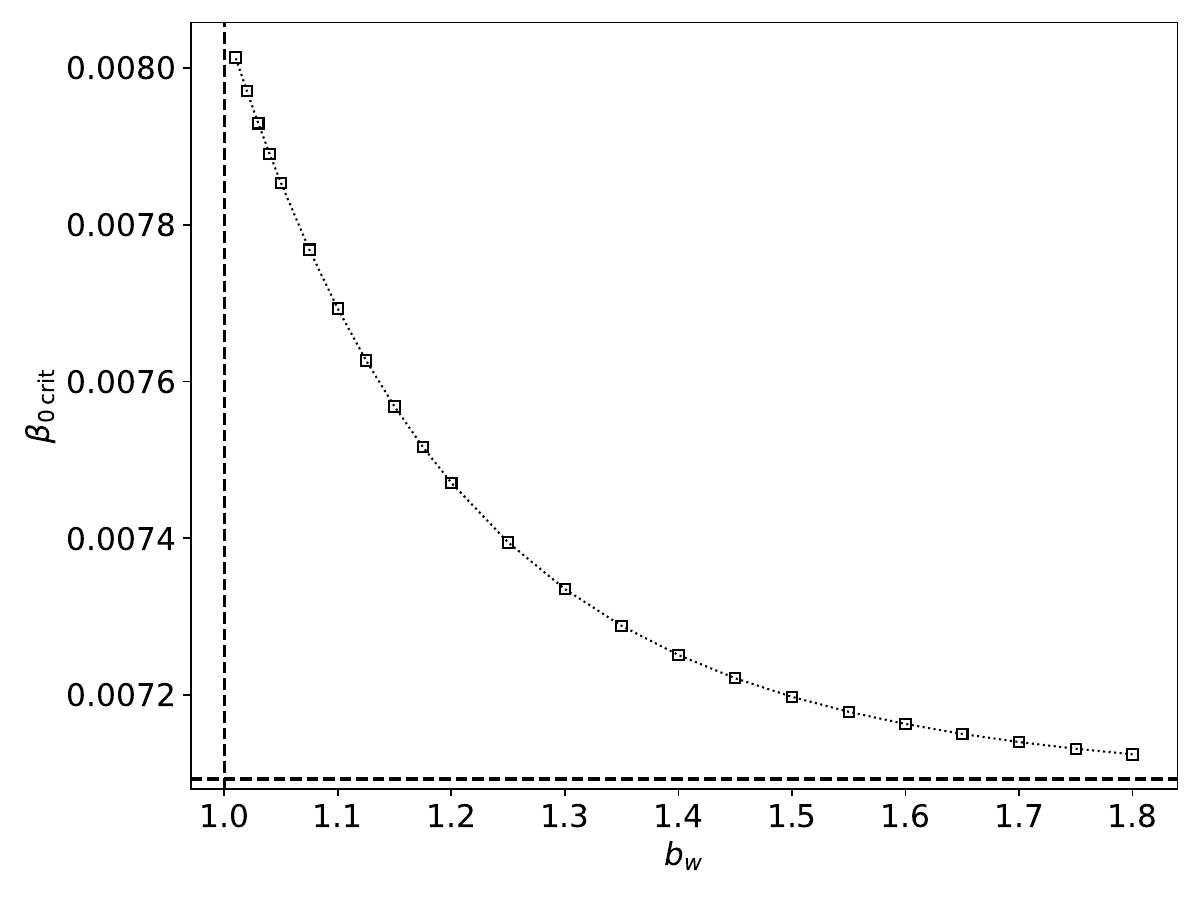}}
\caption{Variation of the critical central beta value at marginal ideal stability, $\beta_{0\,{\rm crit}}$, with the relative radius, $b_w$, of the ideal wall surrounding the plasma in 
a circular cross-section plasma equilibrium characterized by $q_0=0.8$, $q_a= 2.8$, $p_p=2.0$, and
$a=0.2$. The horizontal dashed line corresponds to the no-wall  marginal stability boundary, $\beta_{0\,{\rm crit}}=0.00708$. The vertical dashed line corresponds to $b_w=1.0$.  \label{fig22}}
\end{figure}

Figure~\ref{fig22} shows how the critical central beta value at marginal ideal stability, $\beta_{0\,{\rm crit}}$, varies with the
relative radius, $b_w$, of an ideal wall that is situated in the vacuum region surrounding the plasma.  It is clear that the
presence of the wall has a stabilizing effect on the ideal mode.   However, unlike the case discussed in Sect.~\ref{stabw1}, a close-fitting wall is incapable of
completely stabilizing the ideal mode. In other words,  $\beta_{0\,{\rm crit}}$ remains finite as $b_w\rightarrow 1$. This is in accordance with cylindrical theory, according to which a close-fitting ideal
wall is capable of stabilizing an ideal external-kink mode, but not an ideal internal-kink mode.\cite{freidberg,wesson}
Note that the TJ code cannot calculate the ideal $\delta W$ for a fixed-boundary (i.e., $b_w=1$)  plasma because, in this
case, the vacuum energy, $\delta W_v$, is infinite. However, a fixed-boundary plasma is an idealization. No real tokamak
plasma possesses an ideal wall that lies exactly at the plasma boundary: there is always a small vacuum gap
between the boundary and the wall. We conclude that, in practice, the relatively simple second test for ideal stability
discussed in Sect.~\ref{determine} is capable of determining the ideal stability of both internal-kink and external-kink
modes, even when the wall surrounding the plasma is very close to the plasma boundary. In other words, the
relatively complicated first test for the ideal stability of fixed-boundary modes discussed in Sect.~\ref{determine}
seems unnecessary. 

\section{Summary and Conclusions}\label{conc}
This paper documents three major improvements to the TJ toroidal tearing stability code.\cite{tj} The first improvement is a repurposing of the
ideal-MHD solutions in the outer-region in order to determine the eigenvalues of the ideal perturbed potential energy matrix. (See Appendix~\ref{ideal}.)
This new capability enables TJ to accurately identify ideal stability boundaries. The second improvement allows for the presence of
a perfectly conducting wall in the vacuum region surrounding the plasma. (See Appendix~\ref{vacuum}.) This new capability permits the TJ code 
to calculate the stabilizing effect of a perfectly conducting wall on both ideal and resistive modes. The third improvement is the addition of
a resistive layer model that enables TJ to determine the  growth-rates and rotation frequencies of the various tearing modes to which a
given plasma equilibrium is subject. (See Appendices~\ref{lmodel} and \ref{asym}.) The layer model incorporates 
plasma resistivity, plasma inertia, electron and ion diamagnetic flows, the decoupling of the electron and ion fluids on lengthscales below the ion sound radius, perpendicular
momentum transport, parallel and perpendicular energy transport, and the stabilizing effect of average magnetic field-line curvature.

Section~\ref{benchmark} describes a successful benchmarking exercise of the TJ code against the STRIDE\,\cite{aglas1} toroidal tearing mode code.

Finally, in Sect.~\ref{invest}, the new capabilities of the TJ code are used to investigate the stability of tearing modes in tokamak plasmas as
an ideal stability boundary, associated with either an external-kink or an internal-kink mode, is  approached. A number of conclusion can be drawn
from this investigation. 

The first conclusion is that, when using the asymptotic matching approach to determining tearing mode stability, it is necessary to
solve both halves of the problem, and then combine them together, in order to gain a complete picture. In other words, only calculating the tearing
stability matrix from the ideal-MHD solution in the outer region can be misleading because, when the resistive layer solutions in the
various segments of the inner region are incorporated into the calculation, it turns out that modes with larger tearing stability indices do not
necessarily have  larger growth-rates. 

The second conclusion is that a realistic resistive layer model must be employed in the calculation. 
In particular, the conventional single-fluid resistive-MHD layer model is completely inadequate, because it fails to take into account the fact that
the ion and electron fluid velocities are significantly different from one another as a consequence of diamagnetic flows. As is well known,
diamagnetic flows strongly influence both the growth-rates and rotation frequencies of tearing modes and resistive-kink modes (which is
what tearing modes become when they are very unstable).\cite{ara} Another problem with a  single-fluid resistive-MHD model is that,
unless it takes parallel thermal transport into account, it significantly overestimates the stabilizing effect of average magnetic
field-line curvature on tearing modes.\cite{lut,con1}  Of course, parallel thermal transport in tokamak plasmas is
non-diffusive in nature, because the mean-free-path between collisions is very much longer than the dimensions of the plasma.\cite{hel,hel1}
This fact  needs to be taken into account in the layer model, otherwise the calculation of the stabilizing  effect of average magnetic
field-line curvature will give incorrect results. (See Sect.~\ref{match}.)

The third conclusion is that, for a given toroidal mode number, a tokamak plasma is subject to as many linearly independent tearing modes as
there are rational surfaces in the plasma. Suppose that there are $K$ surfaces. The $K$ tearing modes are coupled together by the ideal-MHD solution in the outer region (as a
consequence of the Shafranov shifts and shaping of equilibrium magnetic flux-surfaces). However, under normal circumstances,
when the layer solutions in the inner region are taken into account, 
this coupling is overridden by sheared E-cross-B and diamagnetic flows, giving rise to $K$ decoupled tearing modes that only
reconnect magnetic flux at one particular rational surface in the plasma, and whose growth-rates and rotation frequencies are determined by
the layer solution at this surface.\cite{am1} 

The fourth conclusion is that, as an ideal stability boundary associated with either an ideal external-kink or an internal-kink
mode is approached, all of the elements of the tearing stability matrix tend to infinity, and the eigenfunctions of the $K$ decoupled tearing modes
all morph into the eigenfunction of the marginally-stable ideal mode. However, the growth-rate
and rotation frequency of each ideal-tearing-mode is determined by the layer physics at the rational surface at which it reconnects
magnetic flux. We find that the growth-rate of the ideal-tearing-mode whose reconnection surface lies closest to the edge of the plasma
 is the one that tends to a very large value as the stability boundary is approached, even when the boundary is that associated
with an internal-kink mode. 

The fifth conclusion is as follows. The off-diagonal elements of the tearing stability index parameterize the coupling of the
various tearing modes due to the ideal-MHD solution in the outer region. The fact that these elements tend to infinity as
an ideal stability boundary is approached implies that the coupling becomes very strong in this limit. Hence, as the boundary is
approached, we would eventually expect  a bifurcation in the equilibrium plasma  rotation profile to be triggered  by the nonlinear electromagnetic
torques that develop at the rational surfaces.\cite{am1,bif} This bifurcation would cause the various ideal-tearing-modes to
lock together to form a single ideal mode that is more unstable than any of the original modes. Note, however,  that this is a strictly nonlinear effect. 

The final conclusion is that the relatively simple second test for ideal stability, described in Sect.~\ref{determine}, that involves calculating
the eigenvalues of the sum of the perturbed ideal plasma potential energy and vacuum energy matrices,\cite{dcon} is perfectly adequate for
determining the ideal stability boundaries associated with both external-kink and internal-kink modes, even when an ideal wall lies
very close to the plasma boundary. The only case that the second test cannot deal with is the unrealistic case of determining
the ideal stability boundary of an internal-kink mode when the wall exactly corresponds to the plasma boundary. This conclusion  suggests
that the much more complicated first test for the ideal stability of fixed-boundary modes,\cite{dcon} described in Sect.~\ref{determine}, is generally unnecessary. 

\section*{Acknowledgements}
This research was directly funded by the U.S.\ Department of Energy, Office of Science, Office of Fusion Energy Sciences, under  contract DE-SC0021156. 
The author would like to thank Daniel Burgess, Nikolas Logan, Jong-Kyu~Park and J.-M.~Lee for help with getting the STRIDE code to
run reliably. 

\section*{Data Availability Statement}
The digital data used in the figures in this paper can be obtained from the author upon reasonable request. The TJ and TEAR codes are freely 
available at {\tt https://github.com/rfitzp/TJ}. 

\appendix

\section{General Definitions}
\subsection{Normalization}\label{norm}
Unless otherwise specified, all lengths in this paper are normalized to  the major radius of the plasma magnetic axis, $R_0$. All magnetic field-strengths
are normalized to the  toroidal field-strength at the magnetic axis, $B_0$. All currents are normalized to $B_0\,R_0/\mu_0$. All current densities are normalized to $B_0/(\mu_0\,R_0)$.  All plasma pressures are normalized to $B_0^{\,2}/\mu_0$.
All toroidal electromagnetic torques are normalized to $B_0^{\,2}\,R_0^{\,3}/\mu_0$. All energies are normalized to $B_0^{\,2}\,R_0^{\,3}/\mu_0$. 

\subsection{Coordinates}\label{coord}
Let $R$, $\phi$, $Z$ be right-handed cylindrical coordinates whose Jacobian 
is $ (\nabla R\times \nabla\phi\cdot\nabla Z)^{-1} = R$, 
and whose symmetry axis corresponds to the symmetry axis of the axisymmetric toroidal plasma equilibrium. 

Let $r$, $\theta$, $\phi$ be right-handed flux-coordinates whose
Jacobian is
\begin{equation}\label{jac}
{\cal J}(r,\theta)\equiv (\nabla r\times \nabla\theta\cdot\nabla\phi)^{-1}= r\,R^{\,2}.
\end{equation}
Note that $r=r(R,Z)$ and $\theta=\theta(R,Z)$. 
The magnetic axis corresponds to $r=0$, and the plasma-vacuum interface to $r=a$. Here, $a\ll 1$ is the effective inverse aspect-ratio of the plasma. 

\subsection{Plasma Equilibrium}\label{equilb}
Consider a tokamak plasma equilibrium whose magnetic field takes the form
\begin{equation}
{\bf B}(r,\theta) = f(r)\,\nabla\phi\times \nabla r + g(r)\,\nabla\phi = f\,\nabla(\phi-q\,\theta)\times \nabla r,
\end{equation}
where
$q(r) = r\,g/f$ is the safety-factor. Note that ${\bf B}\cdot\nabla r=0$, which implies that $r$ is a magnetic flux-surface label.
Furthermore,   $B^r=0$, $B^\theta= f/{\cal J}$ and  $B^\phi =f\,q/{\cal J}$. Here, superscript/subscript denote contravariaint/covariant
vector components. 

Equilibrium force balance requires that
$ \nabla P={\bf J}\times {\bf B}$, 
where $P(r)$ is the equilibrium scalar plasma pressure, and ${\bf J}=\nabla\times {\bf B}$ the equilibrium plasma current density. 

\subsection{Perturbed Magnetic Field}\label{perturbed}
All perturbed quantities are assumed to vary toroidally as $\exp(-{\rm i}\,n\,\phi)$,
where the positive integer  $n$ is the toroidal mode number of the perturbation. 
According to Eqs.~(47), (78), (99), and (101) of Ref.~\onlinecite{tj}, the radial and toroidal components of the perturbed magnetic field are written 
\begin{align}\label{a3}
{\cal J}\,b^r &={\rm i}\,\psi(r,\theta)= \sum_m \psi_m(r)\,\exp(\,{\rm i}\,m\,\theta),\\[0.5ex]
b_\phi &= x(r,\theta)= n\,z(r,\theta) = n\sum_m z_m(r) \,\exp(\,{\rm i}\,m\,\theta),
\end{align}
where
\begin{equation}
z_m(r) = \frac{Z_m(r) + k_m\,\psi_m(r)}{m-n\,q}.
\end{equation}
Here, the (not necessarily positive) integers $m$ are poloidal mode numbers, and the sum is over all mode numbers included in the calculation. 
Furthermore, 
$k_m(r)$ is real, and is specified in Eq.~(100) Ref.~\onlinecite{tj}. 

\subsection{Behavior in Vicinity of Rational Surface}\label{rational}
Suppose that there are $K$ rational surfaces in the plasma. Let the $k$th surface be located at $r=r_k$, and possess the resonant poloidal
mode number $m_k$. By definition, $m_k-n\,q(r_k)=0$. 
According to Sect.~V.B of Ref.~\onlinecite{tj},  the non-resonant poloidal harmonics of the solutions to the
ideal-MHD equations in the outer region  are continuous across the surface. On the other hand, the tearing-parity components of the resonant poloidal
harmonics of the solutions behave locally as
\begin{align}
\psi_{m\,k}(r_k+x) &= A_{L\,k}\,|x|^{\nu_{L\,k}} + A_{S\,k}\,|x|^{\nu_{S\,k}},\\[0.5ex]
Z_{m\,k}(r_k+x)&= \frac{\nu_{L\,k}}{L_0}\,A_{L\,k}\,|x|^{\nu_{L\,k}}+ \frac{\nu_{S\,k}}{L_0}\,A_S\,|x|^{\,\nu_{S\,k}},
\end{align}
where
\begin{align}
\nu_{L\,k} &= \frac{1}{2}-\sqrt{-D_{I\,k}},\\[0.5ex]
\nu_{S\,k} &= \frac{1}{2}+\sqrt{-D_{I\,k}},\\[0.5ex]
D_{I\,k}&= - \left[\frac{2\,(1-q^2)}{s^2}\,r\,\frac{dP}{dr}\right]_{r_k} -\frac{1}{4},\label{di}\\[0.5ex]
L_0 &= -\left(\frac{L_{m_k}^{\,m_k}}{m_k\,s}\right)_{r_k},\\[0.5ex]
L_{m_k}^{\,m_k}(r) &= m_k^2\,c_{m_k}^{\,m_k}(r) + n^2\,r^2,\\[0.5ex]
c_{m_k}^{\,m_k}(r) &=\oint|\nabla r|^{-2}\,\frac{d\theta}{2\pi},
\end{align}
and $s(r)= d\ln q/d\ln r$. Here, the tearing-parity component is such that $\psi_{m_k}(r_k-x)=\psi_{m_k}(r_k+x)$ and $Z_{m_k}(r_k-x)=Z_{m_k}(r_k+x)$. 
Moreover, $A_{L\,k}$ is termed the coefficient of the ``large'' solution, whereas $A_{S\,k}$ is the coefficient of the ``small'' solution. Furthermore, $D_{I\,k}$ is the ideal
Mercier interchange parameter (which needs to be negative to ensure stability to localized interchange modes),\cite{mercier,ggj,ggj1} and $\nu_{L\,k}$ and $\nu_{S\,k}$
are termed the Mercier indices. 

It is helpful to define the quantities\,\cite{tj}
\begin{align}\label{Psidef}
{\mit\Psi}_k&= r_k^{\,\nu_{L\,k}}\left(\frac{\nu_{S\,k}-\nu_{L\,k}}{L_{m_k}^{\,{m_k}}}\right)^{1/2}_{r_k} A_{L\,k},\\[0.5ex]
{\mit\Delta\Psi}_k &= r_k^{\,\nu_{S\,k}}\left(\frac{\nu_{S\,k}-\nu_{L\,k}}{L_{m_k}^{\,m_k}}\right)^{1/2}_{r_k} 2\,A_{S\,k},\label{edpp}
\end{align}
at each rational surface in the plasma. Here, the complex parameter ${\mit\Psi}_k$ is a measure of the reconnected helical magnetic flux at the $k$th rational surface, whereas
the complex parameter ${\mit\Delta\Psi}_k$ is a measure of the strength of a localized current sheet that flows parallel to the equilibrium magnetic field at the surface. 
The net toroidal electromagnetic torque acting on the plasma is\,\cite{tj}
\begin{equation}\label{torque}
T_\phi = 2\pi^2\,n\,\sum_{k=1,K}{\rm Im}({\mit\Psi}_k^\ast\,{\mit\Delta\Psi}_k).
\end{equation}

\subsection{Tearing Mode Dispersion Relation}\label{disp}
According to Sect.~VIII.D of Ref.~\onlinecite{tj}, the tearing mode dispersion relation takes the form
\begin{equation}\label{dispersion}
\sum_{k'=1,K}\left(E_{kk'}-{\mit\Delta}_k\,\delta_{kk'}\right){\mit\Psi}_{k'} = 0
\end{equation}
for $k=1,K$, 
where $E_{kk'}$ is an Hermitian matrix determined from the solution of the ideal-MHD equations in the outer region, and ${\mit\Delta}_k\equiv {\mit\Delta\Psi}_k/{\mit\Psi}_k$ is
a complex quantity that characterizes the tearing-parity response of the resistive layer at the $k$th rational surface to the outer solution. 
In general, ${\mit\Delta}_k$ is a function of the growth-rate and phase-velocity of the reconnected helical magnetic flux at the surface. 

\section{Calculation of Ideal Stability}\label{ideal}

\subsection{Plasma Perturbation}
The perturbed magnetic field associated with an ideal perturbation is written\,\cite{freidberg,ideal}
\begin{equation}
{\bf b} = \nabla\times (\bxi\times{\bf B}),
\end{equation}
where $\bxi$ is the plasma displacement.  According to Eqs.~(2), (25), (26), and
(101) of Ref.~\onlinecite{tj}, 
\begin{align}\label{e9}
{\cal J}\,b^r&=\left(\frac{\partial}{\partial\theta}-{\rm i}\,n\,q\right)y = {\rm i}\,\psi(r,\theta),
\end{align}
where
\begin{align}\label{e10}
y(r,\theta)&= f\,\xi^r.
\end{align}
Furthermore,  Eqs.~(38), (39),  (48), and (78) of Ref.~\onlinecite{tj} yield
\begin{align}
b_\theta = -\frac{\alpha_g}{{\rm i}\,n}\left(\frac{\partial}{\partial\theta} - {\rm i}\,n\,q\right)y +\alpha_p\,R^{\,2}\,y +{\rm i}\,\frac{\partial z}{\partial\theta},
\end{align}
where
\begin{align}
\alpha_p(r) &= \frac{r\,dP/dr}{f^2},\label{ap}\\[0.5ex]
\alpha_g (r)&= \frac{dg/dr}{f}.\label{ag}
\end{align}
Thus,
\begin{equation}\label{e15}
{\bf B}\cdot{\bf b} -\xi^r\,\frac{dP}{dr} = B^\theta\,b_\theta+B^\phi\,b_\phi - \xi^r\,\frac{dP}{dr}=
\frac{{\rm i}\,f}{\cal J}\left(\frac{\partial}{\partial\theta}- {\rm i}\,n\,q\right)\left(\frac{\alpha_g}{n}\,y+z\right).
\end{equation}

\subsection{Plasma Potential Energy}
The perturbed ideal-MHD  force operator in the plasma takes the well-known form\,\cite{freidberg,ideal}
\begin{equation}
{\bf F}(\bxi)= \nabla({\mit\Gamma}\,P\,\nabla\cdot\bxi) - {\bf B}\times(\nabla\times {\bf b})+\nabla(\bxi\cdot\nabla P)+{\bf J}\times  {\bf b},
\end{equation}
where ${\mit\Gamma}=5/3$ is the plasma ratio of specific heats. 
The perturbed  plasma potential energy in the region lying between the magnetic  flux-surfaces whose labels are $r_1$ and $r_2$ can be written\,\cite{ideal}
\begin{align}
\delta W_{12} &= \frac{1}{2}\int_{r_1}^{r_2}\oint\oint\left[{\mit\Gamma}\,P\,(\nabla\cdot\bxi^\ast)\,(\nabla\cdot\bxi)+ \nabla\times (\bxi^\ast\times {\bf B})\cdot {\bf b}
+(\nabla\cdot\bxi^\ast)\,(\bxi\cdot\nabla P)\right.\nonumber\\[0.5ex]&\phantom{=}
\left.+{\bf J}\times \bxi^\ast\,\cdot{\bf b}\right]{\cal J}\,dr\,d\theta\,d\phi.
\end{align}

Now,
\begin{align}
{\mit\Gamma}\,P\,(\nabla\cdot\bxi^\ast)\,(\nabla\cdot\bxi)&=\nabla\cdot({\mit\Gamma}\,P\,\bxi^\ast\,\nabla\cdot\bxi)-\bxi^\ast\cdot\nabla(
{\mit\Gamma}\,P\,\nabla\cdot\bxi),\\[0.5ex]
 \nabla\times (\bxi^\ast\times {\bf B})\cdot {\bf b}&= \nabla\cdot[(\bxi^\ast\times {\bf B})\times {\bf b}] + \bxi^\ast\times {\bf B}\cdot\nabla\times {\bf b},\\[0.5ex]
 (\nabla\cdot\bxi^\ast)\,(\bxi\cdot\nabla P)&=\nabla\cdot(\bxi^\ast\,\bxi\cdot\nabla P) -\bxi^\ast\cdot\nabla(\bxi\cdot\nabla P),
\end{align}
so
\begin{align}
\delta W_{12} &= \frac{1}{2}\int_{r_1}^{r_2}\oint\oint\left\{\nabla\cdot[{\mit\Gamma}\,P\,\bxi^\ast\,\nabla\cdot\bxi+ (\bxi^\ast\times {\bf B})\times {\bf b}+\bxi^\ast\,\bxi\cdot\nabla P]\right.\nonumber\\[0.5ex]&\left.-\bxi^\ast\cdot[ \nabla({\mit\Gamma}\,P\,\nabla\cdot\bxi) - {\bf B}\times(\nabla\times {\bf b})+\nabla(\bxi\cdot\nabla P)+{\bf J}\times  {\bf b}]\right\}{\cal J}\,dr\,d\theta\,d\phi,
\end{align}
which yields
\begin{align}
\delta W_{12} &= \frac{1}{2}\left(\oint\oint{\cal J}\,\nabla r\cdot[{\mit\Gamma}\,P\,\bxi^\ast\,\nabla\cdot\bxi+ (\bxi^\ast\times {\bf B})\times {\bf b}+\bxi^\ast\,\bxi\cdot\nabla P]\,d\theta\,d\phi\right)_{r_1}^{r_2}\nonumber\\[0.5ex]
&-\frac{1}{2}\int_{r_1}^{r_2}\oint\oint \bxi^\ast\cdot{\bf F}(\bxi)\,{\cal J}\,dr\,d\theta\,d\phi,
\end{align}
or
\begin{align}
\delta W_{12} &= \frac{1}{2}\left[\oint\oint{\cal J}\,\xi^{r\,\ast}\left({\mit\Gamma}\,P\,\nabla\cdot\bxi -{\bf B}\cdot{\bf b} + \xi^r\,\frac{dP}{dr}\right)d\theta\,d\phi\right]_{r_1}^{r_2}\nonumber\\[0.5ex]&\phantom{=}-\frac{1}{2}\int_{r_1}^{r_2}\oint\oint\bxi^\ast\cdot{\bf F}(\bxi)\,{\cal J}\,dr\,d\theta\,d\phi,
\end{align}
where use has been made of the fact that ${\bf B}\cdot\nabla r = 0$.  However, in the TJ code, the marginally-stable ideal-MHD  plasma perturbation in the outer region is calculated assuming that ${\bf F}(\bxi)={\bf 0}$ and
$\nabla\cdot\bxi=0$.\cite{tj}   Thus, making use of Eqs.~(\ref{e10}) and (\ref{e15}), we obtain 
\begin{equation}
\delta W_{12} = \frac{1}{2}\int_{r_1}^{r_2}\left[-{\rm i}\,y^\ast\left(\frac{\partial}{\partial\theta}-{\rm i}\,n\,q\right)\left(\frac{\alpha_g}{n}\,y+z\right)d\theta\,d\phi\right]_{r_1}^{r_2},
\end{equation}
which reduces to
\begin{equation}
\delta W_{12}= \pi^2\left[\sum_m\,y_m^\ast\,(m-n\,q)\left(\frac{\alpha_g}{n}\,y_m+z_m\right)\right]_{r_1}^{r_2},
\end{equation}
where $y(r,\theta)=\sum_m y_m(r)\,\exp(\,{\rm i}\,m\,\theta)$. 

Now, according to Eq.~(98) of Ref.~\onlinecite{tj}, 
\begin{align}\label{epsi}
y_m(r) &= \left(\frac{\psi_m}{m-n\,q}\right)_r.
\end{align}
Thus, we get 
\begin{equation}
\delta W_{12}=\pi^2\left(\sum_m\psi_m^\ast\,\chi_m\right)_{r_1}^{r_2}
\end{equation}
where
\begin{align}\label{e29}
\chi_m(r)&=
 \left(\frac{\tilde{k}_m\,\psi_m+Z_m}{m-n\,q}\right)_r,\\[0.5ex]
\tilde{k}_m(r) &=\left( k_m + \frac{\alpha_g}{n} \right)_r= \left[\frac{\alpha_g\,(m\,q\,c_m^{\,m}+n\,r^2)+\alpha_p\,m\,d_m^{\,m}}{m^2\,c_m^{\,m}+n^2\,r^2}\right]_r,\label{e30}
\end{align}
and use has been made of Eq.~(100) Ref.~\onlinecite{tj}. Here, $\tilde{k}_m(r)$ is real, and
\begin{align}
d_m^{\,m}(r) &=\oint|\nabla r|^{-2}\,R^{\,2}\,\frac{d\theta}{2\pi}.
\end{align}

\subsection{Magnetic Axis}
Let the $\psi_m(r)$ and the $\chi_m(r)$ be solutions of the ideal-MHD equations in the outer region that are well behaved at $r=0$. 
Such solutions vary as $r^{|m|}$. (For the special case $m=0$, $\chi_0\sim1$ and $\psi_0=0$.) It follows that\,\cite{tj}
\begin{equation}
\left(\sum_m\psi_m^\ast\,\chi_m\right)_{0}=0.
\end{equation}
Hence, 
\begin{equation}\label{e1}
\delta W_p(r) =\pi^2\left(\sum_m\psi_m^\ast\,\chi_m\right)_r
\end{equation}
is the perturbed plasma potential energy in the region lying between the magnetic axis and the magnetic flux-surface whose label is $r$. 

\subsection{Rational Surfaces}
At the $k$th rational surface, $r=r_k$, the non-resonant components of $\psi_m$ and $\chi_m$ are continuous,  the 
large solution is absent (i.e., ${\mit\Psi}_k=0$), but the small solution is present  (i.e., ${\mit\Delta\Psi}_k=0$). (See Sect.~\ref{rational}.)
It is easily shown that $\sum_m \psi_m^\ast\,\chi_m$ remains finite at $r=r_k$ (which would not be the case if ${\mit\Psi}_k\neq 0$), and\,\cite{tj}
\begin{equation}
\left(\sum_m\psi_m^\ast\,\chi_m\right)_{r_{k-}}^{r_{k+}}=0.
\end{equation}
In other words, there is no contribution to the perturbed potential energy from the surface. Thus, Eq.~(\ref{e1}) holds even when the region between the
magnetic axis and the flux-surface whose label is $r$ contains rational surfaces. 
Thus, the total plasma potential energy is 
\begin{equation}\label{a32}
\delta W_p =\pi^2\left(\sum_m\psi_m^\ast\,\chi_m\right)_{a_-},
\end{equation}
where $r=a_-$ corresponds to an equilibrium magnetic flux that lies just inside the plasma-vacuum interface. 

\subsection{Toroidal Electromagnetic Torque}
The flux-surface averaged toroidal electromagnetic torque acting on the plasma   is\,\cite{tj,ideal}
\begin{align}\label{b27}
T_\phi& = {\rm i}\,n\,\pi^2\left(\sum_m\frac{Z_m^\ast\,\psi_m-\psi_m^\ast\,Z_m}{m-n\,q}\right)_{a_-}= {\rm i}\,n\,\pi^2\left(\sum_m
\chi_m^{\,\ast}\,\psi_m - \psi_m^{\,\ast}\,\chi_m\right)_{a_-}\nonumber\\[0.5ex]
&= 2\,n\,{\rm Im}(\delta W_p),
\end{align}
where use has been made of Eqs.~(\ref{e29}) and (\ref{e1}). 
However, according to Eq.~(\ref{torque}), this torque must be zero for ideal solutions, which are characterized by ${\mit\Psi}_k=0$ (i.e., zero reconnected magnetic
flux at the various rational surfaces in the plasma).\cite{tj}
It follows that
 the total plasma potential energy, $\delta W_p$, is necessarily a real quantity. 
 
\subsection{Marginally-Stable Ideal Eigenfunctions}\label{sideal}
We can construct a complete set of  marginally-stable ideal eigenfuctions in the outer region from the ideal-MHD solutions in the outer region calculated by the TJ code as follows:
\begin{align}
\psi_{m\,m'}^i(r)&= \psi_{m\,m'}^a(r)-\sum_{k=1,K}\psi_{m\,k}^u(r)\,{\mit\Pi}_{k\,m'}^a,\\[0.5ex]
Z_{m\,m'}^i(r)&= Z_{m\,m'}^a(r)-\sum_{k=1,K}Z_{m\,k}^u(r)\,{\mit\Pi}_{k\,m'}^a.
\end{align}
 Here, $m$ labels the poloidal harmonic of the solution,  $m'$ labels the solution itself, and $k$ labels the various rational surfaces in the plasma. There are as many solutions as there are poloidal harmonics included in the calculation. All other quantities are defined in Sects.~VIII.B--VIII.D of Ref.~\onlinecite{tj}. 
 It follows
from the definitions of the various quantities in the previous two equations  that the reconnected magnetic  fluxes at the rational surfaces in the plasma
associated with the ideal eigenfunctions are all zero (i.e., ${\mit\Psi}_k=0$ for all $k$). We can also write
\begin{equation}\label{e39}
\chi^i_{m\,m'}(r) = \left(\frac{\tilde{k}_{m}\,\psi^i_{m\,m'}+Z_{m\,m'}^i}{m-n\,q}\right)_r,
\end{equation}
where use has been made of Eq.~(\ref{e29}).

\subsection{Plasma Energy Matrix}\label{penergy}
We can write a general ideal solution just inside the plasma-vacuum interface as
\begin{align}
\psi_m(a) &= \sum_{m'}\psi_{m\,m'}^i(a)\,\alpha_{m'},\\[0.5ex]
\chi_m(a_-) &= \sum_{m'}\chi_{m\,m'}^i(a_-)\,\alpha_{m'},
\end{align}
where the $\alpha_m$ are arbitrary complex coefficients.  Note that the $\psi_m(r)$ are continuous across the plasma-vacuum interface, whereas the $\chi_m(r)$ are generally discontinuous. 
The previous two equations can be written more succinctly as
\begin{align}\label{e41}
\underline{\psi}&= \underline{\underline{\psi}}_{\,i}\,\underline{\alpha},\\[0.5ex]
\underline{\chi}&= \underline{\underline{\chi}}_{\,i}\,\underline{\alpha},\label{e42}
\end{align}
where $\underline{\psi}$ is the column vector of the $\psi_m(a)$ values, $\underline{\chi}$ is the column vector of the $\chi_m(a_-)$ values,
$\underline{\underline{\psi}}_{\,i}$ is the matrix of the $\psi_{mm'}^i(a)$ values, $\underline{\underline{\chi}}_{\,i}$ is the matrix of the $\chi_{mm'}^i(a_-)$ values,
and $\underline{\alpha}$ the column vector of the $\alpha_m$ values. 
Equation~(\ref{a32}) then becomes
\begin{equation}
\delta W_p=\pi^2\,\underline{\psi}^\dag\,\underline{\chi},
\end{equation} 
or
\begin{equation}\label{e44}
\delta W_p =\pi^2\, \underline{\alpha}^\dag\,\underline{\underline{\psi}}^{\dag}_{\,i}\,\underline{\underline{\chi}}_{\,i}\,\underline{\alpha}.
\end{equation}
The fact that $\delta W_p$ is real implies that
 \begin{equation}\label{e47}
 \underline{\underline{\psi}}^{\dag}_{\,i}\,\underline{\underline{\chi}}_{\,i}= \underline{\underline{\chi}}^{\dag}_{\,i}\,\underline{\underline{\psi}}_{\,i}.
 \end{equation}

Let us define the ``perturbed plasma potential energy matrix'', $\underline{\underline{W}}_{\,p}$, such that 
\begin{equation}\label{e48}
\underline{\underline{\chi}}_{\,i} = \underline{\underline{W}}_{\,p}\,\underline{\underline{\psi}}_{\,i}.
\end{equation}
 It is easily seen that
 \begin{align}
 \underline{\underline{\psi}}^{\dag}_{\,i}\,\underline{\underline{\chi}}_{\,i}= \underline{\underline{\psi}}^{\dag}_{\,i}\,\underline{\underline{W}}_{\,p}\,
 \underline{\underline{\psi}}_{\,i},\\[0.5ex]
 \underline{\underline{\chi}}^{\dag}_{\,i}\,\underline{\underline{\psi}}_{\,i}= \underline{\underline{\psi}}^{\dag}_{\,i}\,\underline{\underline{W}}_{\,p}^{\dag}\,
 \underline{\underline{\psi}}_{\,i}.
 \end{align}
 Making use of Eq.~(\ref{e47}), we obtain
 \begin{equation}
 \underline{\underline{\psi}}^{\dag}_{\,i}\,\underline{\underline{W}}_{\,p}\,
 \underline{\underline{\psi}}_{\,i}=
 \underline{\underline{\psi}}^{\dag}_{\,i}\,\underline{\underline{W}}_{\,p}^{\dag}\,
 \underline{\underline{\psi}}_{\,i},
 \end{equation}
 which implies that $\underline{\underline{W}}_{\,p}$ is an Hermitian matrix. 
  
\subsection{Total Potential Energy}
The total perturbed potential energy of the plasma-vacuum system can be written\,\cite{freidberg,ideal}
\begin{equation}\label{e79}
\delta W= \delta W_p + \delta W_v,
\end{equation}
where
\begin{equation}
\delta W_v = \frac{1}{2}\int_{a+}^\infty\oint\oint {\bf b}^\ast\cdot{\bf b} \,{\cal J}\,dr\,d\theta\,d\phi
\end{equation}
is the perturbed potential energy of the surrounding vacuum.\cite{freidberg,ideal} Here, $r=a_+$ corresponds to an equilibrium magnetic flux-surface that lies just outside the plasma-vacuum interface. 

\subsection{Vacuum Potential Energy}\label{vac}
In the vacuum region, we can write the curl- and divergence-free perturbed magnetic field in the form 
${\bf b} = {\rm i}\,\nabla V$,
where
$\nabla^2 V =0$.
Hence, the vacuum potential energy is
\begin{align}
\delta W_v &= \frac{1}{2}\int_{a+}^\infty\oint\oint\nabla V\,\cdot\nabla V^\ast\,{\cal J}\,dr\,d\theta\,d\phi\nonumber\\[0.5ex]
&= \frac{1}{2}\int_{a+}^\infty\oint\oint  \nabla\cdot(V\,\nabla V^\ast)\,{\cal J}\,dr\,d\theta\,d\phi=-\frac{1}{2}\left(\oint\oint {\cal J}\,\nabla r\cdot\nabla V^\ast\,V\,d\theta\,d\phi\right)_{a_+},
\end{align}
assuming that $V(r,\theta)\rightarrow 0$ as $r\rightarrow \infty$ if there is no wall surrounding the plasma, or  $\nabla V\cdot d{\bf S}=0$ if a perfectly-conducting wall
surrounds the plasma. 
But, Eq.~(210) of Ref.~\onlinecite{tj} implies that 
\begin{equation}
{\cal J}\,\nabla V\cdot\nabla r = \psi,
\end{equation}
so we deduce that
\begin{equation}\label{e87}
\delta W_v = -\frac{1}{2}\left(\oint\oint \psi^\ast\,V\,d\theta\,d\phi\right)_{a_{+}} =- \pi^2\left(\sum_m \psi_m^\ast\,V_m\right)_{a_{+}},
\end{equation}
where
$V(r,\theta)= \sum_m V_m(r)\,\exp(\,{\rm i}\,m\,\theta)$. 
However, making use of Eq.~(214) of Ref.~\onlinecite{tj}, we get
\begin{equation}\label{e88}
\delta W_v =-\pi^2\left(\sum_m\psi_m^\ast\,\chi_m\right)_{a_+},
\end{equation}
where
\begin{equation}\label{e89}
\chi_m=\frac{Z_m}{m-n\,q}.
 \end{equation}
 Note that the previous equation is consistent with Eq.~(\ref{e29}) because, according to Eq.~(\ref{e30}),  $\tilde{k}_m=0$ in the vacuum region, given that $\alpha_g=\alpha_p=0$ in the vacuum. 

Combining Eqs.~(\ref{a32}), (\ref{e79}),  and (\ref{e88}), we deduce that
\begin{equation}\label{e89a}
\delta W = \pi^2\left(\sum_m\psi_m^\ast\,J_m\right)_\epsilon,
\end{equation}
where
\begin{equation}\label{e90}
J_m= -\left[\chi_m\right]_{a_-}^{a_+}.
\end{equation}

\subsection{Boundary Current Sheet}
Now, $\alpha_p=\alpha_g=0$ at the plasma-vacuum interface, assuming that the  equilibrium current is zero there,
which implies that $k_m=\tilde{k}_m=0$ at the interface, where use has been made of Eq.~(\ref{e30}),
as well as Eq.~(100) of Ref.~\onlinecite{tj}.  It follows from Eqs.~(78) and (99) of Ref.~\onlinecite{tj}, combined with Eq.~(\ref{e89}),  that 
\begin{equation}\label{e91}
x_m= n\,\chi_m
\end{equation}
at the plasma-vacuum interface, where $x(r,\theta)=\sum_m x_m(r)\,\exp(\,{\rm i}\,m\,\theta)$. 
Suppose that there is a perturbed current sheet on the interface. 
Thus, if ${\bf K}$ is the current sheet density then Eqs.~(66) and (67) of Ref.~\onlinecite{tj} suggest that
\begin{align}\label{eb59}
{\cal J}\,K_m^{\,\theta} &=  n\,J_m,\\[0.5ex]
{\cal J}\,K_m^{\,\phi} &=m\,J_m,\label{eb60}
\end{align}
where use has been made of Eqs.~(\ref{e90}) and (\ref{e91}). 
Let us write
\begin{equation}\label{e94}
{\bf K} = {\rm i}\,\nabla J\times \nabla r,
\end{equation}
where $J(\theta,\phi)=J(\theta)\,\exp(-{\rm i}\,n\,\phi)$, which ensures that $\nabla\cdot{\bf K}=0$. It follows from (A8) and (A9) of Ref.~\onlinecite{tj} that
\begin{align}
{\cal J}\,K^\theta&= {\rm i}\,\frac{\partial J}{\partial \phi},\\[0.5ex]
{\cal J}\,K^\phi &=-{\rm i}\,\frac{\partial J}{\partial\theta}.
\end{align}
Hence, if we expand the previous two equations as Fourier series in $\theta$ then we reproduce Eqs.~(\ref{eb59}) and (\ref{eb60}), which implies
 that the $J_m$ defined in Eq.~(\ref{e90}) are the Fourier components of the $J(\theta)$ function introduced in Eq.~(\ref{e94}). 

As described in Ref.~\onlinecite{ideal}, the marginally-stable ideal eigenfunctions   feature a perturbed current sheet on the vacuum-plasma interface because they do not satisfy the
perturbed  pressure-balance boundary condition.
However, these current sheets are entirely fictitious. The true ideal eigenfunctions (which, unlike the marginally-stable ideal eigenfunctions, take plasma inertia into account)
satisfy the pressure-balance boundary condition (and, therefore, have no associated current sheets). Current sheets are permitted because the marginally-stable ideal eigenfunctions
are ``trial solutions'' used to determine ideal stability, rather than actual physical solutions. 

\subsection{Determination of Ideal Stability}\label{stab}
According to Eq.~(215) of Ref.~\onlinecite{tj}, combined with Eq.~(\ref{e89}), 
\begin{equation}
\chi_m(a_+)=\sum_{m'}\,H_{m\,m'}\,\psi_{m'}(a),
\end{equation}
where the ``vacuum matrix'', $H_{mm'}$, is Hermitian. (See Sect.~VI.G of Ref.~\onlinecite{tj}.)
Hence, it follows from Eq.~(\ref{e90}) that
\begin{equation}
J_m = \chi_m(a_-)-\sum_{m'} H_{m\,m'}\,\psi_m(a).
\end{equation}
Making use of Eq.~(\ref{e89a}), we can write
\begin{equation}
\delta W =\pi^2\, \underline{\psi}^\dag\,\underline{J},
\end{equation}
where 
\begin{equation}
\underline{J} = \underline{\chi}+ \underline{\underline{W}}_{\,v}\,\underline{\psi}.
\end{equation}
Here,  $\underline{J}$ is the column vector of the $J_m$ values, 
and the ``perturbed vacuum energy matrix'', $\underline{\underline{W}}_{\,v}$,  is the Hermitian matrix of the $-H_{m\,m'}$ values. 
Making use of Eqs.~(\ref{e41}), (\ref{e42}), and (\ref{e48}), we get 
\begin{equation}
\underline{J} = \underline{\underline{W}}\,\underline{\psi},
\end{equation}
and
\begin{equation}
\delta W = \underline{\alpha}^\dag\,\underline{\underline{\psi}}^{\dag}_{\,i}\,(\underline{\underline{\chi}}_{\,i} + \underline{\underline{W}}_{\,v}\,\underline{\underline{\psi}}_{\,i})\,\underline{\alpha}=  \underline{\alpha}^\dag\,\underline{\underline{\psi}}^{\dag}_{\,i}\,\underline{\underline{W}}\,\underline{\underline{\psi}}_{\,i}\,\underline{\alpha},
\end{equation}
where
\begin{equation}
\underline{\underline{W}}=\underline{\underline{W}}_{\,p}+\underline{\underline{W}}_{\,v}.
\end{equation}

Note that $\underline{\underline{W}}_{\,p}$ and $\underline{\underline{W}}_{\,v}$ are both Hermitian, so $\underline{\underline{W}}$ is also
Hermitian. Thus the ``perturbed total energy matrix'', $\underline{\underline{W}}$, possesses real eigenvalues and orthonormal eigenvectors, $\underline{\beta}_m$. Let $(\underline{\beta}_{m'})_{m} = \beta_{m\,m'}$, and let $\underline{\underline{\beta}}$ be the matrix of the $\beta_{mm'}$ values. It follows that  
$\underline{\underline{\beta}}^\dag\,\underline{\underline{\beta}}= \underline{\underline{1}}$. 
We conclude that there are as many linearly independent ideal eigenmodes of the plasma as there are poloidal harmonics included in the calculation. The $m$th
eigenmode has the associated energy
\begin{equation}
\delta W_m = \delta W_{p\,m} + \delta W_{v\,m},
\end{equation}
where 
\begin{align}
\delta W_m &= \underline{\beta}_m^{\,\dag}\,\underline{\underline{W}}\, \underline{\beta}_m,\\[0.5ex]
\delta W_{p\,m} &= \underline{\beta}_m^{\,\dag}\,\underline{\underline{W}}_{\,p}\, \underline{\beta}_m,\\[0.5ex]
\delta W_{v\,m} &= \underline{\beta}_m^{\,\dag}\,\underline{\underline{W}}_{\,v}\, \underline{\beta}_m.
\end{align}
Note that $\delta W_m$, $\delta W_{p\,m}$ and $\delta W_{v\,m}$ are all real quantities. Moreover, $\delta W_{p\,m}$ and $\delta W_{v\,m}$ can be interpreted as the
plasma and vacuum contributions to $\delta W_m$, respectively. Finally, 
\begin{equation}
\delta W = \sum_m |\hat{\alpha}_m|^2\,\delta W_m,
\end{equation}
where $\underline{\hat{\alpha}}= \underline{\underline{\beta}}^\dag\,\underline{\underline{\psi}}_{\,i}\,\underline{\alpha}$. 
As is well known, the plasma is ideally unstable if $\delta W<0$ for any possible ideal peturbation.\cite{freidberg,ideal}
Thus, given that the $\hat{\alpha}_m$ are arbitrary, we deduce that if any of the $\delta W_m$ are negative then the plasma is ideally unstable.

\section{Vacuum Solution}\label{vacuum}
\subsection{Toroidal Coordinates}
Let $\mu$, $\eta$, $\phi$ be right-handed toroidal coordinates defined such that (see Sect.~\ref{coord})
\begin{align}
R &= \frac{\sinh\mu}{\cosh\mu-\cos\eta},\\[0.5ex]
Z&=\frac{\sin\eta}{\cosh\mu-\cos\eta}.
\end{align}
The scale-factors of the toroidal coordinate system are
\begin{align}
h_\mu&=h_\eta= \frac{1}{\cosh\mu-\cos\eta}\equiv h,\\[0.5ex]
h_\phi &= \frac{\sinh\mu}{\cosh\mu-\cos\eta} = h\,\sinh\mu.
\end{align}
Moreover, 
\begin{equation}
{\cal J}' \equiv (\nabla\mu\times\nabla\eta\cdot\nabla\phi)= h^3\,\sinh\mu.
\end{equation}

\subsection{Perturbed Magnetic Field}
The curl-free perturbed magnetic field in the vacuum region is written ${\bf b} = {\rm i}\,\nabla V$,
where
$\nabla^2 V =0$.
The most general solution to Laplace's equation is
\begin{align}
V(z,\eta)&= \sum_m (z-\cos\eta)^{1/2}\,U_m(z)\,{\rm e}^{-{\rm i}\,m\,\eta}, \\[0.5ex]
U_m(z) &= p_m\,\hat{P}_{|m|-1/2}^{\,n}(z)+q_m\,\hat{Q}_{m-1/2}^{\,n}(z),
\end{align}
where  $z=\cosh\mu$, the $p_m$ and $q_m$ are arbitrary complex coefficients, and 
\begin{align}\label{e21}
\hat{P}_{|m|-1/2}^{\,n}(z) &= \cos(|m|\,\pi)\,\frac{\sqrt{\pi}\,\Gamma(|m|+1/2-n)\,a^{\,|m|}}{2^{\,|m|-1/2}\,|m|!}\,P_{|m|-1/2}^{\,n}(z),\\[0.5ex]
\hat{Q}_{|m|-1/2}^{\,n}(z)&= \cos(n\,\pi)\,\cos(|m|\,\pi)\,\frac{2^{\,|m|-1/2}\,|m|!}{\sqrt{\pi}\,\Gamma(|m|+1/2+n)\,a^{\,|m|}}\,Q_{|m|-1/2}^{\,n}(z).\label{e22}
\end{align}
Here,  the $P_{m-1/2}^{\,n}(z)$  and $Q_{|m|-1/2}^{\,n}(z)$ are toroidal functions,\cite{abrama}  and $\Gamma(z)$ is a
gamma function.\cite{abramb}

\subsection{Toroidal Electromagnetic Angular Momentum Flux}
The outward flux of toroidal angular momentum across a constant-$z$ surface is
\begin{align}
T_\phi(z) &= -\oint\oint {\cal J}' \,b_\phi\,b^{\,\mu}\,d\eta\,d\phi\\[0.5ex]
&={\rm i}\,n\,\pi^2\sum_{m}(p_m\,q_m^\ast-q_m\,p_m^{\ast})\,(z^2-1)\,{\cal W}(\hat{P}_{|m|-1/2}^{\,n},\hat{Q}_{|m|-1/2}^{\,n}),
\end{align}
where ${\cal W}(f,g)\equiv f\,dg/dz-g\,df/dz$. 
But,\cite{morse}
\begin{align}
{\cal W}(\hat{P}_{|m|-1/2}^{\,n},\hat{Q}_{|m|-1/2}^{\,n})&= 
\cos(n\,\pi) \,\frac{\Gamma(|m|+1/2-n)}{\Gamma(|m|+1/2+n)}\,{\cal W}(P_{|m|-1/2}^{\,n},Q_{|m|-1/2}^{\,n})\nonumber\\[0.5ex]
&= \frac{1}{1-z^2}, 
\end{align}
where use has been made of Eqs.~(\ref{e21}) and (\ref{e22}), 
so
\begin{equation}\label{e30x}
T_\phi(z) = 2\pi^2\,n\sum_m {\rm Im}(q_m^\ast\,p_m).
\end{equation}
Note that $T_\phi$ is independent of $z$, as must be the case because there are no angular momentum sources in the vacuum region.

\subsection{Solution in Vacuum Region}
In the large-aspect ratio limit, $r\ll 1$, it can be demonstrated that\,\cite{morse}
\begin{align}\label{e25t}
z&\simeq \frac{1}{r},\\[0.5ex]
z^{\,1/2}\,\hat{P}^{\,n}_{-1/2}(z) &\simeq \frac{1}{2}\ln\left(\frac{8\,z}{\zeta_n}\right),\\[0.5ex]
z^{1/2}\,\hat{P}^{\,n}_{|m|-1/2}(z) &\simeq \frac{\cos(|m|\,\pi)\,(a\,z)^{|m|}}{|m|},\label{ety}\\[0.5ex]
z^{1/2}\,\hat{Q}^{\,n}_{|m|-1/2}(z) &\simeq \frac{\cos(|m|\,\pi)\,(a\,z)^{-|m|}}{2},\\[0.5ex]
\zeta_n &= \exp\left(\sum_{j=1,n}\frac{2}{2\,j-1}\right).\label{e29t}
\end{align}
Note that Eq.~(\ref{ety}) only applies to $|m|>0$. 

The plasma-vacuum interface lies at $r=a$. Suppose that the plasma is surrounded by a perfectly conducting wall at $r=b_w\,a$, where $b_w\geq 1$.  In the
vacuum region, $a\leq r\leq b_w\,a$,  lying between the plasma and the wall, we can write
\begin{align}\label{e32a}
\underline{V}(r)&= \underline{\underline{{\cal P}}}(r)\,\underline{p}+ \underline{\underline{{\cal Q}}}(r)\,\underline{q},\\[0.5ex]
\underline{\psi}(r)&= \underline{\underline{{\cal R}}}(r)\,\underline{p}+ \underline{\underline{{\cal S}}}(r)\,\underline{q},\label{e33a}
\end{align}
where $\underline{V}(r)$ is the vector of the $V_m(r)$ values (see Sect.~\ref{vac}), $\underline{\psi}(r)$ is the vector of the $\psi_m(r)$ values [see Eq.~(\ref{a3})], $\underline{\underline{{\cal P}}}(r)$ is the
matrix of the
\begin{equation}
{\cal P}_{mm'}(r)=\oint_{r}(z-\cos\eta)^{1/2}\,\hat{P}_{|m'|-1/2}^{\,n}(z)\,\exp[-{\rm i}\,(m\,\theta+m'\,\eta)]\,\frac{d\theta}{2\pi}
\end{equation}
values, 
$\underline{\underline{{\cal Q}}}(r)$ is the
matrix of the
\begin{equation}
{\cal Q}_{mm'}(r)=\oint_{r}(z-\cos\eta)^{1/2}\,\hat{Q}_{|m'|-1/2}^{\,n}(z)\,\exp[-{\rm i}\,(m\,\theta+m'\,\eta)]\,\frac{d\theta}{2\pi}
\end{equation}
values, $\underline{\underline{{\cal R}}}(r)$ is the matrix of the 
\begin{align}\label{e354}
{\cal R}_{mm'}(r) &=\oint_{r}
\left\{\left[\frac{1}{2}\,(z-\cos\eta)^{-1/2}\,\hat{P}_{|m'|-1/2}^{\,n}(z)+(z-\cos\eta)^{1/2}\,\frac{d\hat{P}_{|m'|-1/2}^{\,n}}{dz}\right]{\cal J}\,\nabla r\cdot \nabla z
\right.\nonumber\\[0.5ex]&
\left.\phantom{=}+\left[\frac{1}{2}\,(z-\cos\eta)^{-1/2}\,\sin\eta-{\rm i}\,m'\,(z-\cos\eta)^{1/2}\right]\hat{P}_{|m'|-1/2}^{\,n}(z)\,{\cal J}\,\nabla r\cdot \nabla \eta
\right\}\nonumber\\[0.5ex] &
\phantom{=}\times\exp[-{\rm i}\,(m\,\theta+m'\,\eta)]\,\frac{d\theta}{2\pi}
\end{align}
values, 
$\underline{\underline{{\cal S}}}(r)$ is the matrix of the 
\begin{align}\label{e355}
{\cal S}_{mm'}(r) &=\oint_{r}
\left\{\left[\frac{1}{2}\,(z-\cos\eta)^{-1/2}\,\hat{Q}_{|m'|-1/2}^{\,n}(z)+(z-\cos\eta)^{1/2}\,\frac{d\hat{Q}_{|m'|-1/2}^{\,n}}{dz}\right]{\cal J}\,\nabla r\cdot \nabla z
\right.\nonumber\\[0.5ex]&
\left.\phantom{=}+\left[\frac{1}{2}\,(z-\cos\eta)^{-1/2}\,\sin\eta-{\rm i}\,m'\,(z-\cos\eta)^{1/2}\right]\hat{Q}_{|m'|-1/2}^{\,n}(z)\,{\cal J}\,\nabla r\cdot \nabla \eta
\right\}\nonumber\\[0.5ex] &
\phantom{=}\times\exp[-{\rm i}\,(m\,\theta+m'\,\eta)]\,\frac{d\theta}{2\pi}
\end{align}
values, $\underline{p}$ is the vector of the $p_m$ coefficients, and  $\underline{q}$ is the vector of the $q_m$ coefficients. Here, the
subscript $r$ on the integrals indicates that they are taken at constant $r$. 

\subsection{Toroidal Electromagnetic Torque}
According to Eq.~(\ref{b27}), (\ref{e89}), and Eq.~(214) of Ref.~\onlinecite{tj}, 
the net toroidal electromagnetic torque acting on the plasma  is
\begin{align}\label{e38}
T_\phi&= -2\pi^2\,n\,{\rm Im}(\underline{V}^\dag\,\underline{\psi})= -\pi^2\,n\,(\underline{V}^\dag\,\underline{\psi}-\underline{\psi}^\dag\,\underline{V}).
\end{align}
However, this torque must equal the flux of toroidal angular momentum into the vacuum region, so Eq.~(\ref{e30x}) implies that
\begin{equation}\label{e39x}
T_\phi= 2\pi^2\,n\,{\rm Im}(\underline{q}^\dag\,\underline{p})= \pi^2\,n\,(\underline{q}^\dag\,\underline{p}- \underline{p}^\dag\,\underline{q}).
\end{equation}
Now, Eqs.~(\ref{e32a}), (\ref{e33a}), and (\ref{e38}) give 
\begin{align}
T_\phi&= -\pi^2\,n\,\left[
\underline{p}^\dag\,(\underline{\underline{\cal P}}^\dag\,\underline{\underline{\cal R}}
- \underline{\underline{\cal R}}^\dag\,\underline{\underline{\cal P}})\,\underline{p}
+\underline{p}^\dag\,(\underline{\underline{\cal P}}^\dag\,\underline{\underline{\cal S}}
-\underline{\underline{\cal R}}^\dag\,\underline{\underline{\cal Q}})\,\underline{q}\right.\nonumber\\[0.5ex]
&\phantom{=}\left.- \underline{q}^\dag\,(\underline{\underline{\cal S}}^\dag\,\underline{\underline{\cal P}}
- \underline{\underline{\cal Q}}^\dag\,\underline{\underline{\cal R}})\,\underline{p}
+\underline{q}^\dag\,(\underline{\underline{\cal Q}}^\dag\,\underline{\underline{\cal S}}
- \underline{\underline{\cal S}}^\dag\,\underline{\underline{\cal R}})\,\underline{q}
\right].
\end{align}
The previous equation is consistent with Eq.~(\ref{e39x}) provided that
\begin{align}\label{e41x}
\underline{\underline{\cal P}}^\dag\,\underline{\underline{\cal R}}&= \underline{\underline{\cal R}}^\dag\,\underline{\underline{\cal P}},\\[0.5ex]
\underline{\underline{\cal Q}}^\dag\,\underline{\underline{\cal S}}&= \underline{\underline{\cal S}}^\dag\,\underline{\underline{\cal Q}},\label{e42x}\\[0.5ex]
\underline{\underline{\cal P}}^\dag\,\underline{\underline{\cal S}}- \underline{\underline{\cal R}}^\dag\,\underline{\underline{\cal Q}}&=\underline{\underline{1}}.\label{e43}
\end{align}

The previous three equations can be combined with Eqs.~(\ref{e32a}) and (\ref{e33a}) to give 
\begin{align}
\underline{p} &= \underline{\underline{\cal S}}^{\,\dag} \,\underline{V} - \underline{\underline{\cal Q}}^{\,\dag}\,\underline{\psi},\\[0.5ex]
\underline{q} &= -\underline{\underline{\cal R}}^{\,\dag} \,\underline{V} + \underline{\underline{\cal P}}^{\,\dag}\,\underline{\psi}.
\end{align}
However, the previous two equations are only consistent with Eqs.~(\ref{e32a}) and (\ref{e33a}) provided 
\begin{align}\label{e46}
\underline{\underline{\cal Q}}\,\underline{\underline{\cal P}}^\dag&= \underline{\underline{\cal P}}\,\underline{\underline{\cal Q}}^\dag,\\[0.5ex]
\underline{\underline{\cal R}}\,\underline{\underline{\cal S}}^\dag&= \underline{\underline{\cal S}}\,\underline{\underline{\cal R}}^\dag,\label{e47x}\\[0.5ex]
\underline{\underline{\cal P}}\,\underline{\underline{\cal S}}^{\dag}- \underline{\underline{\cal Q}}\,\underline{\underline{\cal R}}^{\dag}&=\underline{\underline{1}}.\label{e48x}
\end{align}
Note that Eqs.~(\ref{e41x})--(\ref{e43}) and (\ref{e46})--(\ref{e48x}) hold throughout the vacuum region. 

\subsection{Ideal-Wall Matching Condition}
If the wall is perfectly conducting then  $\underline{\psi}(b_w\,a)=0$. 
It follows from Eq.~(\ref{e33a}) that
\begin{equation}
\underline{q} = \underline{\underline{I}}_b\,\underline{p},
\end{equation}
where
\begin{equation}
 \underline{\underline{I}}_b=- \underline{\underline{\cal S}}_{\,b}^{-1}\,\underline{\underline{\cal R}}_{\,b}
 \end{equation}
 is termed the ``wall matrix''.
 Here, $\underline{\underline{\cal S}}_{\,b}= \underline{\underline{\cal S}}(b_w\,a)$, et cetera. Equation~(\ref{e47x}) ensures that $ \underline{\underline{I}}_b$
 is Hermitian. It immediately follows from Eq.~(\ref{e39x}) that $T_\phi=0$. In other words, zero net toroidal electromagnetic torque is exerted on a plasma
 surrounded by a perfectly conducting wall. As described in Ref.~\onlinecite{tj}, the fact that $T_\phi=0$ ensures that the tearing stability matrix, $E_{kk'}$, introduced in
 Sect.~\ref{disp}, is Hermitian. 
 
  Making use of Eqs.~(\ref{e32a}) and (\ref{e33a}),  the matching condition at the plasma-vacuum interface  for a perfectly-conducting wall becomes 
 \begin{equation}
 \underline{V}(a_+)= \underline{\underline{H}}\,\underline{\psi}(a),
 \end{equation}
 where 
 \begin{equation}
 \underline{\underline{H}}= (\underline{\underline{\cal P}}_{\,a}+\underline{\underline{\cal Q}}_{\,a}\,\underline{\underline{I}}_{\,b})\,(\underline{\underline{\cal R}}_{\,a}+\underline{\underline{\cal S}}_{\,a}\,\underline{\underline{I}}_{\,b})^{-1}
 \end{equation}
 is  the vacuum matrix introduced in Sect.~\ref{stab}. 
  Here, $\underline{\underline{\cal P}}_{\,a}= \underline{\underline{\cal P}}(a_+)$, et cetera. 
Making use of Eqs.~(\ref{e41x})--(\ref{e43}), it is easily demonstrated that
 \begin{equation}\label{e53}
 \underline{\underline{H}}-\underline{\underline{H}}^\dag =- [(\underline{\underline{\cal R}}_{\,a}+\underline{\underline{\cal S}}_{\,a}\,\underline{\underline{I}}_{\,b})^{-1}]^{\dag}\,(\underline{\underline{I}}_{\,b} -\underline{\underline{I}}^\dag_{\,b})\,  (\underline{\underline{\cal R}}_{\,a}+\underline{\underline{\cal S}}_{\,a}\,\underline{\underline{I}}_{\,b})^{-1}.
\end{equation}
Thus,  the vacuum matrix, $\underline{\underline{H}}$, is Hermitian because  the wall matrix, $\underline{\underline{I}}_{\,b}$, is Hermitian. 
 
\subsection{Model Wall Matrix}
 Equations~(\ref{e25t})--(\ref{e29t}), (\ref{e354}), and (\ref{e355}) suggest that
 \begin{align}
 \underline{\underline{{\cal R}}}_{\,b}& = \underline{\underline{{\cal R}}}_{\,a} \,\underline{\underline{\rho}}^{\,-1},\\[0.5ex]
 \underline{\underline{{\cal S}}}_{\,b} &= \underline{\underline{{\cal S}}}_{\,a} \,\underline{\underline{\rho}},
 \end{align}
 where
 \begin{align}
 \rho_{mm'} &= \delta_{mm'}\,\rho_m,\\[0.5ex]
 \rho_0 &= 1+\ln b_w,\\[0.5ex]
 \rho_{m\neq 0} &= b_w^{\,|m|}.
 \end{align}
 Hence,
 \begin{equation}
 \underline{\underline{I}}_{\,b}= - \underline{\underline{\rho}}^{\,-1\,\dag}\,\underline{\underline{\cal S}}_{\,a}^{-1}\,\underline{\underline{\cal R}}_{\,a}\,\underline{\underline{\rho}}^{\,-1}.
 \end{equation}
 Note that our model wall matrix, $\underline{\underline{I}}_b$, is Hermitian given that $\underline{\underline{\cal S}}_{\,a}^{-1}\,\underline{\underline{\cal R}}_{\,a}$
 is Hermitian. [See Eq.~(\ref{e47x}).] Our model wall matrix allows us to smoothly interpolate between a plasma with no wall
 (which corresponds to $b_w\rightarrow\infty$ and $\underline{\underline{H}}= \underline{\underline{\cal P}}_{\,a}\,\underline{\underline{\cal R}}_{\,a}^{-1}$),\cite{tj}
 and a fixed boundary plasma (which corresponds to $b_w=1$ and $\underline{\underline{H}}^{-1}= \underline{\underline{0}}$). 
 
\section{Resistive Layer Model}\label{lmodel}
\subsection{Introduction}
 The resistive layer model implemented in the TJ code is the three-field model described in Ref.~\onlinecite{diff}. This model is a further development of
 the models derived in Refs.~\onlinecite{fw} and \onlinecite{cole}, and is ultimately based on the four-field model of Ref.~\onlinecite{hkm}. 
 The model includes plasma resistivity, plasma inertia, electron and ion diamagnetic flows, the decoupling of ion and electron flows on lengthscales below the ion sound
 radius, and perpendicular momentum and energy transport. The three fields in the model are the (normalized) perturbed helical magnetic flux, $\psi$, the (normalized)
 perturbed electrostatic potential, $\phi$, and the (normalized) perturbed electron number density, $N$. The model makes use of a low-$\beta$
 approximation that decouples the (normalized) parallel ion velocity, $V$,  from the system of equations (thereby converting a four-field model into
 a three-field model). Reference~\onlinecite{lee} explains how the fourth field can be incorporated into the model. We choose not to do this
 because the fourth field is only important at high-$\beta$, and an aspect-ratio expanded equilibrium is, by definition, a relatively low-$\beta$
 equilibrium. 
 
\subsection{Plasma Parameters}\label{ppara}
The values of the following plasma parameters must be specified in order to calculate layer quantities:
\begin{itemize}
\item $B_0$ - the toroidal magnetic field-strength at the magnetic axis (T).
\item $R_0$ - the major radius of the magnetic axis  (m).
\item $n_0$ - the  electron number density at the magnetic axis (${\rm m}^{-3}$).
\item $\alpha$ - the electron density profile is assumed to be $n_e(r)=n_0\,(1-r^2/a^2)^\alpha$.
\item $T_{e\,{\rm edge}}$ - the electron temperature at the plasma-vacuum interface (eV).
\item $Z_{\rm eff}$  - the (assumed spatially uniform) effective ion charge number.
\item $M$ - the ion mass number.
\item $\chi_\perp$ - the (assumed spatially uniform) perpendicular momentum/energy diffusivity (${\rm m}^2/{\rm s}$).
\end{itemize}

\subsection{Rational Surface Parameters}\label{rpara}
Let the $k$th rational surface  lie at $r=r_k$, and possess the resonant poloidal mode number $m_k$. 
We can define  $q_k = q(r_k)$,  $s_k= s(r_k)$, 
$n_{e\,k} = n_0\,(1-r_k^{\,2}/a^2)^\alpha$, $g_k = g(r_k)$, 
\begin{align}
p_k &= \frac{B_0^{\,2}\,P(r_k),}{\mu_0}\\[0.5ex]
T_{e\,k} &= \frac{p_k}{2\,n_{e\,k}\,e} + T_{e\,{\rm edge}},\\[0.5ex]
\ln{\Lambda}_k &= 24 + 3\,\ln 10-\frac{1}{2}\ln n_{e\,k} + \ln T_{e\,k},\\[0.5ex]
\tau_{ee\,k}&= \frac{6\sqrt{2}\,\pi^{3/2}\,\epsilon_0^{\,2}\,m_e^{1/2}\,T_{e\,k}^{\,3/2}}{\ln{\Lambda}_k\,e^{5/2}\,n_{e\,k}},\\[0.5ex]
\sigma_{\parallel\,k} &= \frac{\sqrt{2}+13\,Z_{\rm eff}/4}{Z_{\rm eff}\,(\sqrt{2}+Z_{\rm eff})}\,\frac{n_{e\,k}\,e^2\,\tau_{ee\,k}}{m_e},\\[0.5ex]
L_{s\,k} &= \frac{R_0\,q_k}{s_k},\\[0.5ex]
V_{A\,k} &= \frac{B_0\,g_k}{(\mu_0\,n_{e\,k}\,M\,m_p)^{1/2}},\\[0.5ex]
d_{i\,k} &= \left(\frac{M\,m_p}{n_{e\,k}\,e^2\,\mu_0}\right)^{1/2},\\[0.5ex]
\beta_k &= \frac{5\,P(r_k)}{3\,g_k^{\,2}},\\[0.5ex]
\hat{d}_{\beta\,k} &= \left(\frac{\beta_k}{1+\beta_k}\right)^{1/2}\,\frac{d_{i\,k}}{R_0\,r_k},\\[0.5ex]
\omega_{\ast\,k} &= \frac{m_k\,B_0}{\mu_0\,e\,n_{e\,k}\,R_0^{\,2}\,g_k\,r_k}\,\frac{dP(r_k)}{dr}, \\[0.5ex]
\tau_{H\,k} &= \frac{L_{s\,k}}{m_k\,V_{A\,k}},\\[0.5ex]
\tau_{R\,k} &= \mu_0\,R_0^{\,2}\,r_k^{\,2}\,\sigma_\parallel(r_k),\\[0.5ex]
\tau_{\perp\,k} &= \frac{R_0^{\,2}\,r_k^{\,2}}{\chi_\perp},
\end{align}
where $e$ is the magnitude of the electron charge, $m_e$ the electron mass, and $m_p$ the proton mass. 
Note that we are assuming that the electrons and ions have the same temperature, as we would expect to be the case in a fusion reactor. 
Here, $d_{i\,k}$ is the collisionless skin-depth, $\hat{d}_{\beta\,k}$ the normalized ion sound radius, $\omega_{\ast\,k}$ the diamagnetic frequency, 
$\tau_{H\,k}$ the hydromagnetic timescale, $\tau_{R\,k}$ the resistive diffusion timescale, and $\tau_{\perp\,k}$ the perpendicular momentum/energy diffusion timescale,
at the $k$th rational surface. 

Layer physics at the $k$th rational surface is governed by the following normalized parameters:
$\hat{\gamma}_k= \tau_k\,\gamma_k$, 
$Q_{e\,k }= - \tau_k\,\omega_{\ast\,k}/2$, 
$Q_{i\,k }=  \tau_k\,\omega_{\ast\,k}/2$, 
$D_k = S_{k}^{\,1/3}\,\hat{d}_{\beta\,k}/\sqrt{2}$, 
and $P_{k} = \tau_{R\,k}/\tau_{\perp\,k}$, 
where $S_k = \tau_{R\,k}/\tau_{H\,k}$ and 
$\tau_k = S_k^{\,1/3} \,\tau_{H\,k}$.   Here, $S_k$ is the Lundquist number at the $k$th rational surface, and $\gamma_k$ is the complex growth-rate of the tearing mode seen in the E-cross-B frame (i.e., a frame that co-rotates with the
particle guiding centers) at the $k$th rational surface.

\subsection{Resistive Layer Equations}\label{rlayer}
The following resistive layer equations, which govern the response of the plasma at the $k$th rational surface to the ideal-MHD solution in the outer region,  are obtained by linearizing the three-field model of Ref.~\onlinecite{diff}:
\begin{align}\label{e80y}
(\hat{\gamma}_k+{\rm i}\,Q_{e\,k})\,\psi&= - {\rm i}\,X\left(\phi-N\right) + \frac{d^{\,2}\psi}{d X^2},\\[0.5ex]
(\hat{\gamma}_k+{\rm i}\,Q_{i\,k})\,\frac{d^{\,2}\phi}{d X^2}&= - {\rm i}\,X\,\frac{d^{\,2}\psi}{d X^2}+ P_{k}\,\frac{d^{\,4}}{d X^4}\!\left(\phi + N\right),\\[0.5ex]
\hat{\gamma}_k\,N&= - {\rm i}\,Q_{e\,k}\,\phi  - {\rm i} \,D_k^{\,2}\,X\,\frac{d^{\,2}\psi}{d X^{2}}
+ P_{k}\,\frac{d^{\,2} N}{d X^{2}},\label{e82y}
\end{align}
where $X=S_k^{\,1/3}\,(r-r_k)/r_k$. Here, it is assumed that $S_k\gg 1$.  The asymptotic behavior of the tearing-parity [i.e., $\psi(-X)=\psi(X)$, $\phi(-X)= -\phi(X)$, and $N(-X)=-N(X)$] solution of the layer
equations is such that
\begin{align}\label{e47y}
\psi(X)&\rightarrow  \psi_0\left[\frac{\hat{\mit\Delta}_k}{2}\,|X| +1 + {\cal O}\left(\frac{1}{X}\right)\right],\\[0.5ex]
\phi(X)&\rightarrow  {\rm i}\,\hat{\gamma}_k\,\psi_0\left[\frac{\hat{\mit\Delta}_k}{2}\,{\rm sgn}(X) +\frac{1}{X} + {\cal O}\left(\frac{1}{X^2}\right)\right]\label{e85y}
\end{align}
as $|X|\rightarrow\infty$, where $\psi_0$ is an arbitrary constant, and 
$\hat{\mit\Delta}_k = S_k^{\,-1/3}\,{\mit\Delta}_k$. Here, ${\mit\Delta}_k$ is the layer response parameter introduced in Sect.~\ref{disp}. 

\subsection{Fourier Transformation}\label{ft}
Equations~(\ref{e80y})--(\ref{e82y}) are most conveniently solved in Fourier transform space.\cite{cole}
Let
\begin{equation}\label{e86a}
\hat{\phi}(p) = \int_{-\infty}^\infty \phi(X)\,{\rm e}^{-{\rm i}\,p\,X}\,dX,
\end{equation}
et cetera. The Fourier transformed linear layer equations become
\begin{align}\label{e314}
(\hat{\gamma}_k+{\rm i}\,Q_{e\,k})\,\hat{\psi}&=\frac{d}{d p}\!\left(\hat{\phi}-\hat{N}\right)-p^2\,\hat{\psi},\\[0.5ex]
(\hat{\gamma}_k+{\rm i}\,Q_{i\,k})\,p^2\,\hat{\phi}&=  \frac{d (p^2\,\hat{\psi})}{d p}- P_k\,p^4\,(\hat{\phi} +\hat{N}),\\[0.5ex]
\hat{\gamma}_k\,\hat{N} &= - {\rm i}\,Q_{e\,k}\,\hat{\phi} -D_k^{\,2}\,\frac{d (p^2\,\hat{\psi})}{d p}
  - P_k\,p^2\hat{N},\label{e316}
\end{align}
where, for a tearing-parity solution, Eq.~(\ref{e85y}) yields\,\cite{ed}
\begin{equation}\label{e94y}
\hat{\phi}(p) =\pi\,\hat{\gamma}_k\,\psi_0\left[\frac{\hat{\mit\Delta}_k}{\pi\,p} + 1+ {\cal O}(p)\right]
\end{equation}
as $p\rightarrow 0$. 

Let 
\begin{align}\label{e95}
Y_e(p)\equiv \hat{\phi}(p)-\hat{N}(p)= \pi\,(\hat{\gamma}_k+{\rm i}\,Q_{e\,k})\,\psi_0\,\hat{Y}_e(p).
\end{align}
 Equations~(\ref{e314})--(\ref{e316}) can be combined to give\,\cite{cole,diff}
\begin{align}\label{e91y}
\frac{d}{dp}\!\left[A(p)\,\frac{d\hat{Y}_e}{dp}\right] - \frac{B(p)}{C(p)}\,p^2\,\hat{Y}_e=0,
\end{align}
where
\begin{align}
A(p) &= \frac{p^2}{\hat{\gamma}_k+{\rm i}\,Q_{e\,k}+p^2},\label{e97a}\\[0.5ex]
B(p)&=\hat{\gamma}_k\,(\hat{\gamma}_k+{\rm i}\,Q_{i\,k})+2\,(\hat{\gamma}_k+{\rm i}\,Q_{i\,k})\,P_k\,p^2 +P_k^{\,2}\,p^4,\\[0.5ex]
C(p) &=\hat{\gamma}_k+{\rm i}\,Q_{e\,k}+ [P_k+
(\hat{\gamma}_k+{\rm i}\,Q_{i\,k})\,D_k^{\,2}]\,p^2 + 2\,P_k\,D_k^{\,2}\,p^4.
\end{align}
Because
\begin{equation}
Y_e(p) =\frac{(\hat{\gamma}_k+{\rm i}\,Q_{e\,k})}{\hat{\gamma}_k}\,\hat{\phi}
\end{equation}
as $p\rightarrow 0$, Eqs.~(\ref{e94y}) and (\ref{e95})
yield the following small-$p$ boundary condition that Eq.~(\ref{e91y}) must satisfy:
\begin{equation}\label{e101}
\hat{Y}_e(p)\rightarrow\frac{\hat{\mit\Delta}_k}{\pi\,p} + 1+ {\cal O}(p)
\end{equation}
as $p\rightarrow 0$.  Equation~(\ref{e91y}) must also satisfy the  physical boundary condition 
\begin{equation}\label{e102}
\hat{Y}_e(p)\rightarrow 0
\end{equation}
as $p\rightarrow\infty$. 
In fact, in the large-$p$/large-$D_k$ limit, $p\gg 1$ and $D_k\gg P_k^{\,1/6}$, the well-behaved solution of Eq.~(\ref{e91y}) takes the form 
\begin{equation}\label{d33}
\hat{Y}_e(p) = Y_0\,p^{\,x_k}\,\exp\left(\frac{-\sqrt{b_k}\,p^2}{2}\right)
\end{equation}
where $Y_0$ is independent of $p$, and
\begin{align}
x_k&= \frac{c_k -\sqrt{b_k}\,(1-\sqrt{b_k}\,a_k)}{2\sqrt{b_k}},\\[0.5ex]
a_k &= -(\hat{\gamma}_k+ {\rm i}\,Q_{e\,k}),\\[0.5ex]
b_k&= \frac{P_k}{2\,D_k^{\,2}},\\[0.5ex]
c_k&= \frac{P_k}{2\,D_k^{\,2}}\left[1+\frac{2\, (\hat{\gamma}_k+{\rm i}\,Q_{i\,k})}{P_k}-\frac{(P_k+[\hat{\gamma}_k+{\rm i}\,Q_{i\,k}]\,D_k^{\,2})}{2\,P_k\,D_k^{\,2}}\right].
\end{align}
On the other hand, in the large-$p$/small-$D_k$ limit, $p\gg 1$ and $D_k\ll P_k^{\,1/6}$, the well-behaved solution of Eq.~(\ref{e91y}) is
\begin{equation}\label{d34}
\hat{Y}_e(p) = Y_0 \,p^{-1}\,\exp\left(x_k\,p - \frac{\sqrt{b_k}\,p^3}{3}\right),
\end{equation}
where
\begin{align}
x_k &= \frac{a_k\,b_k-c_k}{2\sqrt{b_k}},\\[0.5ex]
a_k&=  -(\hat{\gamma}_k+ {\rm i}\,Q_{e\,k}),\\[0.5ex]
b_k&= P_k,\\[0.5ex]
c_k&= -{\rm i}\,(Q_{e\,k}-Q_{i\,k}) + (\hat{\gamma}_k+ {\rm i}\,Q_{i\,k}).
\end{align}

\subsection{Ricatti Transformation}
Equation~(\ref{e91y}) is most conveniently solved by means of a Ricatti transformation.\cite{ric1,ric2}
Let 
\begin{equation}
W(p)= \frac{p}{\hat{Y}_e}\,\frac{d\hat{Y}_e}{dp}.
\end{equation}
Equation~(\ref{e91y}) transforms to give
\begin{equation}\label{e72a}
\frac{dW}{dp} =- \frac{A'}{p}\,W -\frac{W^{\,2}}{p} + \frac{B}{A\,C}\,p^3,
\end{equation}
where
\begin{equation}
A' = \frac{\hat{\gamma}_k+{\rm i}\,Q_{e\,k}-p^2}{\hat{\gamma}_k+{\rm i}\,Q_{e\,k}+p^2}.
\end{equation}
According to Eqs.~(\ref{e101}), (\ref{d33}), and (\ref{d34}), this equation must be solved subject to the boundary condition that
\begin{equation}\label{e63a}
W(p) = x_k-\sqrt{b_k}\,p^2
\end{equation}
at large $p$ and large $D_k$, or
\begin{equation}\label{e63b}
W(p) = -1 +x_k\,p-\sqrt{b_k}\,p^3
\end{equation}
at large $p$ and small $D_k$, and 
\begin{equation}\label{e73a}
W (p)=-1+\frac{\pi\,p}{\hat{\mit\Delta}_k}
\end{equation}
at small $p$. 
The method of solution is to  launch a solution of Eq.~(\ref{e72a})  from large $p$, subject to the  boundary condition (\ref{e63a}) or (\ref{e63b}), as appropriate,  and
integrate it to small $p$. We can then deduce the value of $\hat{\mit\Delta}_k$ from Eq.~(\ref{e73a}). 

\section{Asymptotic Matching Between Outer and Inner Regions}\label{asym}
\subsection{Introduction}
Consider the segment of the inner region that is centered on the $k$th rational surface. Let the surface lie at $r=r_k$, and possess the
resonant poloidal mode number $m_k$. According to Sect.~\ref{rational},   in the limit $r\rightarrow r_k$,  the asymptotic behavior of the resonant, tearing-parity, component of the perturbed helical magnetic 
flux, $\psi$,  [see Eq.~(\ref{a3})] in the outer region is
\begin{equation}\label{ee1}
\psi_{m_k}(x) = A_{L\,k}\,|x|^{-\nu_k} + A_{S\,k}\,|x|^{\,1+\nu_k},
\end{equation}
where $x=r-r_k$, $\nu_k= -1/2+\sqrt{-D_{I\,k}}$, and
\begin{equation}
{\mit\Delta}_k = \frac{2\,r_k^{1+2\,\nu_k}\,A_{S\,k}}{A_{L\,k}}.
\end{equation}
Here, $D_{I\,k}$ is the ideal interchange parameter defined in Eq.~(\ref{di}), and ${\mit\Delta}_k$ is the layer response parameter that appears in the
tearing mode dispersion relation, (\ref{dispersion}). On the other hand, according to Sect.~\ref{rlayer}, the asymptotic behavior of the resistive layer
solution in the segment of the inner region, as it approaches the outer region,  is
\begin{equation}\label{ee3}
\psi_{m_k}(x) = B_{L\,k}+B_{S\,k}\,|x|,
\end{equation}
where
\begin{equation}\label{ee4}
{\mit\Delta}_k = \frac{2\,r_k\,B_{S\,k}}{B_{L\,k}}.
\end{equation}
It can be seen that the  solutions (\ref{ee1}) and (\ref{ee3})  only match if $\nu_k=0$. The reason for the mismatch is because the outer solution takes average magnetic field-line curvature into
account,\cite{ggj,ggj1} whereas the layer solution does not. The aim of this appendix is to describe a method for resolving the mismatch. 

\subsection{Pressure Flattening}
In the immediate vicinity of the rational surface, the equation that governs the resonant component of $\psi$ in the outer region reduces to\,\cite{tj}
\begin{equation}\label{ee5}
\frac{d^{\,2}\psi_{m_k}}{dx^2} =\frac{\nu_k\,(1+\nu_k)}{x^{2}}\,\psi_{m_k},
\end{equation}
where
\begin{equation}\label{dii}
\nu_k\,(1+\nu_k)= -D_{I\,k} -\frac{1}{4}= \left[\frac{2\,(1-q^2)}{s^2}\,r\,\frac{dP}{dr}\right]_{r_k}.
\end{equation}
Following Ref.~\onlinecite{bishop}, let us suppose that the pressure gradient is locally flattened in the vicinity of the rational surface in such a manner that
\begin{equation}
\frac{dP}{dx} = \left(\frac{dP}{dr}\right)_{r_k}\,\frac{x^2}{x^2+\delta_k^{\,2}}.
\end{equation}
Here, it is assumed that $\delta_k\ll r_k$. Equation~(\ref{ee5}) transforms to give
\begin{equation}\label{ee13}
(1+X^2)\,\frac{d^{\,2}\psi_{m_k}}{dX^2}= \nu_k\,(1+\nu_k)\,\psi_{m_k},
\end{equation}
where $X=x/\delta_k$. 

The most general tearing-parity solution of Eq.~(\ref{ee13}) in the limit $|X|\rightarrow 0$  takes the form 
\begin{equation}\label{in}
\psi_{m_k}(X) \simeq \hat{B}_{L\,k}+ \hat{B}_{S\,k}\,|X].
\end{equation}
We can define
\begin{equation}\label{din}
{\mit\Delta}_{k\,{\rm in}} = \left(\frac{r_k}{\delta_k}\right)
\frac{2\,\hat{B}_{S\,k}}{\hat{B}_{L\,k}}.
\end{equation}
On the other hand, the most general tearing-parity solution of Eq.~(\ref{ee13}) in the limit $|X|\rightarrow \infty$  is written 
\begin{align}\label{out}
\psi_{m_k}(X)&= \hat{A}_{L\,k}\,|X|^{-\nu_k}+\hat{A}_{S\,k}\,|X|^{1+\nu_k}.
\end{align}
We can define
\begin{equation}\label{out1}
{\mit\Delta}_{k\,{\rm out}}= \left(\frac{r_k}{\delta_k}\right)^{1+2\,\nu_k}\,\frac{2\,\hat{A}_{S\,k}}{\hat{A}_{L\,k}}.
\end{equation}
A comparision between Eqs.~(\ref{ee1})--(\ref{ee4}) and (\ref{in})--(\ref{out1})  reveals that the solution to Eq.~(\ref{ee13}) successfully interpolates
between the resistive layer solution and the outer solution.  It now remains to determine the relationship between ${\mit\Delta}_{k\,{\rm in}}$ and
${\mit\Delta}_{k\,{\rm out}}$. 

\subsection{Connection Formula}
Suppose that we launch a  ``large''  solution  of Eq.~(\ref{ee13}), $\psi_L(X) = X^{-\nu_k}$, 
from large $X$, and integrate to $X=0$. Let
\begin{align}
\psi_L(0) &= a_{LL},\\[0.5ex]
\frac{d\psi_L(0)}{dx} &= a_{SL}.
\end{align}
Next, suppose that we launch a  ``small'' solution, 
$\psi_S(X) = X^{1+\nu_k}$, 
from large $X$, and integrate to  $X=0$. Let
\begin{align}
\psi_S(0) &= a_{LS},\\[0.5ex]
\frac{d\psi_S(0)}{dx} &= a_{SS}.
\end{align}
The most general solution is
\begin{equation}
\psi_{m_k}(X) = \hat{A}_{L\,k}\,\psi_L(X)+ \hat{A}_{S\,k}\,\psi_S(X).
\end{equation}
It follows that
\begin{align}\label{bin}
\hat{B}_{L\,k} &= a_{LL}\,\hat{A}_{L\,k} + a_{LS}\,\hat{A}_{S\,k},\\[0.5ex]
\hat{B}_{S\,k}&= a_{SL}\,\hat{A}_{L\,k} + a_{SS}\,\hat{A}_{S\,k}.\label{bout}
\end{align}
Thus,  Eqs.~(\ref{din}) and (\ref{out1}) yield the connection formula
\begin{equation}\label{ee20}
\left(\frac{\delta_k}{r_k}\right)\frac{{\mit\Delta}_{k \,{\rm in}}}{2}= \frac{a_{SL} +a_{SS}\,(\delta_k/r_k)^{1+2\,\nu_k}\,({\mit\Delta}_{k\,{\rm out}}/2)}
{a_{LL}  +a_{LS}\,(\delta_k/r_k)^{1+2\,\nu_k}\,({\mit\Delta}_{k\,{\rm out}}/2)}.
\end{equation}
It remains to calculate the coefficients $a_{LL}$, $a_{LS}$, $a_{SL}$, and $a_{SS}$. 

\subsection{Analytic Solution}
Let $\psi_{m\,k} = (1+X^2)\,\phi$ and $z={\rm i}\,X$. Equation~(\ref{ee13}) transforms to give 
\begin{equation}\label{ee38}
(1-z^2)\,\frac{d^{\,2}\phi}{dz^2} - 4\,z\,\frac{d\phi}{dz} - [2-\nu_k\,(1+\nu_k)]\,\phi=0.
\end{equation}
Now, the Legendre functions $Q_{\nu_k}(z)$ and $Q_{-1-\nu_{\,k}}(z)$ satisfy\,\cite{abramc}
\begin{equation}
(1-z^2)\,\frac{d^2 w}{dz^2} - 2\,z\,\frac{dw}{dz} + \nu_k\,(1+\nu_k)\,w = 0.
\end{equation}
Let $w'=dw/dz$. Differentiation of the previous equation yields
\begin{equation}\label{ee40}
(1-z^2)\,\frac{d^2w'}{dz^2} - 4\,z\,\frac{dw'}{dz} - [2-\nu_k\,(1+\nu_k)]\,w'=0.
\end{equation}
It is clear from a comparison of Eqs.~(\ref{ee38}) and (\ref{ee40}) that the two independent solutions of Eq.~(\ref{ee13}) can be written 
$(1+X^2)\,Q_{\nu_k}'({\rm i}\,X)$ and $(1+X^2)\,Q_{-1-\nu_k}'({\rm i}\,X)$, where $'$ denotes differentiation with respect to argument. 

Now,\cite{eda}
\begin{align}
Q_{\nu_k}(z)&= \frac{\pi^{1/2}\,\Gamma(1/2+\nu_k/2)\,{\rm e}^{-{\rm i}\,(\pi/2)\,(1+\nu_k)}}{2\,\Gamma(1+\nu_k/2)}F\left(-\frac{\nu_k}{2},\frac{1}{2}+\frac{\nu_k}{2};\frac{1}{2};z^2\right)\nonumber\\[0.5ex]
&\phantom{=}+\frac{\pi^{1/2}\,\Gamma(1+\nu_k/2)\,{\rm e}^{-{\rm i}\,(\pi/2)\,\nu_k}}{\Gamma(1/2+\nu_k/2)}\,z\,F\left(\frac{1}{2}-\frac{\nu_k}{2},1+\frac{\nu_k}{2};\frac{3}{2};z^2\right),
\end{align}
where $F(a,b;c;z)$ is a hypergeometric function and $\Gamma(z)$ is a gamma function,\cite{abramb,abramd}
which yields
\begin{align}
Q_{\nu_k}'(z)&\simeq\frac{\pi^{1/2}\,\Gamma(1+\nu_k/2)\,{\rm e}^{-{\rm i}\,(\pi/2)\,\nu_k}}{\Gamma(1/2+\nu_k/2)}\nonumber\\[0.5ex]
&\phantom{=}- \frac{\pi^{1/2}\,\nu_k\,(1+\nu_k)\,\Gamma(1/2+\nu_k/2)\,{\rm e}^{-{\rm i}\,(\pi/2)\,(1+\nu_k)}}{2\,\Gamma(1+\nu_k/2)}\,z +{\cal O}(z^2).
\end{align}
So, at small $X$, 
\begin{align}\label{ee52}
(1+X^2)\,Q_{\nu_k}'({\rm i}\,X)&\simeq {\rm e}^{-{\rm i}\,\nu_k\,\pi/2}\left[\frac{\pi^{1/2}\,\Gamma(1+\nu_k/2)}{\Gamma(1/2+\nu_k/2)}
-\frac{\pi^{1/2}\,\nu_k\,(1+\nu_k)\,\Gamma(1/2+\nu_k/2)}{2\,\Gamma(1+\nu_k/2)}\,X\right]. 
\end{align}

Furthermore,\cite{eda}
\begin{align}
Q_{\nu_k}(z) &= \frac{\pi^{1/2}\,{\Gamma}(1+\nu_k)}{2^{1+\nu_k}\,\Gamma(3/2+\nu_k)}\,z^{-1-\nu_k}\,F\left(1+\frac{\nu_k}{2},\frac{1}{2}+\frac{\nu_k}{2};\frac{3}{2}+\nu_k;\frac{1}{z^2}\right),
\end{align}
which yields
\begin{align}
Q_{\nu_k}'(z)\simeq -\frac{\pi^{1/2}\,(1+\nu_k)\,{\Gamma}(1+\nu_k)}{2^{1+\nu_k}\,\Gamma(3/2+\nu_k)}\,z^{-2-\nu_k} + {\cal O}(z^{-4-\nu_k})
\end{align}
So, at large $X$, 
\begin{align}\label{ee55}
(1+X^2)\,Q_{\nu_k}'({\rm i}\,X) \simeq {\rm e}^{-{\rm i}\,\nu_k\,\pi/2}\,\frac{\pi^{1/2}\,(1+\nu_k)\,\Gamma(1+\nu_k)}{2^{1+\nu_k}\,\Gamma(3/2+\nu_k)}\,X^{-\nu_k}.
\end{align}

It is clear from a comparison of Eqs.~(\ref{in}), (\ref{out}), (\ref{bin}), (\ref{bout}), (\ref{ee52}), and (\ref{ee55}) that
$\hat{B}_{L\,k} = a_{LL}\,\hat{A}_{L\,k}$ and 
$\hat{B}_{S\,k}= a_{SL}\,\hat{A}_{L\,k}$,
where
\begin{align}
a_{LL} &= \frac{2^{1+\nu_k}\,\Gamma(1+\nu_k/2)\,\Gamma(3/2+\nu_k)}{(1+\nu_k)\,\Gamma(1/2+\nu_k/2)\,\Gamma(1+\nu_k)},\\[0.5ex]
a_{SL}&=- \frac{2^{\nu_k}\,\nu_k\,\Gamma(1/2+\nu_k/2)\,\Gamma(3/2+\nu_k)}{\Gamma(1+\nu_k/2)\,\Gamma(1+\nu_k)}.
\end{align}
In the limit $\nu_k\rightarrow 0$,
\begin{align}\label{ee32}
a_{LL}&\rightarrow 1,\\[0.5ex]
a_{SL} &\rightarrow - \frac{\nu_k\,\pi}{2}.\label{ee33}
\end{align}

Now, according to Eq.~(\ref{ee52}), 
\begin{align}\label{ee52a}
(1+X^2)\,Q_{-1-\nu_k}'({\rm i}\,X)&\simeq {\rm e}^{\,{\rm i}\,(1+\nu_k)\,\pi/2}\left[\frac{\pi^{1/2}\,\Gamma(1/2-\nu_k/2)}{\Gamma(-\nu_k/2)}-
\frac{\pi^{1/2}\,\nu_k\,(1+\nu_k)\,\Gamma(-\nu_k/2)}{2\,\Gamma(1/2-\nu_k/2)}\,X\right]
\end{align}
at small $X$. 
Furthermore, according to Eq.~(\ref{ee55}), 
\begin{align}\label{ee55a}
(1+X^2)\,Q_{-1-\nu_k}'({\rm i}\,X) \simeq- {\rm e}^{\,{\rm i}\,(1+\nu_k)\,\pi/2}\,\frac{\pi^{1/2}\,\nu_k\,{\Gamma}(-\nu_k)}{2^{-\nu_k}\,\Gamma(1/2-\nu_k)}\,X^{1+\nu_k}
\end{align}
at large $X$. 
A comparison of Eqs.~(\ref{in}), (\ref{out}), (\ref{bin}), (\ref{bout}), (\ref{ee52a}), and (\ref{ee55a}) yields
$\hat{B}_{L\,k} = a_{LS}\,\hat{A}_{S\,k}$ and 
$\hat{B}_{S\,k}= a_{SS}\,\hat{A}_{S\,k}$, 
where
\begin{align}
a_{LS} &= -\frac{\nu_k\,\Gamma(1/2-\nu_k/2)\,\Gamma(1/2-\nu_k)}{2^{1+\nu_k}\,\Gamma(1-\nu_k)\,\Gamma(1-\nu_k/2)},\\[0.5ex]
a_{SS}&= \frac{(1+\nu_k)\,\Gamma(1-\nu_k/2)\,\Gamma(1/2-\nu_k)}{2^{\nu_k}\,\Gamma(1/2-\nu_k/2)\,\Gamma(1-\nu_k)}.
\end{align}
In the limit $\nu_k\rightarrow 0$,
\begin{align}\label{ee38x}
a_{LS}&\rightarrow -\frac{\nu_k\,\pi}{2},\\[0.5ex]
a_{SS} &\rightarrow 1.\label{ee39}
\end{align}

\subsection{Matching Formula}\label{match}
A large aspect-ratio tokamak equilibrium is characterized by $|\nu_k|\ll 1$. In this limit, Eqs.~(\ref{ee20}) (\ref{ee32}), (\ref{ee33}), (\ref{ee38x}), and (\ref{ee39}) yield
\begin{equation}
 {\mit\Delta}_{k\,{\rm out}} = {\mit\Delta}_{k\,{\rm in}}+ \frac{\pi\,\nu_k\,r_k}{\delta_k}.
 \end{equation}
It remains to specify the pressure-flattening width, $\delta_k$. Following Refs.~\onlinecite{lut} and \onlinecite{con1}, we identify this width with the critical layer width above
which parallel thermal transport forces the electron temperature to be a flux-surface function. This critical layer width is given by\,\cite{hel,hel1}
\begin{align}
\delta_{d\,k}&= \sqrt{8}\left(\frac{\chi_\perp}{\chi_\parallel}\right)^{1/4}
\frac{r_k}{(r_k\,s_k\,n)^{1/2}},\\[0.5ex]
\chi_\parallel &= \frac{\chi_\parallel^{\rm smfp} \,\chi_\parallel^{\rm lmfp}}{\chi_\parallel^{\rm smfp}+\chi_\parallel^{\rm lmfp}},\\[0.5ex]
\chi_\parallel^{\rm smfp} &= \frac{1.581\,\tau_{ee\,k}\,v_{t\,e\,k}^{\,2}}{1+ 0.2535\,Z_{\rm eff}},\\[0.5ex]
\chi_\parallel^{\rm lmfp}&= \frac{2\,R_0\,v_{t\,e\,k}\,r_k}{\pi^{1/2}\,n\,s_k\,\delta_{d\,k}},\\[0.5ex]
v_{t\,e\,k} &= \left(\frac{2\,T_{e\,k}}{m_e}\right)^{1/2}.
\end{align}
The previous five equations can be solved via iteration. In fact, if we choose $\delta_k= \delta_{d\,k}/(2\sqrt{\pi})$, identify ${\mit\Delta}_{k\,{\rm out}}$ with the
${\mit\Delta}_k$ appearing in the tearing mode dispersion relation (\ref{dispersion}), identify ${\mit\Delta}_{k\,{\rm in}}$ with the $S_k^{1/3}\,\hat{\mit\Delta}_k$
obtained from the solution of the layer equations (see Appendix.~\ref{lmodel}), and identify $\nu_k$ with minus the resistive interchange stability parameter,\cite{ggj,ggj1}
\begin{equation}\label{dr}
D_{R\,k} = \left(\frac{2\,q^2}{s^2}\,r\,\frac{dP}{dr}\right)_{r_k}\left[1-\frac{1}{q_k^{\,2}}+ \frac{q_k^{\,2}\,s_k}{r_k^{\,4}}\int_0^{r_k}
\left(\frac{r^3}{q^2} -2\,R_0^{\,2}\,r^2\,\frac{dP}{dr}\right)dr\right]
\end{equation}
 [rather than $-D_{I\,k}-1/4$---see Eq.~(\ref{dii})],\cite{kot}
then we obtain
\begin{equation}
{\mit\Delta}_k = S_k^{1/3}\,\hat{\mit\Delta}_k - \sqrt{2}\,\pi^{3/2}\,D_{R\,k}\,\frac{r_k}{\delta_{d\,k}}.
\end{equation}
This formula reproduces one given (but, unfortunately,  not derived) in Ref.~\onlinecite{lut} that has been successfully benchmarked against the XTOR code. The final term on the right-hand side
of the previous equation represents the stabilizing effect of average magnetic field-line curvature on tearing modes in tokamak plasmas. 

It should be understood that the local flattening of the pressure profile considered in this section is not real, but rather a proxy for modeling the influence of large parallel thermal conductivity on the so-called 
Glasser curvature stabilization term.\cite{ggj,ggj1,con1} Furthermore,  the neglect of parallel transport in the resistive layer model
described in Appendix~\ref{lmodel} is justified on the assumption that the layer width is much less than the flattening width, $\delta_{k}$. 

\section{Comparison between TJ and STRIDE Codes}\label{comp}
Suppose that there are $K$ rational surfaces in the plasma. 
The TJ toroidal tearing stability code calculates a $K$-dimensional square tearing stability matrix whose elements are denoted $E_{kk'}$, for $k,k'=1,K$, from the
ideal-MHD solution in the outer region.\cite{tj} Similarly, the STRIDE toroidal tearing mode stability code calculates a $K$-dimensional square tearing stablility
matrix whose elements are denoted ${\mit\Delta}_{kk'}$, for $k,k'=1,K$.\cite{aglas1} As described in Ref.~\onlinecite{tj}, the $E_{kk'}$ matrix is necessarily Hermitian,
otherwise toroidal electromagnetic angular momentum would not be conserved. On the other hand, the ${\mit\Delta}_{kk'}$ matrix is not generally Hermitian. 
Fortunately, it is possible to transform the ${\mit\Delta}_{kk'}$ into an Hermitian matrix, whose elements are denoted $\hat{\mit\Delta}_{kk'}$, 
that is equivalent to the $E_{kk'}$ matrix calculated by the TJ code. 

The normalized  equilibrium poloidal magnetic flux is defined as ${\mit\Psi}(r)= \int_0^r f(r')\,dr'$. Let ${\mit\Psi}_a={\mit\Psi}(a)$ be the total flux enclosed in the plasma. 
If
\begin{align}
\rho(r) &= \frac{r\,f}{{\mit\Psi}_a},\\[0.5ex]
f_{L\,k}& = \left[\rho^{\,\nu_{L\,k}-1}\left(\frac{\nu_{S\,k}-\nu_{L\,k}}{L_{m_k}^{\,m_k}}\right)^{1/2}\,s\,m_k\,{\mit\Psi}_a\right]_{r_k},\\[0.5ex]
f_{S\,k}& = \left[\rho^{\,\nu_{S\,k}-1}\left(\frac{\nu_{S\,k}-\nu_{L\,k}}{L_{m_k}^{\,m_k}}\right)^{1/2}\,s\,m_k\,{\mit\Psi}_a\right]_{r_k},
\end{align}
where all quantities are defined in Sects.~\ref{coord}, \ref{equilb}, and \ref{rational}, then
\begin{equation}
\hat{\mit\Delta}_{kk'} = \cos[(k+k')\,\pi]\,f_{S\,k}\,{\mit\Delta}_{kk'}/f_{L\,k'}.
\end{equation}
Note that the STRIDE calculation must be performed with PEST coordinates (because these are the coordinates used by TJ). Finally, the
STRIDE code needs to be run using the following non-default parameters: \verb|tol_r| = \verb|tol_nr| $=10^{-12}$ and \verb|sas_flag| = \verb|False|.

\newpage

\end{document}